\documentclass[11pt,english]{article}
\usepackage{import}

\usepackage[utf8]{inputenc}

\usepackage[a4paper,       
            bindingoffset=0.0in,
            left=0.75in,
            right=0.75in,
            top=0.75in,
            bottom=0.75in,
            footskip=.25in]{geometry}

\usepackage{xcolor}
\usepackage{amsmath}
\usepackage{centernot}  
\usepackage{amsthm}    
\usepackage{amssymb}  
\usepackage{bm}       
\usepackage{moreverb}
\usepackage{accents}
\usepackage{epstopdf}
\usepackage{float}
\restylefloat{table}
\usepackage{breakcites}
\usepackage{blindtext}
\usepackage[parfill]{parskip}
\usepackage{multirow} 
\usepackage{booktabs} 
\usepackage{threeparttable} 
\usepackage{array}          
\usepackage{hyperref}
\usepackage{authblk}
\usepackage[backend=bibtex,style=numeric, sorting=none]{biblatex} 
\usepackage{breqn}
\usepackage{soul}
\usepackage{xcolor}
\usepackage{algorithm}
\usepackage{algpseudocode}
\usepackage{enumitem} 

\usepackage{makecell}

\usepackage{caption}

\usepackage{tcolorbox}

\usepackage{titlesec}
\titleformat{\paragraph}
{\normalfont\normalsize\bfseries}{\theparagraph}{1em}{}
\titlespacing*{\paragraph}
{0pt}{3.25ex plus 1ex minus .2ex}{1.5ex plus .2ex}

\title{\LARGE{ Latent class multivariate probit and latent trait models for evaluating test accuracy without a gold standard: A simulation study }}

\author[1]{Enzo Cerullo}
\author[2]{Sean Pinkney}
\author[1]{Alex J. Sutton}
\author[1]{Tim Lucas}
\author[1]{Nicola J. Cooper}
\author[3]{Hayley E. Jones}

\affil[1]{\small{Biostatistics Research Group, Department of Health Sciences, University of Leicester, Leicester, UK}}
\affil[2]{Center of Excellence, Omnicom Media Group, New York City, USA}
\affil[3]{Population Health Sciences, Bristol Medical School, University of Bristol, UK}

\date{}

\begin{document}

\maketitle

\section*{\large{ Abstract}}
In the context of an imperfect gold standard, latent class modelling can be used to estimate accuracy of multiple medical tests. 
However, the conditional independence (CI) assumption is rarely thought to be clinically valid. 
Two models accommodating conditional dependence are the latent class multivariate probit (LC-MVP) and latent trait models. 
Despite LC-MVP’s greater flexibility - modelling full correlation matrices versus the latent trait’s restricted structure - 
the latent trait has been more widely used. No simulation studies have directly compared these two models.

We conducted a comprehensive simulation study comparing both models across five data generating mechanisms: 
CI, low-heterogeneity (latent trait-generated), and high-heterogeneity (LC-MVP-generated) correlation structures. 
We evaluated multiple priors, including novel constrained correlation priors using Pinkney’s method that preserves prior interpretability.
Models were fit using our BayesMVP R package (\url{https://github.com/CerulloE1996/BayesMVP/}), 
which achieves GPU-like speed-ups on these inherently serial models.

The LC-MVP model demonstrated superior overall performance.
Whilst the latent trait model performed acceptably on its own generated data, 
it failed for high-heterogeneity structures, sometimes performing worse than the CI model. 
The CI model did badly for most dependent structures. We also found ceiling effects: 
high sensitivities reduced the importance of correlation recovery, 
explaining paradoxes where models achieved good performance despite poor correlation recovery.

Our results strongly favour LC-MVP for practical applications. 
The latent trait model’s severe consequences under realistic correlation structures make it a more risky choice. 
However, LC-MVP with custom correlation constraints and priors provides a safer,
more flexible framework for test accuracy evaluation without a perfect gold standard.

\section*{\large{Keywords}}
Simulation study, 
test accuracy,
imperfect gold standard, 
latent class, 
multivariate probit, 
latent trait. 

*Corresponding Author

Email address: enzo.cerullo@bath.edu







\newpage

\section{ Background}
\label{section_background}
Evaluating test accuracy without a perfect gold standard is challenging. 
A common approach is to use latent class models (LCMs). 
These models estimate the sensitivity, specificity and disease prevalence by modelling the overlap between multiple test results,
without requiring an assumption that any one of the tests used is perfect. 
It is assumed that all the tests are measuring the same underlying disease, 
where each individual is modelled as belonging to one of two classes - 'diseased' or 'non-diseased'.
A widely-used variation of LCMs for test accuracy is a model which we will refer to in this paper as the "traditional" LCM (TLCM). 
This model was proposed by Hui \& Walter in 1980 \supercite{Hui1980} - 
and assumes that test results are conditionally independent 
(CI) - that is, conditional on the true disease status, the test results are uncorrelated to each other. 
This model was then later extended by Vacek \supercite{Vacek1985} to model the conditional dependence (CD) between tests. 
It is important to model this dependence, since the CI assumption is rarely - if ever - thought to be valid in clinical practice. 
Unfortunately, TLCMs have some major limitations. 
For instance:
(i) Modelling the CD between three or more tests is very difficult, and generally does not work very well\supercite{Wang2017, Lurier_et_al_2021_Thibaut,Torrance_Rynard_1997_CD_DTA}; 
(ii) To our knowledge, they cannot be extended to model ordinal tests (e.g. questionnaires for depression) in any coherent way, especially whilst also
modelling CD, as noted by others \supercite{Xu2013}.
However, note that in this paper we will be focusing just on binary test accuracy data - 
we discuss future work as well as our future developments for ordinal tests in the discussion (section \ref{section_discussion}).

More advanced latent class models can be used to more appropriately model CD when there are three or more more tests, 
two of which include the latent class multivariate probit 
(LC-MVP; Xu et al, 2009\supercite{Xu2009}; Xu et al, 2013\supercite{Xu2013}; 
Ueabersax et al, 1996\supercite{Uebersax}; Cerullo et al, 2022\supercite{cerullo_meta_ord}) - 
which is a latent class extension of the multivariate probit (MVP) regression model for correlated binary data 
(Chib \& Greenberg, 1998 \supercite{Chib_Greenberg_MVP_1998} ; Greene, 2012 \supercite{Greene2012})
and the latent trait model
(Qu et al, 1996 \supercite{Qu1996}, Hadgu \& Qu \supercite{Qu1998}, 1998\ Dendukuri \& Joseph, 2001 \supercite{dendukuri2001}).
Unlike the TLCM and its variations, which use aggregate-level data, both the LC-MVP and latent trait models use the individual-level binary data.
For test accuracy research, the latent trait model has been used much more than the LC-MVP model.
This is despite the fact that the LC-MVP model is more flexible (particularly when there are 4 or more tests) -
as it models the full within-class correlation matrices - whereas the latent trait model assumes a restricted correlation matrix by using less parameters.

An issue both the latent trait and the LC-MVP models have (compared to the TLCM) is that they are:
(i) computationally intensive, and therefore often not practical to fit; and 
(ii) there has until now been a lack of software available to efficiently fit these models.  
To address this, we implemented both the LC-MVP and latent trait models using an efficient, 
adaptive Hamiltonian Monte Carlo 
(HMC \supercite{DUANE_et_al_1987_hybrid_Monte_Carlo_HMC_original_paper, betancourt_2018_conceptual_intro_to_HMC}) algorithm, 
which is up to several orders of magnitude more efficient than Stan\supercite{Carpenter2017}, 
and over 10-fold more efficient than the highly optimised, models-specific Mplus software\supercite{mplus, MplusAutomation} at fitting these models. 
These GPU-like speed-ups (for a model which fundamentally cannot be efficient on a GPU because of substantial serial dependencies) relative to Stan 
were achieved in several ways, such as:
(i) using manual-gradients for the log-likelihood, 
(ii) cache-aware chunking, which splits up the gradient function into "chunks",
and dynamically detects and adjusts the number of chunks according to the users CPU’s L3 cache size;
(iii) custom AVX-512/AVX2 SIMD intrinsics math functions;
We implemented these methods into an R package - BayesMVP (\url{https://github.com/CerulloE1996/BayesMVP/}) - 
which enabled us to conduct the simulation study in this paper.

It is not known which of the two aforementioned models - the latent trait and the LC-MVP - generally performs better, 
or more specifically when one model might perform better than the other.
This is because - to our knowledge - to date there have been no simulation studies directly comparing the two models. 
This could be partly or entirely due to the aforementioned computational issues.
The only simulations studies on these two models - for the context of test accuracy - 
either look at their performance in isolation across just a few limited scenarios \supercite{Xu2009, Xu2013}, 
or compare them to other models but have other important limitations  \supercite{Keddie2023, koukounari_lca_2021}.  
To be more specific, in their simulation study 
(which was done as part of a paper which proposed a frequentist LC-MVP model with general correlation structures), 
Xu et al \supercite{Xu2009} compared the general LC-MVP model to an LC-MVP model with a so-called "block structure" 
(not a typical block structure since in their model, the correlations can still vary within each block), 
and to a (very) restricted variation of the latent trait model proposed by Hadgu \& Qu \supercite{Qu1998}, and - unsurprisingly - 
they found the LC-MVP models to perform best. 
However, we do not think this is a particularly fair or relevant comparison. 
The variation of the latent trait model they used forces all of the correlations in each latent class to be identical,
and indeed they note this limitation, but then do not mention why they did not compare it to the original, 
more flexible "standard" latent trait model\supercite{Qu1996}.
Then, in 2013 Xu et al \supercite{Xu2013} also compared an ordinal extension of the LC-MVP model to the TLCM 
and the log-linear LCM\supercite{Xu2013}, and found the LC-MVP performed best on conditionally dependent data. 
However, this study was very limited, since they only considered one data generating mechanism. 
Furthermore, both of the aforementioned studies\supercite{Xu2013, Xu2009} used frequentist methods. 
This implicitly implies flat priors on all parameters, which severely limits the use of these models since they have identifiability issues, 
particularly with smaller $N$ and/or when trying to estimate all parameters, so often unrealistic restrictions need to be made.

For the latent trait model, there has been some more extensive simulation studies conducted; 
namely Keddie et al, 2023\supercite{Keddie2023} and Koukounari et al, 2021\supercite{koukounari_lca_2021}. 
The former\supercite{Keddie2023} evaluated the performance of the latent trait model across a range of assumptions 
(e.g. CD in disease-positive only, disease-negative only, or both and of equal magnitude in both classes).
However, their study had some major limitations (which were mostly noted by the authors), including: 
(i) They used the popular, but poorly-performing and inefficient parameterisation of the latent trait model 
(see section \ref{section_models_LC_LT_latent_trait_LC_MVP_relationship} for more details on alternative parametrisations), 
which led to poor convergence and/or a high proportion of divergences for the most flexible model they looked 
at, which modelled CD in both classes; limiting their conclusions for this model, 
and overall it's likely that this limited the number of scenarios they could consider as it would simply take too long; 
(ii) The prior they used on the "random-effect" parameters ($\text{gamma}(1,1)$) implies a median and $95\%$ prior interval of 
$0.23 ~ (0.00, 0.82)$ on the within-study correlations, 
but the authors do not appear to have considered this and did not justify why it was appropriate to only consider this prior, 
especially one with such a low prior median and extremely thick tail;
(iii) overall, the conclusions they could draw from their study were somewhat limited - 
in particular they concluded that assuming CI (when CD exists) can result in biased estimates of accuracy (and poor coverage) - 
but this was already known. 
Koukounari et al, 2021 \supercite{koukounari_lca_2021} conducted an extensive study on the latent trait and compared it to other models, 
including the "finite-mixture" LCM and a standard CI LCM\supercite{Hui1980, Vacek1985}.
However, despite being a very well-designed study, unfortunately after inspecting their code the latent trait model was implemented incorrectly;
namely, the "random-effect" parameter - which must not vary between tests for CD to be modelled - 
did in fact vary between tests. 
This means any conclusions they draw about this model cannot be interpreted.

To address these issues, in this paper we:
(i) performed an extensive simulation study that evaluated both the latent trait and LC-MVP models across a wide range of prior distributions;
(ii) compared these models to one another using "approximately equivalent" priors to enable fair comparison;
(iii) implemented Pinkney et al's (Pinkney et al, 2024\supercite{Sean_Pinkney_2024_shortnoteflexiblecholesky}) 
novel correlation matrix parameterisation method into our BayesMVP R package\supercite{Cerullo_BayesMVP_2025}.
This provides users with substantial modelling flexibility,
allowing users to restrict the element-wise correlations of the LC-MVP model to be be positive (or within any specified interval), 
without distorting the prior distribution - unlike other methods.
It also allows us to make a much fairer comparison between the LC-MVP and latent trait models, 
since the latter inherently assumes positive correlations. 
This paper is structured as follows. 
In section \ref{section_models} ("Models"), we define and give background knowledge of the latent trait and LC-MVP models, and  
in section \ref{section_design} ("Design"), we give a detailed overview of the design of our simulation study, 
which was planned using the "ADEMP" structure (Morris et al, 2019\supercite{morris_using_2019}).
Then, in section \ref{sections_implementation} ("Implementation \& computing"), we discuss how we implemented the models, and in
section \ref{section_results} ("Results"), we present the results of our simulation study. 
Finally, in section \ref{section_discussion} ("Discussion"), we discuss the results we present in section 5, as well as future work which needs to be done.

\section{ Models}
\label{section_models}
For each of the two models below, 
we will assume that $ n \in \{1, \hdots, N\}$ is an index for individual 
and $ t \in \{1, \hdots, T \}$ an index for each test, 
and each individual has been assessed all $T$ tests
(note that missing data is easy to incorporate for either model, but is beyond the scope of this paper), 
so that there are a total of $NT$ observed data points.
Each individual has an observed binary outcome vector of length $T$ defined by:
$ \underline{Y}_{n}'$ = $\left(Y_{n,1} ~ \hdots ~ Y_{n,T} \right) $, 
where each 
$ Y_{n, t} \in \{0, 1\} $.
Furthermore, let $ d = d_{n} \in \{1, 2\} $ denote the latent class each individual belongs to, 
where $ d = 0 $ denotes the disease-free class and $ d = 1 $ denotes the diseased class.
Finally, we will use \underline{underlining} to denote column vectors 
(e.g., $ \underline\gamma = \left(\gamma_{1}, \hdots, \gamma_{N} \right) $ 
in section \ref{section_models_LC_LT_latent_trait} below) and \textbf{bold font} to denote matrices 
(e.g., the correlation matrices of the LC-MVP model - $\boldsymbol{\Omega}^{[d]}$ - 
in section \ref{section_models_LC_MVP_multivariate_probit} below). 
\subsection{ The latent class latent trait model}
\label{section_models_LC_LT_latent_trait}
The latent trait model is defined as follows. The individual-level sensitivity, specificity, 
and false-positive rate (Fp = 1 - specificity) for individual $n$ and test $t$ are given by:
\begin{equation}
\begin{aligned}
\text{Se}_{n, t} & = 
\Phi\left( {a_{t}^{[2]}}  +   {b_{t}^{[2]}} \cdot  \gamma_{n} \right) ~   \\
\text{Fp}_{n, t} & = 
\Phi\left( {a_{t}^{[1]}}  +   {b_{t}^{[1]}} \cdot  \gamma_{n} \right) ~ ,  \gamma_{n} \sim \text{normal}\left(0, 1\right).  
\end{aligned}
\label{LC_LT_standard_param_individual_level_Se_and_Sp}
\end{equation}
Where $ \Phi\left( \cdot \right) $ denotes the cumulative distribution function (CDF) of the standard normal distribution, and 
$   \gamma_{n} \sim \text{normal}\left(0, 1\right)   $
is the so-called "random-effect" term which varies from subject-to-subject. 
Note that it is important to ensure that $\underline\gamma$ only varies between individuals and not between tests; 
otherwise, the conditional dependence between tests will not be modelled correctly.

The log-likelihood for the latent trait model (when parameterised the standard way as above) is given by:
\begin{equation}
\log{L\left(\underline\Theta |\mathbf{Y}\right)} =
\sum_{n=1}^{N}
\log{\left[ ~ p \prod_{t = 1}^{T}   \Phi\left( {a_{t}^{[2]}}  +   {b_{t}^{[2]}} \cdot  \gamma_{n} \right)  ~
+ ~(1-p)        \prod_{t = 1}^{T}   \Phi\left( {a_{t}^{[1]}}  +   {b_{t}^{[1]}} \cdot  \gamma_{n} \right)  ~  \right]}
\label{LC_LT_standard_log_likelihood}
\end{equation}
Where $p $ denotes the latent class membership probability of the second class (i.e. the disease prevalence), 
and $\underline\Theta$ denotes the vector of model parameters. The \textbf{log-posterior} function is then given by:
$
\log{\pi\left(\underline\Theta |\mathbf{Y}\right)} =
\log{L\left(\underline\Theta |\mathbf{Y}\right)}  + 
\log{\pi\left(\underline\Theta \right)} 
$,
where $ \pi\left(\underline\Theta \right) $ is the density of the prior distributions.
We will assume independent priors and discuss the choice of $ \pi\left(\underline\Theta \right) $ in section  \ref{section_design_methods}.
The overall test accuracy measures for this model are given by:
\begin{equation}
\begin{aligned}
\text{Se}_{t} & = 
\Phi\left( \frac{ {a_{t}^{[2]}}  }{  \sqrt{1 +   \left({b_{t}^{[2]}}\right)^{2} }  }   \right) ~ , ~ 
\text{Sp}_{t} & = 
\Phi\left( \frac{ - {a_{t}^{[1]}}  }{  \sqrt{1 +   \left({b_{t}^{[1]}}\right)^{2} }  }   \right) ~   \\  
\end{aligned}
\label{LC_LT_standard_param_summary_level_Se_and_Sp}
\end{equation}

An issue which is seldom mentioned in the literature is that despite being a relatively simple model to implement - 
using e.g. JAGS (Plummer et al, 2003 \supercite{JAGS_ref_Plummer_2003}) or Stan (Carpenter et al, 2017 \supercite{stan}) - 
in practice this model (as parameterised above) is far from efficient. 
This is because the elements of the high-dimensional random-effect parameter vector ($ \underline\gamma $) tend to be highly correlated to each other. 
Hence, a re-parameterisation of this model is needed, which we discuss in section \ref{section_models_LC_LT_latent_trait_LC_MVP_relationship} below.

\subsubsection{ Prior modelling}
\label{section_models_LC_LT_latent_trait_priors}
One option for setting priors in the latent trait model is to set them directly on each $ a_{t}^{[d]} $ and $ b_{t}^{[d]} $. 
However, we do not recommend this approach since, 
as can be seen from equation \ref{LC_LT_standard_param_summary_level_Se_and_Sp}, 
the summary accuracy is made up of two parameters, making prior modelling very challenging. 
Another option, which is what we do in the simulation study of this paper
(see section \ref{section_design}), is to set priors directly on each:
$
{\theta_{t}^{[d]}}  =   {a_{t}^{[d]}}  / \sqrt{1 +   \left({b_{t}^{[d]}}\right)^{2} }    
$
and 
$ {b_{t}^{[d]}}  $.
This allows one to set priors directly on the summary-level accuracy parameters, making prior modelling much easier.

We recommend two possible ways of incorporating the aforementioned prior model:
(i) keep the original parameterisation, making  $  {\theta_{t}^{[d]}} $ a \textit{transformed} model parameter, 
then set a prior directly on $  {\theta_{t}^{[d]}} $, 
but include the necessary Jacobian adjustment to account for the change in volume of the transformation from
$  {a_{t}^{[d]}} \mapsto {\theta_{t}^{[d]}} $, which is given by:
$ 
J\left( {a_{t}^{[d]}} \mapsto {\theta_{t}^{[d]}}  \right) = 
\frac{\partial  {\theta_{t}^{[d]}} }{\partial {a_{t}^{[d]}} } = 1 / \sqrt{ 1 +  \left({b_{t}^{[d]}}\right)^{2} } 
$.
(ii) re-parameterise the model in terms of 
$  {\theta_{t}^{[d]}} $ and  $  {b_{t}^{[d]}} $, 
rather than 
$  {a_{t}^{[d]}} $ and  $  {b_{t}^{[d]}} $, making  $  {\theta_{t}^{[d]}} $ a model \textit{parameter}.
In this paper (and hence in our simulation study), we went with \textbf{option (i)}. 
This is because option (ii) - although easy to implement in Stan (or JAGS) - 
is more difficult than option (i) to implement with manual-gradient algorithms, 
since it complicates the gradient derivation (implemented in our BayesMVP R package\supercite{Cerullo_BayesMVP_2025})
when combined with the GHK parameterisation we use (see section \ref{section_models_LC_MVP_GHK_algorithm}).

Finally, we complete the latent trait model specification by specifying the prior model - 
$ \pi\left( \Theta \right) $ - 
specifically, we will set priors on each
$ \theta_{t}^{[d]} $ and $  b_{t}^{[d]} $. 
In this paper, we will use independent normal priors such that: 
$ 
\pi \left(   \theta_{t}^{[d]}  \right) = 
\text{normal\_PDF}\left( 
\theta_{t}^{[d]}  | 
\mu_{ \pi ( \theta_{t}^{[d]}  )  }, 
\sigma_{  \pi ( \theta_{t}^{[d]}  ) }
\right)
$, where
$ \mu_{ \pi ( \theta_{t}^{[d]}  )  } $ and 
$ \sigma_{  \pi ( \theta_{t}^{[d]}  ) } $
denote the prior mean and SD of $ \theta_{t}^{[d]}  $, respectively. 
Alternatively, we can write this more concisely (albeit less formally) as:
$
\theta_{t}^{[d]} \sim 
\text{normal}\left( 
\mu_{ \pi ( \theta_{t}^{[d]}  )  }, \sigma_{  \pi ( \theta_{t}^{[d]}  ) }  
\right)
$,
and similarly for $ b_{t}^{[d]} $:
$
b_{t}^{[d]} \sim 
\text{normal}\left( 
\mu_{ \pi ( b_{t}^{[d]}  )  }, \sigma_{  \pi ( b_{t}^{[d]}  ) }  
\right)
$
(for the remainder of this paper, we will use this more concise notation). 
We discuss the choice of 
$
\{ 
\mu_{ \pi ( \theta_{t}^{[d]}  )  }, 
\sigma_{ \pi ( \theta_{t}^{[d]}  )  }, 
\mu_{ \pi ( b_{t}^{[d]}  )  }, 
\sigma_{ \pi ( b_{t}^{[d]}  )  }
\} 
$
in section \ref{section_design_methods}.

\subsubsection{ Covariates}
\label{section_models_LC_LT_latent_trait_covariates}
In our simulation study, we do not incorporate covariates into either the latent trait model nor the LC-MVP model. 
However, we nonetheless discuss ways to implement them as they are often of interest.
For the latent trait model, there are (at least) three ways to implement covariates:
(i) keep the original model parameterisation - i.e. in terms of 
$  {a_{t}^{[d]}} $ and  $  {b_{t}^{[d]}} $ - 
and put covariates on $  {a_{t}^{[d]}} $ or $  {b_{t}^{[d]}} $, or on both parameters 
(as one could argue that putting them on just one parameter is not a good idea, since the accuracy is made up of two parameters); or
(ii) re-parameterise the model in terms of $  {\theta_{t}^{[d]}} $ and  $  {b_{t}^{[d]}} $ - 
which enables modelling the accuracy and conditional dependence separately -  and then put covariates on $  {\theta_{t}^{[d]}} $; or
(iii) re-parameterise the model as an LC-MVP model (see section \ref{section_models_LC_LT_latent_trait_LC_MVP_relationship}), 
perhaps using the GHK parameterisation (as we do in our simulation study and R package), 
and then set covariates as you would for the LC-MVP model (see section \ref{section_models_LC_MVP_multivariate_probit} for details). 
We recommend either option (ii) and (iii) over option (i), as they make prior modelling and interpretation much easier than (i). 
Taking model efficiency into account, we recommend option (iii) the most.

\subsection{ The latent class multivariate probit (LC-MVP) model}
\label{section_models_LC_MVP_multivariate_probit}
In the general LC-MVP (and also the standard, non-latent class MVP), 
we assume there are $ K $ covariates; specifically we assume that each individual has an $ T $ x $ K $ design matrix $ \mathbf{X}_{n} $ 
(where $K = \sum_{t=1}^{T} K_{t}$ is the total number of covariates and $K_t$ is the number of covariates for the $t$-th outcome),
which is formed by combining each univariate design row-vector for each outcome of individual $n$ 
- 
$ \{ \underline{(X_{n, 1})}', \hdots, \underline{(X_{n, T})}' \}$ -
where each $\underline{(X_{n, t})}' $ is of dimension $ 1 $ x $ K_{t} $ and is defined as: 
$ \underline{({X_{n, t}})}' = \left(X_{n, t}^{1}  \hdots  X_{n, t}^{K_{t}} \right) $
- 
diagonally, such that:
\begin{displaymath}
\begin{aligned}
\mathbf{X_n} = 
\begin{pmatrix}
 ( {X_{n, 1}}' ) & 0 & \cdots &   0 \\ 
 0 & \ddots &   ~ & \vdots \\ 
 \vdots & ~ & \ddots &  0  \\ 
 0  & \cdots & 0 &  ( {X_{n, T}}'  )
\end{pmatrix}
\\ 
\end{aligned}
\end{displaymath}
Let $\underline{\beta}^{[d]}$ denote the full covariate vector of length $K$ in latent class $c$, 
which is formed by stacking all of the individual covariate vectors for outcome $t$
-
$\underline{\beta}_{t}^{[d]} =  \left(\beta_{t}^{1 [d]} ~ \hdots ~ \beta_{t}^{K_{t} [d] } \right) $ 
-
on top of one another so that:
$\underline{\beta}^{[d]} = \left(\underline{\beta}_{1}^{[d]} \hdots \underline{\beta_{T}}^{[d]} \right)  $,
where 
$\underline{\beta}_{t}^{[d]}$ is of dimension $ 1 $ x $ K_{t} $ and 
$\underline{\beta}^{[d]}$ is of dimension $ 1 $ x $ K $.
Note that for the context of test accuracy, it is somewhat common to assume no covariates \supercite{Xu2009, Xu2013, Uebersax, dendukuri2001}, 
in which case $K = T$, and therefore $\mathbf{X_n}$ will be a  $T$ x $T$ diagonal matrix with 1's on the diagonal (i.e., the identity matrix). 
This will be the case for our simulation study.
That being said, it is not rare for there to be covariates, 
and including them in the model can be useful to determine the relationship between test accuracy with patient characteristics, 
hence why we defined the model in this more general form (furthermore, our R package we developed - 
BayesMVP \supercite{Cerullo_BayesMVP_2025} - can include covariates).

We augment each observed data vector $\underline{Y}_{n}'$  with a continuous latent vector 
$\underline{Z}_{n}'$ = $\left(Z_{n,1} ~ \hdots ~ Z_{n,K_{T}} \right)$ 
which is conditional on latent class $d$, 
and has a truncated multivariate normal distribution with mean vector $\underline{X} \underline\beta^{[d]}$, 
and a $T$-dimensional variance-covariance matrix $\mathbf{\Omega}^{[d]}$ - restricted to a correlation matrix (for identifiability):
\begin{equation}
\begin{aligned}
\underline{Z}_{n} \sim \text{multi\_normal}\left(\mathbf{X_n} \underline{\beta}^{[d]}, \mathbf{\Omega}^{[d]} \right)
\end{aligned}
\label{augmented_data_truncated_mv_normal}
\end{equation}
Note that  $\underline{Z}_{n}$ is truncated depending on the observed value $\underline{Y}_{n}$, such that:  
${Y}_{n,t} = 1$ $\implies$ $  {Z}_{n,t} > 0   $, and 
${Y}_{n,t} = 0$ $\implies$ $  {Z}_{n,t} < 0   $.
Conditional on the true disease status of each individual, 
the probability of observing the test response vector $\underline{Y}_{n}'$ is given by:
\begin{equation}
Pr\left(\underline{Y}_{n} | d = d_{n}, \underline\beta^{[d]}, \boldsymbol\Omega^{[d]} \right) 
=
\int_{ LB\left( Y_{n, 1} \right) }^{  UB\left( Y_{n, 1} \right)   }
\hdots
\int_{ LB\left( Y_{n, T} \right) }^{  UB\left( Y_{n, T} \right)   }
\Phi_{T}\left( x |  \underline\beta^{[d]},   \boldsymbol\Omega^{[d]} \right) dx
\label{lc_mvp_prob_response_vec}
\end{equation}
where $ \Phi_{T}\left( x |  \underline\beta^{[d]},   \boldsymbol\Omega^{[d]} \right) $ 
denotes the cumulative distribution function of a $T$-dimensional multivariate normal distribution, 
with mean vector $ \underline\beta^{[d]} $ and variance-covariance matrix $  \boldsymbol\Omega^{[d]}   $. 
Furthermore, 
$LB\left( Y_{n, t} \right)$ and 
$UB\left( Y_{n, t} \right)$
denote the lower and upper truncation bounds within the $t$-th dimension, 
which is conditional on the observed value $Y_{n, t}$ such that:
$ Y_{n, t} = 0 \implies 
LB\left( Y_{n, t} \right)  = - \infty, ~ UB\left( Y_{n, t} \right) = 0 $, 
and:
$ Y_{n, t} = 1 \implies 
LB\left( Y_{n, t} \right)  = 0, ~ UB\left( Y_{n, t} \right) = + \infty $.

Using equation (\ref{lc_mvp_prob_response_vec}), we can write the \textbf{log-likelihood} function as:
\begin{equation}
\log{L\left(\underline\Theta |\mathbf{Y}\right)} =
\sum_{n=1}^{N}
\log{\left[ ~ p  \cdot Pr\left(\underline{Y}_{n} | c_{n} = 1, \underline\beta^{[d]}, \boldsymbol\Omega^{[d]} \right)  ~
+ ~(1-p) \cdot Pr\left(\underline{Y}_{n} |  c_{n} = 0, \underline\beta^{[d]}, \boldsymbol\Omega^{[d]} \right)  ~  \right]}
\label{lc_mvp_log_likelihood}
\end{equation}
Where $p $ denotes the latent class membership probability of the second class
(i.e. the disease prevalence for test accuracy context), and $\underline\Theta$ denotes the vector of model parameters. 
The \textbf{log-posterior} function is then given by:
$
\log{\pi\left(\underline\Theta |\mathbf{Y}\right)} =
\log{L\left(\underline\Theta |\mathbf{Y}\right)}  + 
\log{\pi\left(\underline\Theta \right)} 
$,
where $\pi\left(\underline\theta \right) $ is the density of the prior distributions. 
We will assume independent priors so that:
$\pi\left(\underline\theta \right) = \pi\left(\underline\beta^{[d]} \right) \cdot \pi\left(\mathbf{\Omega}^{[d]} \right) \cdot \pi\left(p \right) $. 
We discuss the choice of priors in sections \ref{section_models_LC_MVP_prior_modelling} and \ref{section_design_methods}.
Note that the summary estimates of sensitivity and specificity (when there are no covariates i.e., intercept-only) are given by:
\begin{equation}
\begin{aligned}
\text{Se}_{t} & = 
\Phi\left(  {\beta_{t}^{[2]}}    \right) ~ , ~ 
\text{Sp}_{t} & = 
\Phi\left(   - {\beta_{t}^{[1]}}      \right) ~   \\  
\end{aligned}
\label{LC_MVP_standard_param_summary_level_Se_and_Sp}
\end{equation}
Where each $K_{t} = 1$, hence:
$
\underline{\beta}_{t}^{[d]} = 
\left(\beta_{t}^{1 [d]} ~ \hdots ~ \beta_{t}^{K_{t} [d] } \right) = 
\beta_{t}^{1 [d]} = 
\beta_{t}^{[d]}
$.

\subsubsection{ Prior modelling}
\label{section_models_LC_MVP_prior_modelling}
Analogously to the $ \theta_{t}^{[d]} $ parameters in the latent trait model 
(see section \ref{section_models_LC_LT_latent_trait}), 
we will set independent normal priors on the intercept parameters - $ \beta_{t}^{[d]} $ - such that:
$
\beta_{t}^{[d]} 
\sim 
\text{normal}\left( 
\mu_{ \pi ( \beta_{t}^{[d]}  )  }, \sigma_{  \pi ( \beta_{t}^{[d]}  ) }  
\right)
$.
For the within-class correlation matrices ($ \boldsymbol\Omega^{[d]} $), 
a natural choice is the Lewandowski-Kurowicka-Joe (LKJ; Lewandowski et al, 2009\supercite{Lewandowski2009}) distribution - 
such that:
$ \boldsymbol\Omega^{[d]} \sim \text{LKJ}\left( \eta^{[d]} \right) $, 
where $ \eta^{[d]} $ is the shape parameter in latent class $d$, and this distribution is defined by:
$ 
\text{LKJ\_PDF}\left( \boldsymbol\Omega^{[d]} | \eta^{[d]} \right)
\propto 
C \cdot \det\left( \left( {\boldsymbol\Omega^{[d]}}\right)^{\eta^{[d]} - 1} \right)
$, 
where $ C $ denotes the constant of proportionality 
(see Lewandowski et al, 2009 \supercite{Lewandowski2009} for a definition and derivation).

The LKJ distribution is essentially a shifted-and-scaled beta distribution with both shape parameters equal to one another 
(i.e., it is symmetric around 0); hence as $ \eta^{[d]} $ increases, the density becomes more concentrated around 0.
However, it is important to note that:
$
\boldsymbol\Omega^{[d]} \sim \text{LKJ}\left( \eta^{[d]} \right) 
\not\rightarrow
\Omega_{i, j}^{[d]} \sim \text{B}\left( \eta^{[d]}, \eta^{[d]} \right)
$. 
In fact, as shown in Lewandowski et al, 2009\supercite{Lewandowski2009}, 
$
\boldsymbol\Omega^{[d]} \sim \text{LKJ}\left( \eta^{[d]} \right) 
\implies
\Omega_{i, j}^{[d]} \sim \text{B}\left( \eta^{[d]} - 1 + T/2, \eta^{[d]} - 1 + T/2 \right)
$. 
Hence, as the number of tests (i.e., dimension) increases, the prior becomes more concentrated around 0, even if we do not increase the shape parameter. 
This is not a drawback of the LKJ distribution per se; it is caused by the positive-definiteness constraint of correlation matrices in general - 
that is, $\boldsymbol\Omega^{[d]}$ becomes more concentrated around 0 as $T$ increases regardless of the prior we use. 
This is because, as $T$ increases, the chance of a random matrix 
(with unit diagonal and all off-diagonal elements being between 0 and 1) being a valid (i.e., positive-definite) correlation matrix decreases,
with those whose elements are further away from 0 generally having a lower probability of being positive-definite than those who have more elements closer to 0.

Overall, the priors we will consider in this paper for $ \boldsymbol\Omega^{[d]} $, 
as well as the priors for the beta parameters 
(i.e., the choice of each: the choice of each
$
\{ 
\mu_{ \pi ( \beta_{t}^{[d]}  )  }, 
\sigma_{ \pi ( \beta_{t}^{[d]}  )  },
\eta^{[d]}
\} 
$),
are discussed in section \ref{section_design_methods}.


\subsubsection{ Correlation structure \& restrictions}
\label{section_models_LC_MVP_Omega_corr_structure_and_restictions}
Recently, Pinkney et al \supercite{Sean_Pinkney_2024_shortnoteflexiblecholesky} proposed a new method for generating correlation matrices,
which is based on previously proposed methods to simulate random correlation matrices 
(Numpacharoen et al, 2012\supercite{Nump_et_al_Numpacharoen_2012_corr_param}; Madar et al, 2015\supercite{Madar2015_corr_param_Schur}).
We implemented this method into our BayesMVP R package \supercite{Cerullo_BayesMVP_2025}.
This method allows one to restrict all (or a subset of) the correlations to be greater than 0 - 
or even more generally between any lower bound and upper bound (for instance between 0.25 and 0.75);
however, for this paper, we only consider restricting each $ \Omega_{i, j} \in (0, 1) $).

Furthermore, unlike other methods used to date - 
such as the default parameterisation used in the probabilistic programming language Stan (Carpenter et al, 2017\supercite{Carpenter2017}) - 
it allows us to do this without distorting the prior distribution.
Besides modeling flexibility, this method also allows a more "fair" comparison to be made between the LC-MVP and the latent trait models, 
since the latter restricts the correlations to be positive.

\subsubsection{ The Geweke, Hajivassiliou and Keane (GHK) parameterisation}
\label{section_models_LC_MVP_GHK_algorithm}
Sampling from the LC-MVP (or even the standard, non-latent-class MVP model \supercite{Chib_Greenberg_MVP_1998, Greene2012}) is difficult. 
This is because the latent normal variables  ($ \underline{Z_{n}} $) can be highly correlated to each other, 
resulting in high auto-correlation (similarly to the $\gamma_{n}$ in the latent trait model; see section \ref{section_models_LC_LT_latent_trait}).
One way to mitigate this is by using the Geweke, Hajivassiliou and Keane (GHK; \supercite{Hajivassiliou_geweke_GHK_1986}) method, 
which re-parameterises these latent normal variables ($ \underline{Z_{n}} $) as latent uniform variables -
$ \underline{u_{n}} = 
\left(u_{n, 1} \hdots u_{n, T} \right) $
where each 
$ u_{n, t} \sim \text{Uniform}\left(0, 1 \right) $ 
and are uncorrelated to one another. 
More specifically, we express each $ \underline{Z_{n}} $ in terms of conditional univaiate truncated-normal distributions, 
where each $ Z_{n, t} $ is conditional on the previous variable (i.e., $ Z_{n, t - 1} $ ). 
We can achieve this by expressing each $ \underline{Z_{n}} $ in terms of the Cholesky factor of the correlation matrix - 
defined as:
$ \boldsymbol\Omega^{[d]} = \mathbf{L}^{[d]} (\mathbf{L}^{[d]})^{'} $.
Finally, we can then use the inverse-CDF method to generate each $ Z_{n, t} $ from uncorrelated standard uniform variates
$ u_{n, t} \sim \text{Uniform}\left(0, 1\right)$. 
In other words, we express each $ Z_{n, t} $ in terms of the uncorrelated parameters, $ u_{n, t} $, 
and then sample from these parameters instead of $Z_{n, t}$. 
A more detailed and formal description of the GHK parameterisation for the LC-MVP model is given in 
\textcolor{red}{REF ALGORITHM/BayesMVP PAPER HERE WHEN ON ARXIV}.

\subsection{ Relationship between the LC-MVP and latent trait models}
\label{section_models_LC_LT_latent_trait_LC_MVP_relationship}
An often overlooked fact is that - as noted by Xu et al \supercite{Xu2009} - 
the latent trait model is a restricted version the LC-MVP model. 
More specifically, the latent trait model is equivalent to an latent trait model with variance-covariance matrix equal to:
\begin{equation}
\begin{aligned}
   \mathbf{\Sigma}^{[d]} & = \mathbf{I} + \underline{b}^{[d]} {\underline{b}^{[d]}}^{\prime} \\ 
                & = 
            \begin{pmatrix}
             1 + {b_{1}^{[d]}}^{2}           &  b^{[d]}_{1} b^{[d]}_{2}     &  \hdots                         &   b^{[d]}_{1} b^{[d]}_{T}    \\ 
             b^{[d]}_{1} b^{[d]}_{2}         & \ddots                       &                                 &   \vdots     \\ 
             \vdots                          &                              & \ddots                          &   b^{[d]}_{T - 1} b^{[d]}_{T}    \\ 
             b^{[d]}_{1} b^{[d]}_{T}         & \hdots                       &  b^{[d]}_{T - 1} b^{[d]}_{T}    &   1 + {b_{T}^{[d]}}^{2}     \\
             \end{pmatrix}                
\end{aligned}
\label{LC_equiv_to_MVP_var_cov_mtx_Sigma}
\end{equation}
It is not known what impact this restriction may have on parameter recovery. 
We hypothesise that, based on the structure of $ \Sigma^{[d]} $, when correlations are highly varied
(e.g. some 0 correlations and other relatively high) that the latent trait model will perform worse than the LC-MVP model. 
It can be seen from \ref{LC_equiv_to_MVP_var_cov_mtx_Sigma} that each correlation is equal to:
$ 
\Omega^{[d]}_{i, j} =  \frac{ b^{[d]}_{i} b^{[d]}_{j} }{ \sqrt{\left(  1 +  {b_{i}^{[d]}}  \right) \left(  1 +  {b_{j}^{[d]}}  \right)    } } 
$.
Note that an additional restriction that the latent trait model has is that, since each $ b_{i} > 0 $, then each $ \Omega^{[d]}_{i, j} > 0 $.

As mentioned in section \ref{section_models_LC_LT_latent_trait}, 
the latent trait model suffers from very poor sampling efficiency when it is parameterised the "standard" way
(i.e. as in section \ref{section_models_LC_LT_latent_trait}). 
However, using \ref{LC_equiv_to_MVP_var_cov_mtx_Sigma}, 
we can express the log-likelihood of the latent trait model as the log-likelihood of the LC-MVP model,
and hence use the GHK algorithm 
(see section \ref{section_models_LC_MVP_GHK_algorithm}) to more efficiently estimate the latent trait model.

\section{Design}
\label{section_design}
\subsection{Aims}
\label{section_design_aims}
In line with the first "ADEMP" (Aims, data generating mechanisms, estimands, methods, and performance measures) 
recommendation from Morris et al, 2019 \supercite{morris_using_2019}, we describe the aims of our simulation study below.  
This study will have two broad aims, and within each broad aim, there are several smaller aims. These aims are defined as follows:

\begin{enumerate}
    \item For the estimation of test accuracy and disease prevalence in the absence of a perfect gold standard, 
    which model performs better - the LC-MVP, or the latent trait)? More specifically:
    \begin{enumerate}
        \item Does one model consistently perform better than the other?
        \item When does the latent trait perform better or equal to the LC-MVP? Should one ever use the latent trait model, given that the more flexible LC-MVP exists?
        \item How does the performance of each model vary with $N$?
    \end{enumerate}
    \item Specifically for the LC-MVP, which \textbf{priors} should one use? More specifically:
    \begin{enumerate}
        \item Is there such thing as good "default" priors to use for the correlation matrices for the LC-MVP? If so, what should it be?
        \item Should we add structural restrictions on the within-class correlation matrices, such as restricting all correlations to be greater than 0 (like the latent trait model does)?
        Or just restricting a subset to be greater than 0 whilst letting other ones vary freely?
    \end{enumerate}
\end{enumerate}

\subsection{Data generating mechanisms (DGMs)}
\label{section_design_DGMs}
In line with the second point of the "ADEMP" structure from Morris et al\supercite{morris_using_2019}, 
in this section we describe the data generating mechanisms we will use in our simulation study. 
The DGMs are summarised in table \ref{Table:DGMs_summary} .
\begin{table}[H]
\centering
\caption{Characteristics of the data generating mechanisms (DGMs)}
\begin{tabular}{ccccc}
\hline
\textbf{DGM} & \textbf{CI/CD?} & \makecell{Within-\\class\\correlation\\structure?} & \makecell{Sensitivity/\\Specificity\\(\%)} & \makecell{Probable\\clinical\\areas?} \\
\hline
1 & CI & None 
& \makecell{(65, 55, 60, 65, 70) / \\(99, 95, 90, 90, 85)} & \makecell{Arguably\\none} \\
\hline
2 & \makecell{CD in D+, \\CD in D- (50\% of D+)} & \makecell{Latent trait\\structure*}
& \makecell{(65, 55, 60, 65, 70) /\\(99, 95, 90, 90, 85)} & \makecell{Viral \&\\infectious\\diseases***} \\
\hline
3 & \makecell{CD in D+,\\CD in D- (50\% of D+)} & \makecell{Highly\\varied**} 
& \makecell{(65, 55, 60, 65, 70) /\\(99, 95, 90, 90, 85)} & \makecell{Viral \&\\infectious\\diseases***} \\
\hline
4 & \makecell{CD in D+,\\CD in D- (50\% of D+)} & \makecell{Latent trait\\structure*} 
& \makecell{(92.5, 86, 87, 91, 86) /\\(95, 81, 70, 67, 85)} & \makecell{Most****\\other\\diseases} \\
\hline
5 & \makecell{CD in D+,\\CD in D- (50\% of D+)} & \makecell{Highly\\varied**} 
& \makecell{(92.5, 86, 87, 91, 86) /\\(95, 81, 70, 67, 85)} & \makecell{Most****\\other\\diseases} \\
\hline
\multicolumn{5}{l}{\makecell[l]{\textbf{Table 1:} Characteristics of the data generating mechanisms (DGMs)\\
*Latent trait structure following $b_t b_{t'} / \sqrt{(1 + b_t^2)(1 + b_{t'}^2)}$, \\
with correlations between $0.223 - 0.578$ in D+ group (SD = $0.12$), \\
and between $0.069 - 0.255$ in D- group (SD = $0.06$). \\
**Correlations between $0.10 - 0.65$ in D+ group (SD = $0.20$), \\
and between $0.05 - 0.325$ in D- group (SD = $0.10$). \\
***For example tuberculosis, COVID-19, veterinary applications of LCMs, etc. \\
****For instance cardiovascular disease (CVD), type 2 diabetes (T2DM), \\
mental health diseases such as depression, etc.}} \\
\hline
\end{tabular}
\label{Table:DGMs_summary}
\end{table}
The following factors will be common across all DGMs:
\begin{enumerate}
    \item \textbf{Disease prevalence, p}: Fixed at $0.20$. This implies that, for $N = 300$, 
    there will be (on average across simulations) around $60$ and $240$ diseased and non-diseased individuals, respectively. 
    For $N = 3000$, there will be around $600$ diseased and $2400$ non-diseased individuals, respectively.
    \item \textbf{Number of tests, T} (i.e., model dimension): We will assume that there are 5 tests being evaluated across all DGMs. 
\end{enumerate}
\subsubsection{DGMs: sensitivity and specificity}
\label{section_design_DGMs_Se_and_Sp}
The first three DGMs (see table \ref{Table:DGMs_summary}) assume accuracy set 1, which has sensitivities of:
$\{65, 55, 60, 65, 70\}$ and specificities of: $\{99, 95, 90, 90, 85\}$.
This accuracy set is similar to the one used in Keddie et al \supercite{Keddie2023}, 
and is consistent with disease areas which have "reference" tests with near-perfect specificity, 
but much lower sensitivity; such as in viral and infectious disease areas 
(e.g. PCR tests for COVID-19 \supercite{Arevalo_Rodriguez_2020_COVID_PCR_Cochrane}, 
culture tests for tuberculosis \supercite{Steingart_2006_TB_culture_systematic_review}) - 
an area where LCMs are commonly used, relative to other clinical areas.
In particular, we used tuberculosis as a motivating example. 
More specifically, test 1 represents culture - the "reference" or "gold standard" test (hence $99\%$ specificity). 
The other tests represent four index tests which often have a relatively high specificity - 
but lower than the reference test - and often have low-moderate sensitivity 
(e.g. serology tests for tuberculosis \supercite{Steingart_2011_TB_serology_WHO}). 
Note that we put the imperfect reference test as the first test intentionally, 
so that it approximately 'matches' the conditional dependence magnitude and pattern 
(see section \ref{sim_study_plan_DGMs_DGPs_CD_structure_pattern}). 
More specifically, we put it as the first test because - using tuberculosis as the motivating example -
research has suggested that (Dendukuri et al, 2001\supercite{dendukuri2001})
we might expect the within-class correlation to be low between culture and serology tests, 
since they work by a different mechanism (see section \ref{sim_study_plan_DGMs_DGPs_CD_structure_pattern}).
However, the conditional dependence between the other 4 tests - which work via the same or similar mechanisms - would be higher.


The last two DGMs assume accuracy set 2, which has sensitivities of:
$\{92.5, 86, 87, 91, 86\}$ and specificities of: $\{95, 81, 70, 67, 85\}$.
This accuracy set is more consistent with other disease areas, 
such as cardiovascular disease (CVD), type 2 diabetes (T2DM) and mental health diseases such as depression. 
In fact, the test accuracy parameters were motivated by literature on screening tools for depression. 
More specifically, a recent meta-analysis from Levis et al \textbf{REF} suggested that the 2-item Patient Health Questionnaire 
had sensitivities and specificities of $67\% (64\%, 71\%)$ and $91\% (88\%, 94\%)$, respectively (in the general population, 
and for cut-off score of $ \ge 2 $); and the 9-item version had a sensitivity and specificity of:
$86\% (80\%, 90\%)$ and $85\% (82\%, 87\%)$, respectively (also in the general population, at a cut-off score of $ \ge 10 $). 
Another meta-analysis from Vilagut et al, 2016 \supercite{Vilagut_2016_CES_D_depression_MA}
found the sensitivity and specificity (also in the general population) of another screening tool used for depression -
the Center for Epidemiologic Studies, Depression (CES-D) - at a cut-off of 16 to be 87\% (82\%, 92\%) and 70\% (65\%, 75\%), respectively. 
Furthermore, another meta-analysis from Topp et al \supercite{Topp_2015_WHO_5_depression_SR} 
found the average sensitivity and specificity of the WHO-5 (at a cut-off of $ \ge 50 $ across a range of populations) to be 0.86 and 0.81, respectively. 
Motivated by this data, for DGMs 6-7 we used the WHO-5, CES-D, PHQ-2 and the PHQ-9 as tests 2, 3, 4 and 5, respectively. 
This ordering was done intentionally, so that it 'matches' the correlation matrix (see section \ref{sim_study_plan_DGMs_DGPs_CD_structure_pattern}). 
For instance, we put the PHQ-2 and PHQ-9 as the last two tests so that when we model CD, 
these two tests will have the highest correlations compared to any other pair of tests - 
this would be expected given that the PHQ-2 is a subset of the PHQ-9. 
It is important to note that the most optimal way of modelling such data is using the full ordinal data and applying an ordinal LC-MVP (LC-MVOP) model, 
which would allow one to summarise accuracy at each threshold and likely yield better estimates of accuracy and prevalence
(see Cerullo et al, 2022\supercite{cerullo_meta_ord} for an example); 
however, this was out of the scope of this paper, 
and we have also not yet implemented these ordinal models into BayesMVP \supercite{Cerullo_BayesMVP_2025} (currently in development).


\subsubsection{DGMs: Conditional dependence across the latent classes}
\label{section_design_DGMs_CD}
The first DGM will assume CI between all 5 tests, which is arguably very unrealistic, 
but is important to have as a baseline as it allows us to fully assess what happens when we model dependence even when none truly exists. 
On the other hand, DGMs 2-5 will assume CD in both groups, 
but with the CD in the disease-free group being less than in the diseased group;
more specifically, the correlations are exactly half for DGMs \#3 and \#5, 
and for DGMs \#2 and \#4 the correlation in the non-diseased group was obtained by first halving the b parameters, 
and then computing the correlation matrix from these).
This is because CD may still be present in the non-diseased group for some tests in the viral \& infectious diseases category.

This decision was motivated by the fact that in some tests (particularly for disease areas other than viral \& infectious diseases), 
there might be substantial correlation in the non-diseased group, which may reach that of the diseased group. 
For example, using the example of depression; patients testing negative may still have some level of depression which
doesn't meet the criteria for diagnosis, resulting in more highly correlated test results in this class.
Furthermore, the near-zero correlations seen in DGMs \#3 and \#5 (e.g., $0.10 - 0.15$ in the diseased group) are likely to still exist,
but may have a different interpretation in the infectious/viral disease areas.
For example, even though both screening tests will work via the same "mechanism" (i.e., asking patients questions), 
they may target distinct facets of depression; for instance, one may primarily target anhedonia 
(the loss of pleasure in daily activities) whilst another may target primarily mood-related symptoms - 
hence the tests end up picking up distinct subsets of patients.

\subsubsection{DGMs: Conditional dependence (CD) magnitude \& pattern}
\label{section_design_DGMs_CD_magnitude_and_pattern}
As can be seen from table \ref{Table:DGMs_summary}, we will consider three different within-class correlation structures, 
one of which is CI (for DGM \#1). 
The other two structures both assume CD, however the pattern (and magnitude) of the correlations varies between the two structures. 
Namely, the "varied" structure is generated from the latent trait model with:
$ \underline{b} = \left( 0.40, 0.75, 1.10, 0.80, 1.25 \right) $
- which means it can also be generated from the LC-MVP model 
(see section \ref{section_models_LC_LT_latent_trait_LC_MVP_relationship}), 
with the following variance-covariance matrix and corresponding correlation matrix:
\begin{equation}
\label{corr_struct_matrix_varied_d}
\begin{aligned}
{\boldsymbol{\Omega}}_{\text{varied}}^{[2]} = 
\begin{pmatrix}
 1.00     & 0.223   & 0.275   & 0.232    & 0.29   \\ 
 0.223    & 1.00    & 0.444   & 0.375    & 0.469  \\ 
 0.275    & 0.444   & 1.00    & 0.462    & 0.578  \\ 
 0.232    & 0.375   & 0.462   & 1.00     & 0.488  \\
 0.29     & 0.469   & 0.578   & 0.488    & 1.00 
\end{pmatrix}
\end{aligned}
\end{equation}
\begin{equation}
\label{corr_struct_matrix_varied_nd}
\begin{aligned}
{\boldsymbol{\Omega}}_{\text{varied}}^{[1]} = 
\begin{pmatrix}
 1.00     & 0.069   & 0.095   & 0.073    & 0.104  \\ 
 0.069    & 1.00    & 0.169   & 0.130    & 0.186  \\ 
 0.095    & 0.169   & 1.00    & 0.179    & 0.255  \\ 
 0.073    & 0.130   & 0.179   & 1.00     & 0.197  \\
 0.104    & 0.186   & 0.255   & 0.197    & 1.00 
\end{pmatrix}
\end{aligned}
\end{equation}
Note that, for DGMs \#2 and \#4 (latent trait structure DGMs), in the non-diseased class,
we obtained the correlation matrix in \ref{corr_struct_matrix_varied_nd} by halving the b values, and then computing the new 
correlation matrix from these.

On the other hand, for DGMs \#3 and \#5 we will assume a different structure -
which we term the "highly varied" structure (see table \ref{Table:DGMs_summary}). 
This structure can only be generated from the LC-MVP model, not the latent trait model - 
and as the name implies, is more varied than the "varied" structure 
(correlations between $0 - 0.65$ and $0 - 0.325$ in the diseased and non-diseased groups, respectively), 
and has the following matrix:
\begin{equation}
\label{corr_struct_matrix_highly_varied_d}
\begin{aligned}
{\boldsymbol{\Omega}}_{\text{highly\_varied}}^{[2]} = 
\begin{pmatrix}
 1         & 0.10   & 0.10    & 0.10     &   0.10   \\ 
 0.10      & 1      & 0.50    & 0.25     &   0.15   \\ 
 0.10      & 0.50   & 1       & 0.40     &   0.35   \\ 
 0.10      & 0.25   & 0.40    & 1        &   0.65   \\
 0.10      & 0.15   & 0.35    & 0.65     &   1  
\end{pmatrix}
\end{aligned}
\end{equation}
\begin{equation}
\label{corr_struct_matrix_highly_varied_nd}
\begin{aligned}
{\boldsymbol{\Omega}}_{\text{highly\_varied}}^{[1]} = 
\begin{pmatrix}
 1         & 0.05    & 0.05    & 0.05     &   0.05   \\ 
 0.05      & 1       & 0.25    & 0.125    &   0.075   \\ 
 0.05      & 0.25    & 1       & 0.20     &   0.175   \\ 
 0.05      & 0.125   & 0.20    & 1        &   0.325   \\
 0.05      & 0.075   & 0.175   & 0.325    &   1  
\end{pmatrix}
\end{aligned}
\end{equation}
\subsection{ Estimands}
\label{section_design_estimands}
The primary estimands are test sensitivity, specificity, and disease prevalence.
As a secondary measure, we assess correlation matrix recovery - but focus on the diseased class ($\Omega^{[d+]}$).
\subsection{ Methods}
\label{section_design_methods}
In line with the fourth point of the "ADEMP" structure from Morris et al\supercite{morris_using_2019},
in this section we describe the methods we used our simulation study. 
We already described both the latent trait and LC-MVP models in section \ref{section_models}  - 
specifically see sections \ref{section_models_LC_LT_latent_trait} and \ref{section_models_LC_MVP_multivariate_probit}
for a detailed description of the latent trait and LC-MVP models, respectively. 
We will now discuss the priors we considered for each of the two models below, 
as well as a method for making the within-study correlation matrix priors for both models approximately equivalent, 
which also makes it much easier for users to implement priors for these parameters in practice. 
\subsubsection{ Test accuracy priors}
\label{section_design_methods_priors_Se_Sp_prev}
We will assume mostly `vague' priors for sensitivity and specificity, and
moderately informative priors on the sensitivity and specificity of the first ("reference") tests, as follows.  

\paragraph{Test accuracy priors for tests 2-5}
\label{section_design_methods_priors_Se_Sp_prev_for_tests_2_to_5}
For all DGMs (see table \ref{Table:DGMs_summary}), 
for the specificity of tests 2-5 we will assume priors which are roughly equivalent to "flat" priors on the probability scale -
i.e. with a 95\% prior interval of $(0.05, 0.95)$ for tests 2-5 sensitivities and specificities. 
For the LC-MVP model (see section \ref{section_models_LC_MVP_multivariate_probit}), since (in the case of no covariates), 
the "beta" parameters are defined simply as the inverse-probit transform of the sensitivity or false positive rate (1-specificity), 
these priors are achieved by setting:
$  \beta^{[d]}_{t}   \sim   \text{normal}\left(0, 1 \right) $,
and for the latent trait model  (see section \ref{section_models_LC_LT_latent_trait}) by setting:
$  \theta^{[d]}_{t}   \sim   \text{normal}\left(0, 1 \right) $
(for $ t = \{2,3,4,5\} $, and $ d = \{0, 1\} $).

\paragraph{Test accuracy priors for reference test}
\label{section_design_methods_priors_Se_Sp_prev_for_tests_2_to_5}
For the reference test (i.e., test \#1), the sensitivity and specificity priors we use will 
depend on the DGM (see table \ref{Table:DGMs_summary} for DGM details).
More specifically, for DGMs 1-3, where the true sensitivity and specificity is $65\%$ and $99\%$, 
we will use priors which correspond to prior medians and $95\%$ prior intervals of:
$ xxx, (0.43, 0.90) $ 
and
$ 0.99, (0.91, 1.00) $ 
for the sensitivity and specificity of the reference test, respectively.

We achieved this by using the following normal priors:
$  \beta^{[2]}_{1}    \sim   \text{normal}\left(0.385, 0.45 \right) $ (for LC-MVP) and
for the sensitivity of test 1 (true value = $0.60$), and: 
$  \beta^{[1]}_{1}    \sim   \text{normal}\left(-2.33, 0.50 \right) $ (for LC-MVP)
for the specificity of test 1 (true value = $0.99$);
replacing $\beta^{[d]}_{1}$ with $\theta^{[d]}_{1}$ for the latent trait model.

On the other hand, for DGMs 4-5 (see table \ref{Table:DGMs_summary}), 
we will assume priors which imply a prior median and 95\% prior interval of:
$ 0.90, (0.63, 0.99) $ 
for {\textit{both}} the sensitivity  (true value = $0.925$) and specificity  (true value = $0.95$) of test 1, respectively.
For the sensitivity, these are achieved by setting:
$  \beta^{[2]}_{1}   \sim   \text{normal}\left(1.30, 0.50 \right) $ (for LC-MVP) and 
For the specificity, these are achieved by setting:
$  \beta^{[1]}_{1}   \sim   \text{normal}\left(-1.30, 0.50 \right) $ (for LC-MVP) - 
replacing $\beta^{[d]}_{1}$ with $\theta^{[d]}_{1}$ for the latent trait model.

\begin{figure}[H]
    \centering
    \includegraphics[width=15cm]{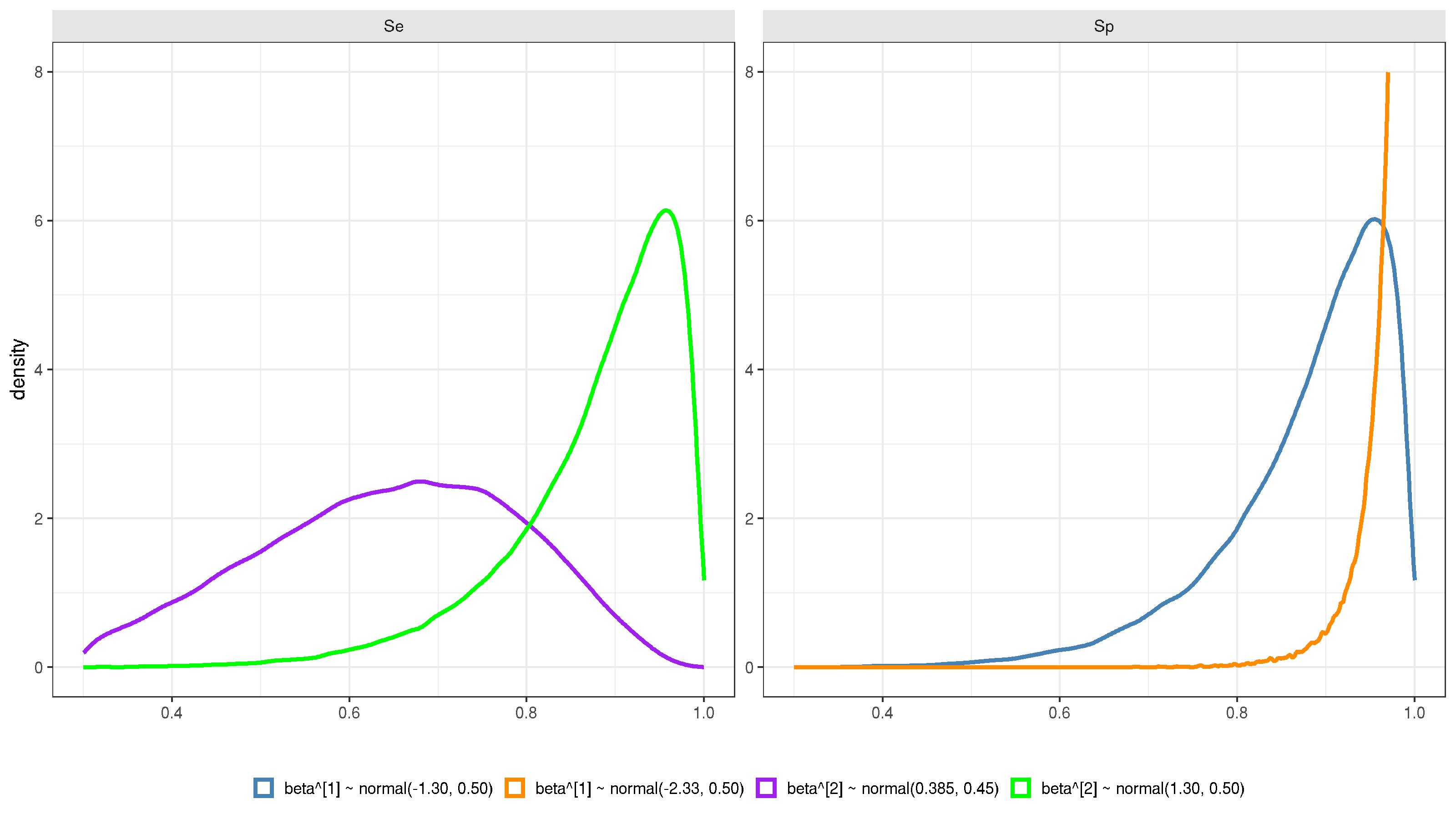}
    \caption{ Plot of informative priors for sensitivity (Se) and specificity (Sp) }
    \label{Figure_Se_Sp_priors}
\end{figure}

We plot these priors in \textbf{figure} \ref{Figure_Se_Sp_priors}.
\subsubsection{ Correlation priors: LC-MVP model}
\label{section_design_methods_priors_MVP_correlation}
For the LC-MVP, the correlation priors (used to address both aims 1 and 2) will come under \textbf{3 broad categories},
with specific prior specifications within each category.
It is important to note that, as mentioned in section \ref{section_models_LC_MVP_prior_modelling}.
all of the priors listed below will vary with the dimension of the problem (T). 

However, for the LKJ prior, one can simulate directly from the distribution as long as the interval is not truncated, in which case:
$ \boldsymbol\Omega^{[d]} \sim \text{LKJ}\left( \eta^{[d]}  \right)$ 
implies that each
$ (\Omega_{i, j}^{[d]} + 1)/2 \sim \text{Beta}\left(\eta + T/2 - 1, \eta + T/2 - 1 \right)$.
On the other hand, for the truncated LKJ distribution (i.e. when each $ \Omega_{i, j} > 0 $), 
the exact marginal is not currently known,;
however, one can also simulate from this directly by using rejection sampling on the non-truncated distribution.

The priors we will consider for the LC-MVP model are 
(note that we obtained 95\% PIs by sampling from a prior-only model using Stan \supercite{stan}):
\begin{enumerate}
    \item Prior set 1: \textbf{Standard LKJ priors} (see section \ref{section_models_LC_MVP_prior_modelling} for more details) - 
    i.e., each $  {\Omega}^{[d]}_{i, j} \in (-1, 1), ~ \boldsymbol{\Omega}^{[d]}  \sim\text{LKJ}\left(\eta^{[d]}\right) $. Note that these are centered around 0. 
    \begin{itemize}
        \item For all DGMs, we will aim to assess this for values of 
        $\{\eta^{[1]}, \eta^{[2]}\}$ in:
        \begin{enumerate}
            \item $\{10, 1.5\}$ - 95\% PI's of $(-0.40, 0.40)$ and $(-0.71, 0.71)$. We will denote this prior set as "LKJ(10, 1.5)" for short.
            \item $\{24, 4\}$   - 95\% PI's of $(-0.27, 0.27)$ and $(-0.55, 0.55)$. We will denote this prior set as "LKJ(24, 4)" for short.
        \end{enumerate}
    \end{itemize}
    \item Prior set 2: \textbf{Truncated LKJ priors} - i.e., each $  {\Omega}^{[d]}_{i, j} \in (0, 1), ~ \boldsymbol{\Omega}^{[d]} \sim\text{LKJ}\left(\eta^{[d]}\right)  $.
        \begin{itemize}
        \item For all DGMs we will aim to assess this prior set for values of:
        $\{\eta^{[1]}, \eta^{[2]}\}$ in:
        \begin{enumerate}
              \item $\{10, 1.5\}$ - 95\% PI's of $ 0.17, (0.01, 0.47)$ and $ 0.37, ( 0.02, 0.82) $. We will denote this prior set as "TruncLKJ(10, 1.5)" for short.
              \item $\{24, 4\}$   - 95\% PI's of $ 0.11, (0.01, 0.33)$ and $ 0.26, ( 0.01, 0.67) $. We will denote this prior set as "TruncLKJ(24, 4)" for short.
        \end{enumerate} 
    \end{itemize}
\end{enumerate}
\begin{figure}[H]
    \centering
    \includegraphics[width=15cm]{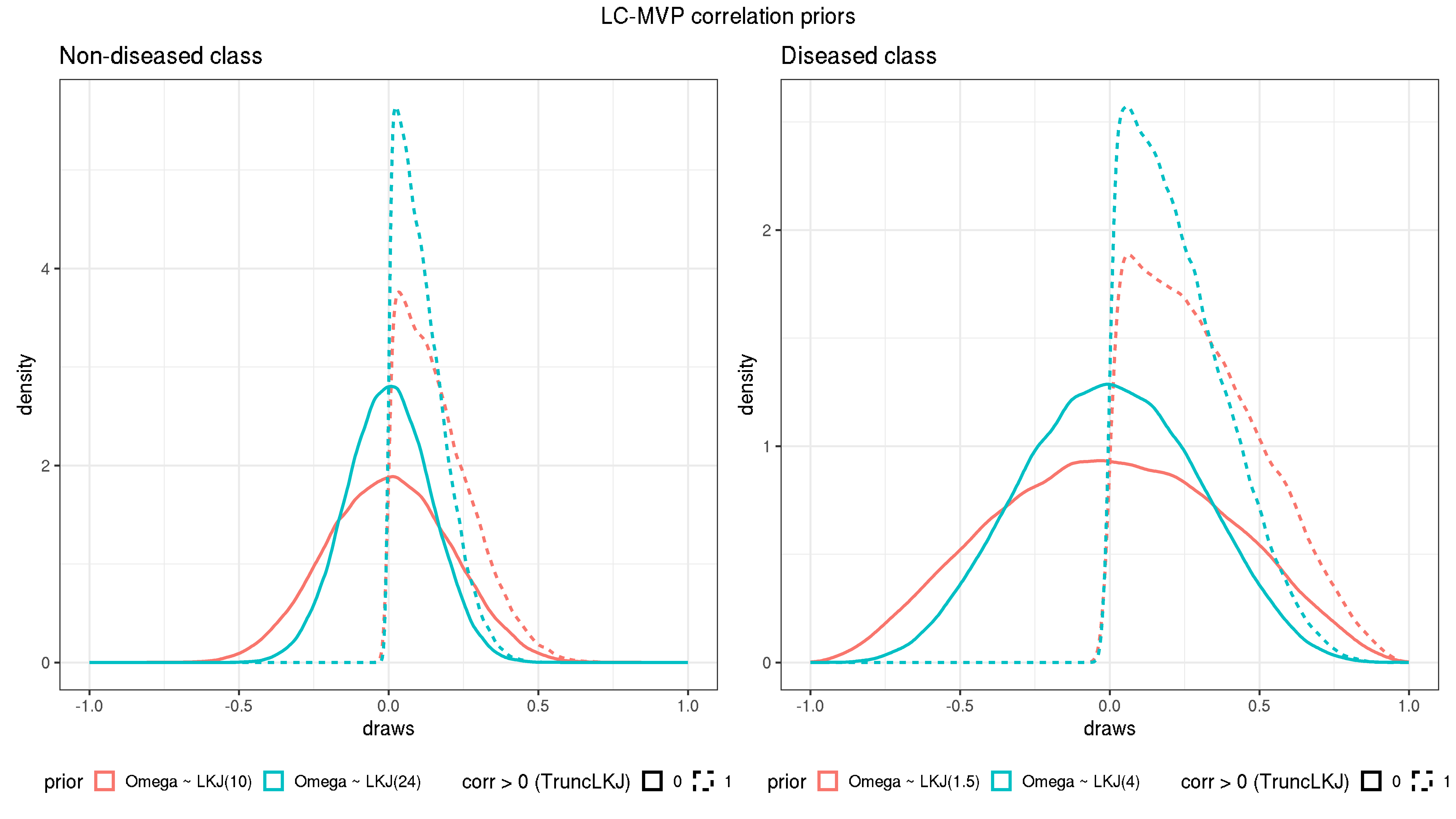}
    \caption{ LC-MVP correlation priors plot (diseased class) }
    \label{Figure_LKJ_and_TruncLKJ_priors}
\end{figure}
\subsubsection{ Correlation priors: latent trait model}
\label{section_design_methods_priors_LT_correlation}
\begin{figure}[H]
    \centering
    \includegraphics[width=15cm]{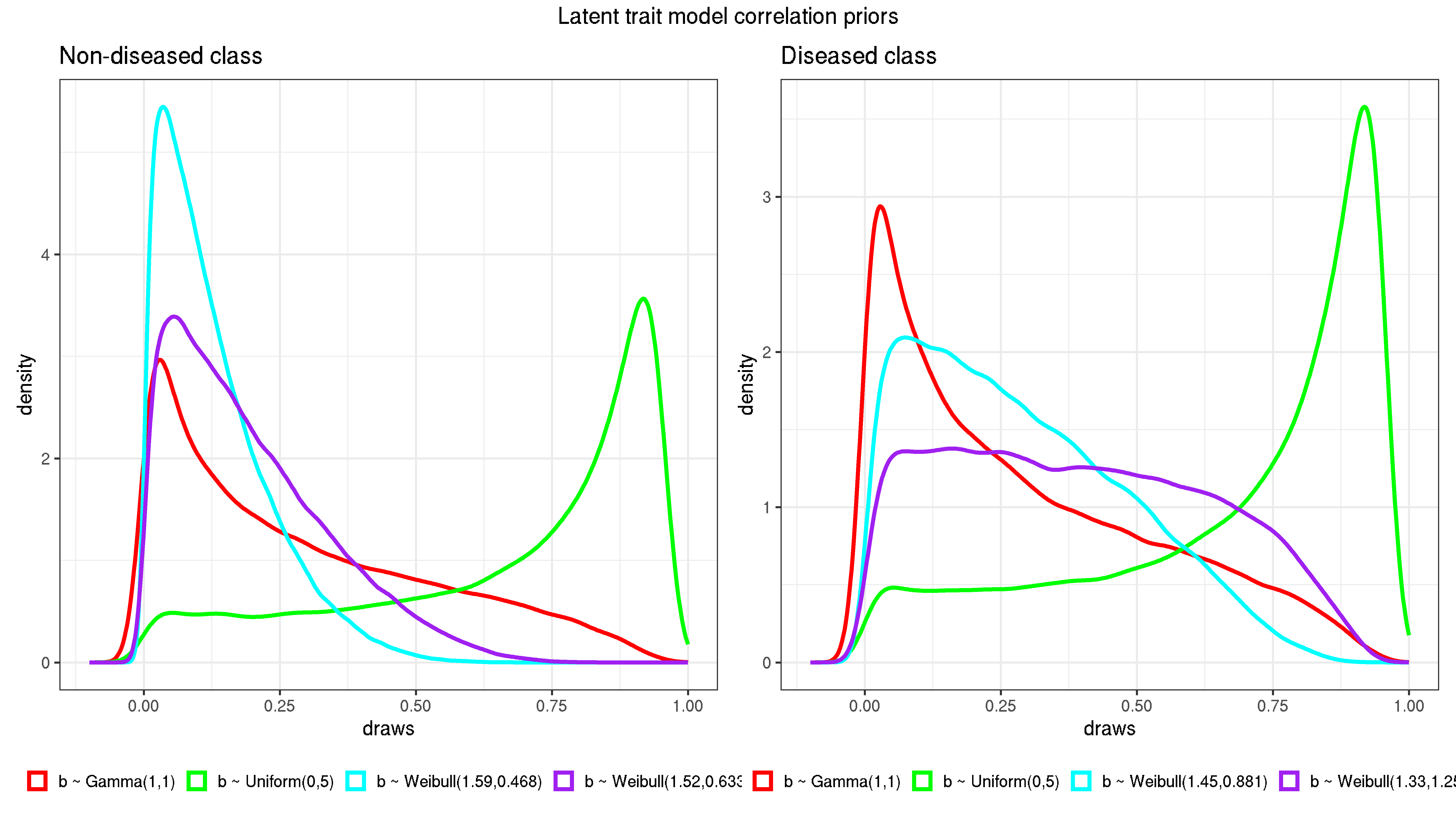}
    \caption{
    Latent trait model priors plot for the $b_{t}^{[d]}$ parameters. \\
    Note: prior density plotted on the correlation scale, i.e.: $b_t b_{t'} / \sqrt{(1 + b_t^2)(1 + b_{t'}^2)}$
    }
    \label{Figure_all_LT_LC_priors}
\end{figure}

For the latent trait model, we will also consider several priors on the within-study correlations (see figure \ref{Figure_all_LT_LC_priors}. 
The first prior is: 
$ b_{t}^{[d]} \sim \text{Gamma}\left(1, 1\right) $ 
- which is the same prior used in Keddie et al \supercite{Keddie2023} - 
which gives a prior median and 95\% PI of 0.23 (0.00, 0.82) on the within-study correlations 
(see section \ref{section_models_LC_LT_latent_trait_LC_MVP_relationship}
for the relationship between the $  b_{t}^{[d]} $ parameters and the implicit within-study correlation matrix).

\begin{figure}[H]
    \centering
    \includegraphics[width=15cm]{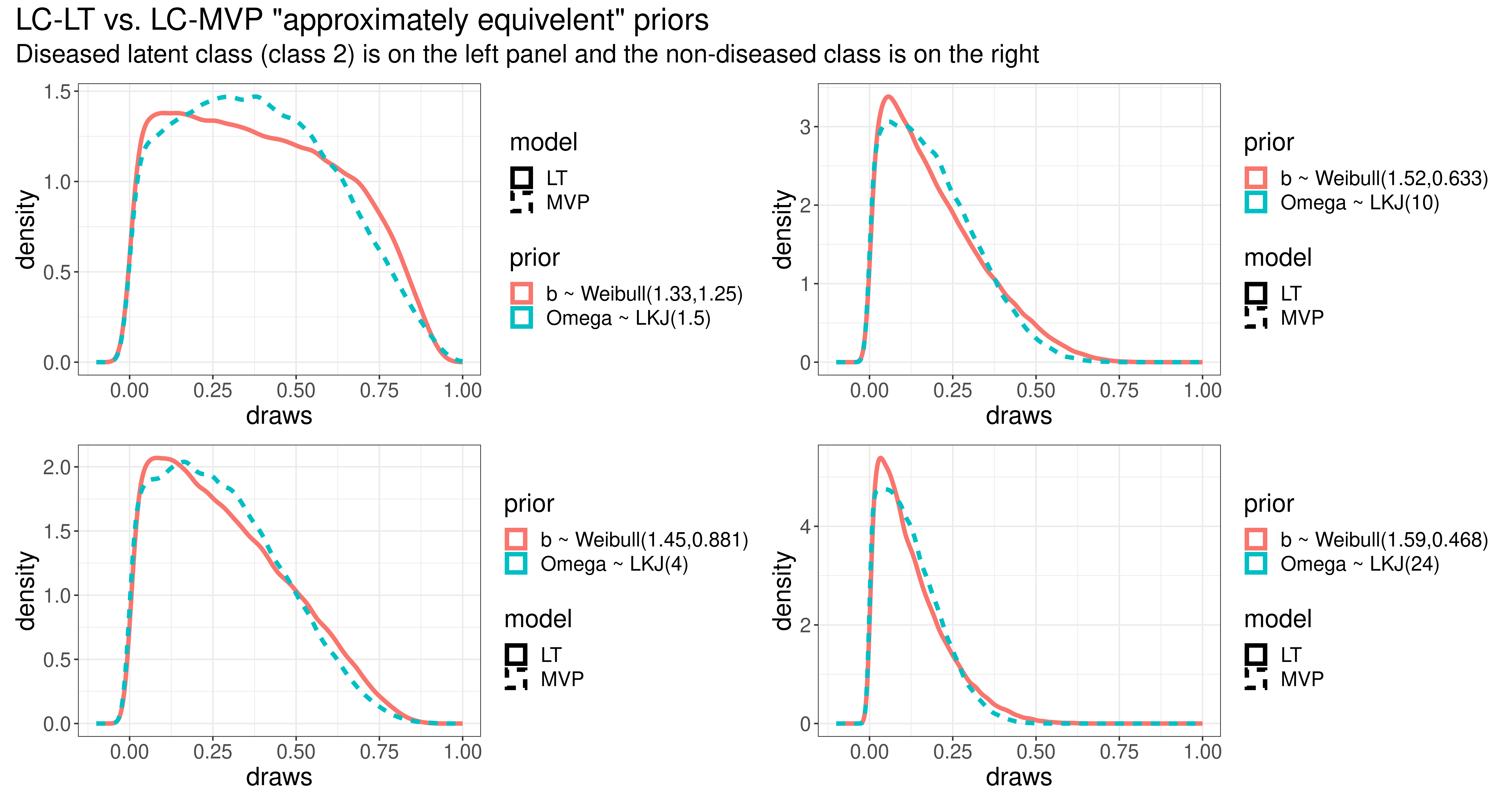}
    \caption{ 
    Latent trait \& LC-MVP TruncLKJ priors plot - "approximately equivalent" correlation priors 
    }
    \label{Figure_LT_LC_vs_LC_MVP_approx_equiv_priors}
\end{figure}

Finally, the last two prior sets we will consider are:
\begin{itemize}
    \item[(i)] $ b_{t}^{[1]} \sim \text{Weibull}\left(1.59, 0.468 \right) $ 
    (prior median for each $\Omega_{i, j}$ = $0.10$ and $95\%$ prior interval of: $(0.01, 0.38)$), and:
    $ b_{t}^{[2]} \sim \text{Weibull}\left(1.45, 0.881 \right)  $ 
    (prior median = $0.26$ and $95\%$ prior interval of: $(0.02, 0.70)$ - 
    approximately equivalent to TruncLKJ(24) and TruncLKJ(4) 
    (defined in section \ref{section_design_methods_priors_MVP_correlation})
    in the non-diseased and diseased classes, respectively.
    We will denote this prior set as "Weibull($\frac{1.59}{1.45}, \frac{0.468}{0.881}$)" for short.
    \item[(ii)] $ b_{t}^{[1]} \sim \text{Weibull}\left(1.52, 0.633 \right) $ 
    (prior median = $0.17$ and $95\%$ prior interval = $(0.01, 0.54)$, and:
    $ b_{t}^{[2]} \sim \text{Weibull}\left(1.33, 1.25 \right) $
    (prior median = $0.38$ and $95\%$ prior interval = $(0.02, 0.84)$ -
    approximately equivalent to the TruncLKJ(10) and TruncLKJ(1.5) 
    (defined in section \ref{section_design_methods_priors_MVP_correlation})
    in the non-diseased and diseased classes, respectively.
    We will denote this prior set as "Weibull($\frac{1.52}{1.33}, \frac{0.633}{1.25}$)" for short.
\end{itemize}
These last two prior sets are "approximately equivalent" to the corresponding LC-MVP TruncLKJ priors 
(see figure \ref{Figure_LT_LC_vs_LC_MVP_approx_equiv_priors}).

To identify latent trait prior parameters that produce correlation distributions approximately equivalent to the LC-MVP's LKJ priors, 
we used a maximum likelihood approach. 
Specifically, we first generated samples from LKJ distributions (e.g., LKJ(1.5), LKJ(10)) representing the correlation distributions in the LC-MVP model.
We then used Stan to find the Weibull distribution parameters ($a$, $b$) for the latent trait model that maximized the likelihood of these samples - 
essentially fitting $\text{LKJ samples} \sim \text{Weibull}(a, b)$ after transforming to the correlation scale.
This approach yielded the approximately equivalent prior specifications shown in Figure \ref{Figure_LT_LC_vs_LC_MVP_approx_equiv_priors}, 
where, for example, correlations from LKJ(10) in the LC-MVP model were well-approximated by Weibull(1.52, 0.633) in the latent trait model.

Note that, we did originally plan to also assess a fourth prior: 
$ b_{t}^{[d]} \sim \text{Uniform}\left(0, 5\right) $,
which has been used and recommended by Dendukuri et al\supercite{dendukuri2001} as a "default" prior because it is supposedly "vague";
however - as can be seen from figure \ref{Figure_all_LT_LC_priors} -
this prior is far from non-informative on the correlations themselves, as it induces a prior median and 95\% prior interval of:
$0.78 ~ (0.05, 0.95)$. 
Hence, this prior will encourage correlations to be higher than they actually are, particularly when $N^{[d]}$ is not large enough to overcome it.
However, the reason why we ended up not using this prior is simply because of MCMC convergence issues - we consistently obtained substantial divergences 
using this prior, which further suggests it likely should not be used as a "default" prior for the latent trait model.

\subsection{ Performance measures}
\label{section_desin_performance_measures}
We assess the following performance measures as suggested in Morris et al\supercite{morris_using_2019}, 
using RMSE of accuracy (sensitivity and specificities) as primary measures:
\begin{enumerate}
    \item \textbf{Root Mean Square Error (RMSE)}:
    $\sqrt{\mathbb{E}\left[(\hat\theta - \theta)^2\right]}$
    \begin{itemize} 
      \item Estimated by: 
               $\sqrt{\frac{1}{N_{sim}} \sum_{i=1}^{N_{sim}} (\hat{\theta}_{i} - \theta)^2}$
      \item Monte Carlo SE is estimated by:
              $\frac{\widehat{\text{RMSE}}}{\sqrt{2 \cdot N_{sim}}}$
   \end{itemize}
   
    \item \textbf{Bias}:
    $\mathbb{E}\left[\hat\theta\right] - \theta$  
    \begin{itemize} 
      \item Estimated by: 
               $\frac{1}{N_{sim}} \sum_{i=1}^{N_{sim}} \hat{\theta}_{i} - \theta$
      \item Monte Carlo SE: as shown above
   \end{itemize}
   
    \item \textbf{Coverage}:
    $\text{Prob} \left( \hat\theta_{\text{low}} \le \theta \le  \hat\theta_{\text{upper}}  \right)$
    \begin{itemize} 
      \item Estimated by: 
               $\frac{1}{N_{sim}} \sum_{i=1}^{N_{sim}}
               \mathbb{I} \left(  \hat\theta_{\text{low}} \le \theta \le  \hat\theta_{\text{upper}}  \right)$
      \item Monte Carlo SE: as shown above
    \end{itemize}
    
    \item \textbf{Interval Width}:
    Mean width of 95\% credible intervals
    \begin{itemize} 
      \item Estimated by: 
               $\frac{1}{N_{sim}} \sum_{i=1}^{N_{sim}} (\hat\theta_{\text{upper},i} - \hat\theta_{\text{low},i})$
    \end{itemize}
\end{enumerate}
\section{ Implementation \& computing using BayesMVP}
\label{sections_implementation}
All models will be run using our BayesMVP R package (Cerullo et al, 2024\supercite{Cerullo_BayesMVP_2025}).
This uses a custom-coded, adaptive HMC algorithm with between-chain adaptation during burnin; 
our algorithm implementation is based on CHESSR-HMC (Hoffman et al, 2021\supercite{Hoffman_et_al_ChEES_2021})
(also with option for SNAPER-HMC\supercite{sountsov_and_Hoffman_2022_focusing_SNAPER_HMC_and_ChEESR}, currently in development),
which is superior to Stan's\supercite{Carpenter2017} NUTS-based (Hoffman et al, 2014\supercite{Hoffman_and_Gelman_2014_NUTS_paper}) algorithm.

Our implementation also has several special computational enhancements which make it so efficient 
for the LC-MVP and latent trait models, such as: 
(i) using manual-gradients for the log-likelihood (but Stan math C++ library autodiff for less intensive parts such as Jacobians and priors) - 
both normally and also all on the log-scale (for the LC-MVP) - to solve issues with 64-bit precision limitations sometimes causing divergences for these models;
(ii) CPU, L3-Cache-aware chunking - which splits up the log-posterior/gradient function into "chunks", 
dynamically detecting and adjusting the number of chunks according to the users CPU’s L3 cache size to maximise efficiency and parallel
scalability;
(iii) between-chain adaptation during burnin, ensuring all chains finish at the same time (solving the common "lagging chains" issue with e.g. Stan);
(iv) custom AVX-512 and AVX2 math functions - such as very fast and accurate approximations to math functions like log, exp, CDF of standard normal, etc.
The above aforementioned factors result in a very efficient algorithm for the LC-MVP and latent trait models
(compared to Stan\supercite{Carpenter2017} and Mplus \supercite{mplus}).
More details are available on the BayesMVP GitHub page\supercite{Cerullo_BayesMVP_2025}, 
and in the associated paper \textcolor{red}{REF}.
This efficient implementation is what made this large simulation study possible.

All models will be run using 4 parallel chains on a local HPC/server computer with 96 cores (AMD EPYC 9654; 96 cores/192 threads) 
and 384GB of RAM.
Furthermore, we will run 24 models at once (so 96 parallel chains total), taking advantage of all 96 cores, 
which will greatly reduce the total computation time.

\subsection{ Adaptive Simulation Size ($N_{sim}$)}
\label{section_adaptive_N_sim}
Rather than fixing $N_{sim}$ a priori, we implemented an adaptive stopping rule that terminated simulations 
when the maximum Monte Carlo standard error for RMSE(Se) fell below $0.25\%$.
This ensured adequate precision while avoiding unnecessary computation.
The algorithm checked convergence after every simulation.
The mean MCSE of the sensitivity RMSE was used as the convergence criterion - 
as it consistently showed higher MCSE than specificity.

\section{Results}
\label{section_results}
\subsection{The ceiling effect paradox}
\label{section_ceiling_effect_paradox}
In this section, we will define the ceiling effect/paradox, 
which we reference multiple times throughout the results section and discussion of this paper.

Throughout our results for DGMs with conditional dependence (i.e. DGMs 2-5), we observe a consistent pattern:
DGMs with higher sensitivity values (i.e., DGMs 4 and 5) achieve better accuracy performance compared to DGMs which have lower 
true sensitivity values (i.e., DGMs 2 and 3) - despite obtaining worse overall correlation recovery.

Essentially, the reason why this occurs is that, with very high sensitivities (true sensitivity range: $86\% - 92.5\%$ for DGM \#5), 
the overwhelming majority of diseased individuals test positive regardless of the correlation structure. 
This creates a ceiling effect - where correlation misspecification has much less impact on accuracy estimation - 
in other words, the model correctly identifies that "more/almost everyone tests positive," even with badly estimated correlations.

In contrast, when we have low-moderate sensitivities (e.g., $55\% - 70\%$ for DGM's \#2 and \#3), there is more substantial variability
in test outcomes within the diseased group. Here, the conditional, between-test correlations more critically determine which combinations 
of test results are likely, directly affecting accuracy estimates.
Hence, misspecified correlations are more likely to lead to worse predictions about test patterns, 
resulting in worse accuracy estimates - even when the correlation bias is lower.
This explains the paradox: DGMs 4-5 achieve better accuracy performance despite worse correlation recovery 
because the higher sensitivity values create such a strong signal that the correlation structure becomes less important. 
On the other hand, DGMs 2-3 - with their much lower sensitivity values operating closer to the middle of the probability scale -
require accurate correlation estimation for good accuracy performance.

\subsection{The inverse problem}
\label{section_latent_trait_inverse_problem}
In this section, we will define the "inverse problem" of the latent trait model,
which we reference multiple times throughout the results section and discussion of this paper.

The latent trait model faces what we term an "inverse problem" when estimating correlations.
Unlike the LC-MVP model which directly parameterizes each correlation element,
the latent trait must determine which $b$ parameters - when transformed through the formula
$\Omega_{i,j}^{[d]} = \frac{b_i^{[d]} b_j^{[d]}}{\sqrt{(1 + (b_i^{[d]})^2)(1 + (b_j^{[d]})^2)}}$ -
will generate the observed correlation structure.

For $T$ tests, this creates a challenging underdetermined system:
the model must recover $T(T-1)/2$ unique correlations from only $T$ parameters.
In our case with $T = 5$ tests, the model must determine 5 $b$ parameters that simultaneously satisfy 10 correlation relationships.
This is "inverse" in the mathematical sense - working backwards from observed patterns to underlying parameters,
analogous to classic inverse problems in physics and engineering where one infers causes from effects.

Furthermore, this problem is compounded by three factors:
\begin{itemize}
    \item The relationship between $b$ parameters and correlations is non-linear and constrained -
          all correlations must be positive since $b_t > 0$ by construction.
    \item These same $b$ parameters also determine accuracy estimates through
          $\text{Se}_t = \Phi(a_t^{[2]}/\sqrt{1 + (b_t^{[2]})^2})$,
          meaning they must simultaneously satisfy both correlation and accuracy constraints.
    \item Third, when data is limited or uninformative (particularly under ceiling effects - see section \ref{section_ceiling_effect_paradox}),
          the likelihood surface becomes extremely flat, making it computationally difficult to identify 
          the correct $b$ values even though a unique solution exists.
          With minimal variation in test outcomes to guide estimation, 
          standard optimization algorithms struggle to navigate toward the true parameters,
          even with the model's parsimony.
\end{itemize}

\subsection{DGM 1 (conditional independence)}
\label{section_results_DGM_1}
\begin{figure}[H]
      \centering
    \includegraphics[width=16cm]{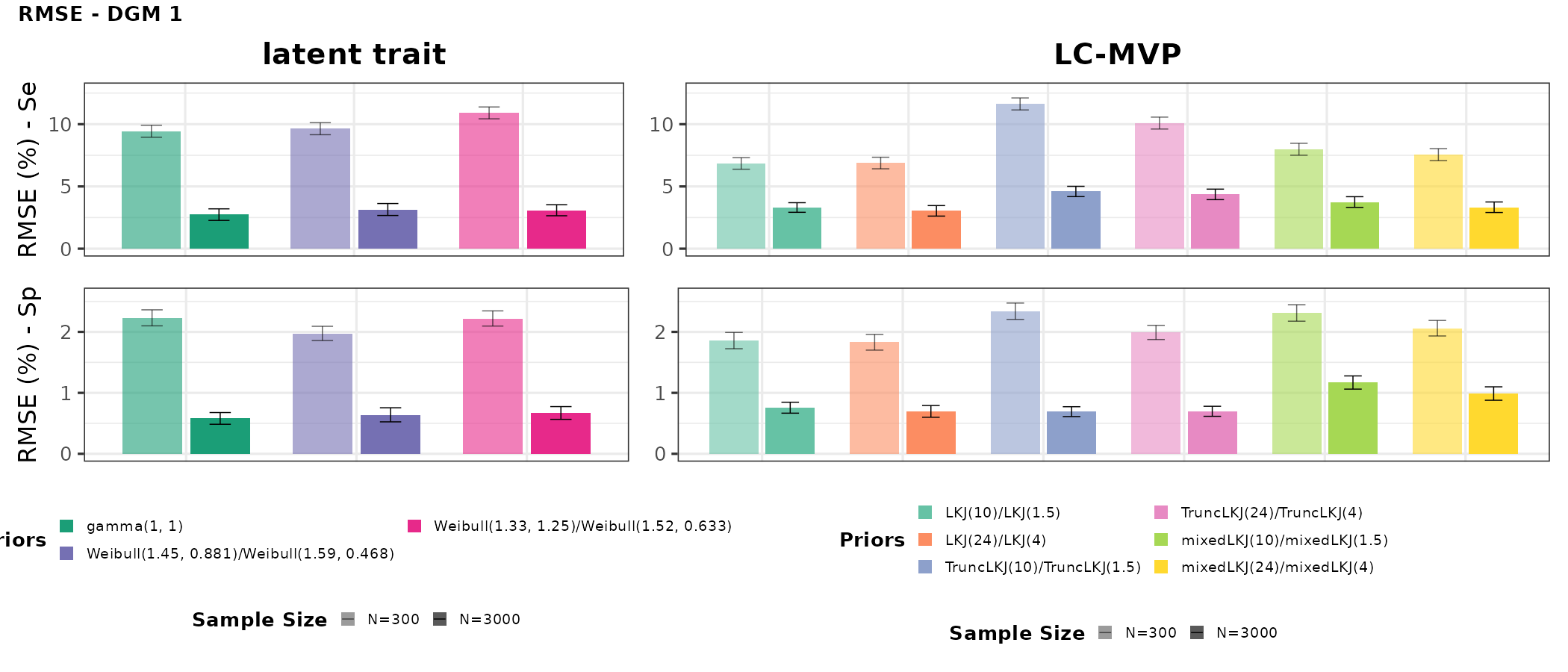}
    \caption{ Simulation study results (Se and Sp) for DGM 1 - RMSE }
    \label{Figure:Sim_study_RMSE_DGM_1_}
\end{figure}
\subsubsection{ Model \& prior performance (DGM \#1) }
\label{section_results_DGM_1_model_and_prior_performance} 
\begin{table}[H]
\footnotesize
\centering
\caption{
Overall performance (both models) for DGM \#1 (Conditional Independence)
(statistically equivalent groups at 95\% CI; ranked by RMSE)
}
\footnotesize
\setlength{\tabcolsep}{4pt}
\begin{tabular}{c|l|l}
\toprule
\textbf{DGM} & \textbf{N=300} & \textbf{N=3000} \\
\midrule
\multicolumn{3}{c}{\textbf{Ranked/ordered by overall RMSE (RMSE(Se) + RMSE(Sp))}} \\
\midrule
1 & \makecell[l]{\underline{Best ($8.71 - 9.61$):} \\ 
               LC-MVP w/ LKJ(10, 1.5) [$8.71, 0.97$], \\
               LC-MVP w/ LKJ(24, 4) [$8.71, 0.66$], \\
               CI [$8.84, 0.85$], \\ 
               LC-MVP w/ mixedLKJ(24, 4) [$9.61, 4.28$]. \\
               \\
               \underline{Worse ($10.3 - 14.0$):} \\ 
               LC-MVP w/ mixedLKJ(10, 1.5) [$10.3, 5.66$], \\
               Latent trait w/ Weibull(1.45) [$11.6$, $7.55$], \\
               Latent trait w/ Gamma(1, 1) [$11.7$, $7.56$], \\
               LC-MVP w/ TruncLKJ(24, 4) [$12.1, 8.15$], \\
               Latent trait w/ Weibull(1.33) [$13.1$, $9.76$], \\
               LC-MVP w/ TruncLKJ(10, 1.5) [$14.0, 10.8$].} 
  & \makecell[l]{\underline{Best ($2.93 - 3.79$):} \\
                CI [$2.93, 0.30$], \\
                Latent trait w/ Gamma(1, 1) [$3.32$, $1.38$], \\
                LC-MVP w/ LKJ(24, 4) [$3.74, 0.29$], \\
                Latent trait w/ Weibull(1.33) [$3.76, 2.32$], \\
                Latent trait w/ Weibull(1.45) [$3.79, 2.14$]. \\            
                \\
                \underline{Worse ($4.06 - 5.29$):} \\
                LC-MVP w/ LKJ(10, 1.5) [$4.06, 0.36$], \\
                LC-MVP w/ mixedLKJ(24, 4) [$4.32, 2.52$]. \\
                LC-MVP w/ mixedLKJ(10, 1.5) [$4.92, 2.92$], \\
                LC-MVP w/ TruncLKJ(24, 4) [$5.06, 3.68$], \\
                LC-MVP w/ runcLKJ(10, 1.5) [$5.29, 4.05$].} \\
\bottomrule 
\end{tabular}
\begin{tablenotes}
\footnotesize
\item Note: \\
For each prior/constraint, both the RMSE and bias are shown in square brackets (i.e.: "prior [RMSE, bias]").
Recall that, for DGM \#1: \\
true sensitivities ($\%$) are: (65, 55, 60, 65, 70);
true specificities ($\%$) are: (99, 95, 90, 90, 85).
\end{tablenotes}
\label{Table:summary_table_overall_both_models_DGM_1}
\end{table}
Table \ref{Table:summary_table_overall_both_models_DGM_1} presents the complete ranking of all priors across both models.
At $N = 300$, we can see that four model/prior combinations tie in the best-performing group (RMSE: $8.71 - 9.61$); 
namely, both LC-MVP LKJ priors (RMSE/bias range: $8.71 - 8.71$/$0.66 - 0.97$), 
CI model (RMSE/bias = $8.84/0.85$), 
and then finally LC-MVP mixedLKJ(24, 4), which had competitive RMSE ($9.61$),
but notably worse bias than the top three model/prior combinations (bias of: $4.28$ vs. $0.66 - 0.97$).
The LC-MVP's standard LKJ priors here perform just as well as the CI model (any differences are well within MCSE for both RMSE and bias).
Furthermore, the remaining three LC-MVP priors (in the "worse" group) - mixedLKJ(10, 1.5) and both TruncLKJ priors - 
perform notably worse than CI (RMSE: $10.3 - 14.0$ vs. $8.84$), 
especially the two TruncLKJ priors (RMSE: $12.1 - 14.0$ vs. $8.84$).

For the latent trait model, we can see that all three latent trait priors end up in the "worse" group 
(RMSE range: $11.6 - 13.1$).
Hence, the overall performance hierarchy is: 
CI $\approx$ LC-MVP with standard LKJ $>$ LC-MVP with constrained priors $>$ all latent trait priors.
This gap is driven by sensitivity estimation, where CI achieves RMSE(Se) of $7.08$, 
LC-MVP's best four priors achieve $6.85 - 7.55$ (see table \ref{Table:summary_table_overall_both_models_DGM_1}),
whilst latent trait lags substantially at $9.44 - 10.9$.

At $N = 3000$ (see right-hand side of table \ref{Table:summary_table_overall_both_models_DGM_1}),
the pattern shifts notably: 
now CI leads (RMSE/bias = $2.93/0.30$) and now all three latent trait priors 
tie statistically in the best group (RMSE range: $3.32 - 3.79$, bias range: $1.38 - 2.32$), 
alongside LC-MVP's LKJ(24, 4) (RMSE/bias = $3.74/0.29$).
It is also important to note that, although in the "worse" group, the LC-MVP with LKJ(10, 1.5) prior actually does quite well
(RMSE = $4.06$), especially with respect to its bias ($0.36$ vs. $0.30$ for CI model).
Meanwhile, LC-MVP's constrained priors perform worse:
mixedLKJ with overall RMSE of $4.32 - 4.92$, and TruncLKJ with $5.06 - 5.29$.
The TruncLKJ priors' poor coverage 
($82 - 84\%$; see figure \ref{Appendix_figure:Sim_study_Coverage_DGM_1_} in appendix \ref{appendix_C_coverage_plots}) 
confirms that forcing positive correlations on zero correlations causes systematic bias that worsens with more precise estimation.
Overall, the hierarchy here at $N = 3000$ is: 
CI $\approx$ LC-MVP LKJ(24, 4) $\approx$ all latent trait priors $>$ 
LC-MVP LKJ(10, 1.5) $>$ 
LC-MVP mixedLKJ $>$ 
LC-MVP TruncLKJ.

The latent trait's competitive performance here likely reflects its parametric efficiency with sufficiently large $N$
(note that $N^{[d+]} \sim 600$ here) - 
estimating only 10 correlation parameters versus LC-MVP's 20.
Hence, with sufficient data for truly independent structure, this two-fold parameter reduction allows 
the latent trait to achieve comparable or better performance than most LC-MVP priors, 
in this conditionally independent case.

\subsubsection{ Correlation analysis (DGM \#1) }
\label{section_results_DGM_1_correlation_analysis} 
\begin{table}[H]
\centering
\caption{
Correlation RMSE and bias for LC-MVP model in diseased group;
DGM \#1 (Conditional Independence)
}
\footnotesize
\begin{tabular}{cclcccc}
\toprule
\textbf{DGM} & \textbf{N} & \textbf{Prior} & \textbf{RMSE(Se)} & \textbf{Total Corr} & \textbf{Test 1 pairs} & \textbf{Tests 2-5 pairs} \\
             &            &                & \textbf{(\%)}     & {RMSE/Bias}         & {RMSE/Bias}          & {RMSE/Bias} \\
\midrule
1 & 300 & LKJ(10, 1.5)      & 6.85 & 1.81/-0.04 & 0.71/-0.03 & 1.10/-0.01 \\
1 & 300 & LKJ(24, 4)        & 6.88 & 1.41/+0.01 & 0.54/0.00 & 0.86/+0.01 \\
1 & 300 & TruncLKJ(10, 1.5) & 11.6 & 3.09/+2.91 & 1.36/+1.31 & 1.72/+1.60 \\
1 & 300 & TruncLKJ(24, 4)   & 10.1 & 2.38/+2.24 & 1.01/+0.96 & 1.37/+1.28 \\
1 & 300 & mixedLKJ(10, 1.5) & 7.99 & 2.34/+1.66 & 0.84/+0.29 & 1.49/+1.36 \\
1 & 300 & mixedLKJ(24, 4)   & 7.55 & 1.81/+1.31 & 0.58/+0.18 & 1.23/+1.13 \\
\midrule
1 & 3000 & LKJ(10, 1.5)      & 3.31 & 0.95/+0.06 & 0.43/+0.02 & 0.51/+0.04 \\
1 & 3000 & LKJ(24, 4)        & 3.04 & 0.87/+0.04 & 0.39/0.00 & 0.48/+0.04 \\
1 & 3000 & TruncLKJ(10, 1.5) & 4.60 & 1.31/+1.20 & 0.52/+0.47 & 0.79/+0.72 \\
1 & 3000 & TruncLKJ(24, 4)   & 4.36 & 1.22/+1.12 & 0.49/+0.45 & 0.74/+0.67 \\
1 & 3000 & mixedLKJ(10, 1.5) & 3.75 & 1.25/+0.20 & 0.62/-0.37 & 0.63/+0.57 \\
1 & 3000 & mixedLKJ(24, 4)   & 3.33 & 1.08/+0.32 & 0.48/-0.23 & 0.60/+0.55 \\
\midrule
\bottomrule
\end{tabular}
\begin{tablenotes}
\footnotesize
\item Note: Values shown as RMSE/bias. True correlations are zero for DGM 1.
\end{tablenotes}
\label{Table:corr_RMSE_LC_MVP_DGM_1}
\end{table}
\begin{table}[H]
\centering
\caption{
Correlation RMSE and bias for latent trait model in diseased group; 
for DGM \#1
}
\footnotesize
\begin{tabular}{cclcccc}
\toprule
\textbf{DGM} & \textbf{N} & \textbf{Prior} & \textbf{RMSE(Se)} & \textbf{Total Corr} & \textbf{Test 1 pairs} & \textbf{Tests 2-5 pairs} \\
             &            &                & \textbf{(\%)}     & {RMSE/Bias}         & {RMSE/Bias}          & {RMSE/Bias} \\
\midrule
1 & 300 & gamma(1, 1) & 9.44 & 1.86/+1.61 & 0.81/+0.71 & 1.05/+0.89 \\
1 & 300 & Weibull($\frac{1.59}{1.45}, \frac{0.468}{0.881}$) & 9.64 & 2.00/+1.85 & 0.87/+0.81 & 1.13/+1.04 \\
1 & 300 & Weibull($\frac{1.52}{1.33}, \frac{0.633}{1.25}$) & 10.9 & 2.56/+2.34 & 1.15/+1.07 & 1.40/+1.27 \\
\midrule
1 & 3000 & gamma(1, 1) & 2.74 & 0.46/+0.37 & 0.19/+0.15 & 0.27/+0.22 \\
1 & 3000 & Weibull($\frac{1.59}{1.45}, \frac{0.468}{0.881}$) & 3.15 & 0.71/+0.62 & 0.27/+0.24 & 0.44/+0.38 \\
1 & 3000 & Weibull($\frac{1.52}{1.33}, \frac{0.633}{1.25}$) & 3.09 & 0.74/+0.63 & 0.29/+0.24 & 0.45/+0.39 \\
\midrule
\bottomrule
\end{tabular}
\begin{tablenotes}
\footnotesize
\item Note: Values shown as RMSE/bias. Positive bias indicates forcing positive correlations where truth is zero.
\end{tablenotes}
\label{Table:corr_rmse_latent_trait_DGM_1}
\end{table}
\begin{table}[H]
\centering
\caption{DGM 1 (Conditional Independence): Correlations between RMSE(Se) and correlation recovery}
\begin{tabular}{clccc}
\toprule
DGM & Sample Size/Model &  
\makecell[l]{ Cor(RMSE(Se), \\ RMSE(Cor) } & 
\makecell[l]{ Cor(RMSE(Se), \\ RMSE(test-1-Cor) } & 
\makecell[l]{ Cor(RMSE(Se), \\RMSE(tests-2-5-Cor) } \\
\midrule
\multirow{4}{*}{1} & N=300, LC-MVP & 0.92 & 0.95 & 0.83 \\
                   & N=300, Latent Trait & 1.00 & 1.00 & 1.00 \\
                   & N=3000, LC-MVP & 0.89 & 0.51 & 0.97 \\
                   & N=3000, Latent Trait & 0.97 & 0.95 & 0.98 \\
\bottomrule
\end{tabular}
\label{Table:corr_patterns_RMSE_both_models_DGM_1}
\end{table}
The correlation analysis may give some insight as to why standard LKJ priors succeed for this DGM 
(tables: \ref{Table:corr_RMSE_LC_MVP_DGM_1},
\ref{Table:corr_rmse_latent_trait_DGM_1}, and
\ref{Table:corr_patterns_RMSE_both_models_DGM_1}).
With true correlations of zero (conditional independence), models must push all correlations toward zero.
Standard LKJ priors, centered at zero, naturally accommodate this 
(achieving correlation RMSE of $1.41 - 1.81$ at $N = 300$, see table \ref{Table:corr_RMSE_LC_MVP_DGM_1}).
In contrast, the latent trait model's inherent positive correlation constraint forces overestimation
(correlation RMSE of $1.86 - 2.56$).
The strong correlations between RMSE(Se) and correlation recovery (r=$0.92 - 1.00$) strongly suggest that 
correlation estimation drives accuracy performance for this DGM.

Paradoxically, for $N = 3000$, the correlation analysis reveals that the latent trait achieves lower correlation RMSE 
($0.46 - 0.74$ for latent trait vs. $0.87 - 0.95$ for LC-MVP with standard LKJ,
see table \ref{Table:corr_patterns_RMSE_both_models_DGM_1}),
yet worse accuracy RMSE/bias.
For this conditionally independent DGM, $b = 0$ would be correct for both correlations and sensitivity estimation,
effectively reducing to the appropriate conditional independence model.
However, with only ${\sim}600$ diseased individuals at $N = 3000$, the model cannot drive $b$ values to zero.
These intermediate $b$ values partially reduce correlations (improving correlation RMSE)
but are incorrect for sensitivity estimation - they're neither zero (correct) 
nor whatever non-zero values the model is forced toward by finite-sample constraints.
Furthermore, since the true sensitivities are high for this DGM, 
this issue is likely amplified by the ceiling effect (see section \ref{section_ceiling_effect_paradox}),
since there is not enough variation in responses in the diseased group to overcome the "inverse problem" of the latent trait model
(we discuss this "inverse problem" more in section \ref{section_latent_trait_inverse_problem}).

The bias patterns explain the performance differences at N=300: LC-MVP with LKJ correctly estimates zero correlations 
(bias: $-0.04$ to $+0.01$) and achieves good RMSE(Se) ($6.85 - 6.88$), 
whilst TruncLKJ forces massive positive bias ($+2.24$ to +2.91), 
and latent trait priors force substantial positive bias ($+1.61$ to $+2.34$), 
both resulting in worse RMSE(Se) ($7.55 - 11.6$ for LC-MVP with constrained priors, $9.44 - 10.9$ for latent trait). 
However, at $N=3000$, all latent trait models, the CI model, and the LC-MVP model with LKJ(24, 4) prior 
(with the LKJ(10, 1.5) only just being outside of statistical significance) - 
all converge to similar RMSE(Se) performance within (or very nearly within) MCSE - 
and the correlation bias differences no longer matter with sufficient data,
likely due to the ceiling effect (see section\ref{section_ceiling_effect_paradox}).

\subsection{DGM's \#3 and \#5 (LC-MVP structure DGM's) }
\label{section_results_LC_MVP_DGM_3_5}
\begin{figure}[H]
      \centering
    \includegraphics[width=16cm]{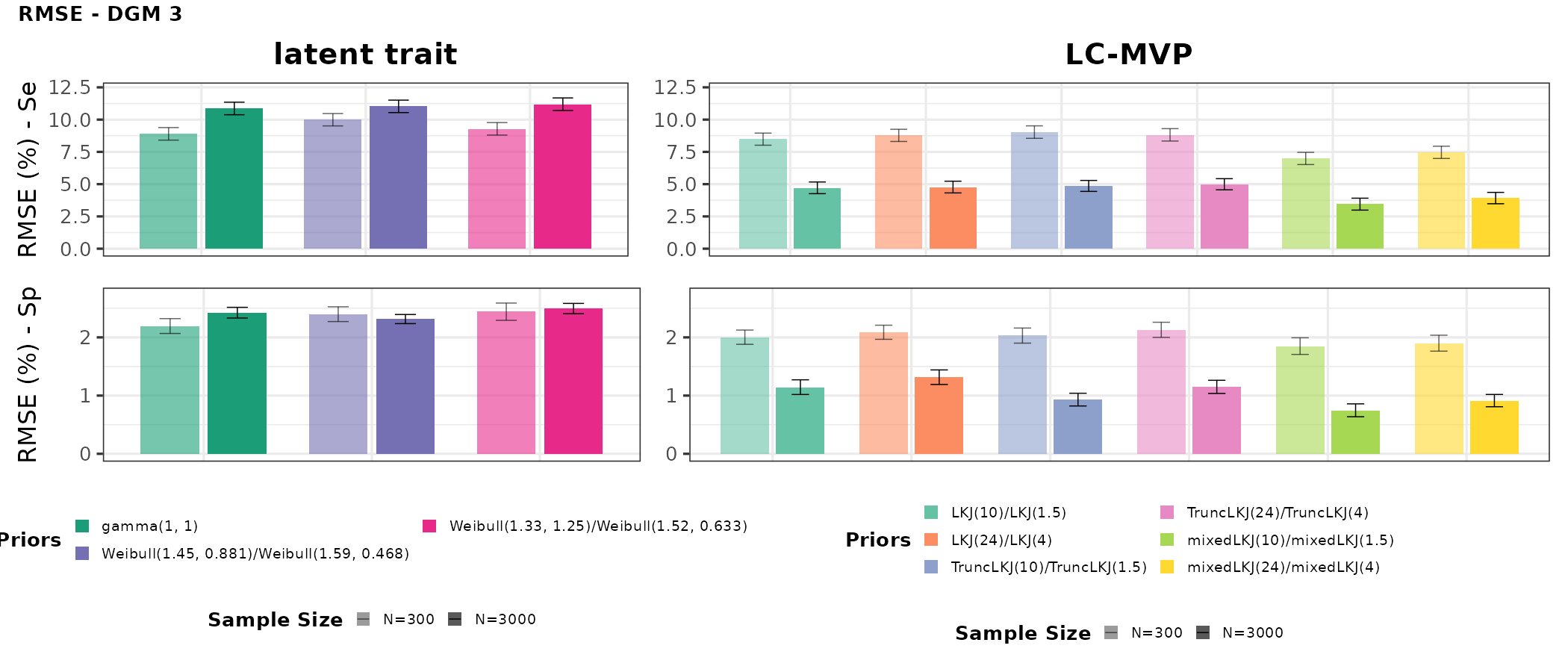}
    \caption{ Simulation study results (Se and Sp) for DGM \#3 - RMSE }
    \label{Figure:Sim_study_RMSE_DGM_3_}
\end{figure}
\begin{figure}[H]
      \centering
    \includegraphics[width=16cm]{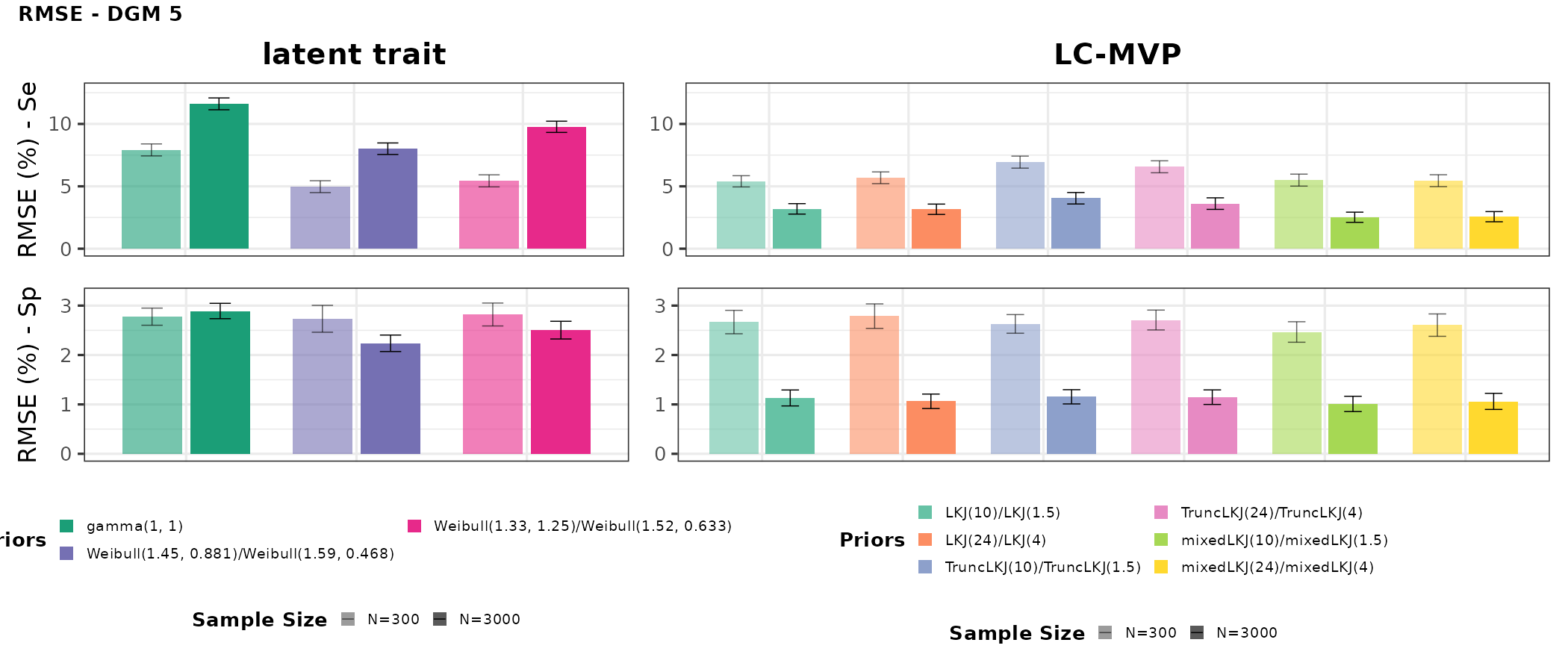}
    \caption{ Simulation study results (Se and Sp) for DGM \#5 - RMSE }
    \label{Figure:Sim_study_RMSE_DGM_5_}
\end{figure}
\subsubsection{ Model \& prior performance (DGM's \#3 \& \#5) }
\label{section_results_DGMs_3_5_model_and_prior_performance} 
\begin{table}[H]
\centering
\caption{
Overall model performance (both models) for DGMs \#3 \& \#5 (LC-MVP structure)
(statistically equivalent groups at 95\% CI; ranked by RMSE)
}
\footnotesize
\setlength{\tabcolsep}{4pt}
\begin{tabular}{c|l|l}
\toprule
\textbf{DGM} & \textbf{N=300} & \textbf{N=3000} \\
\midrule
\multicolumn{3}{c}{\textbf{By Overall RMSE (Se + Sp)}} \\ 
\midrule
3 & \makecell[l]{\underline{Best ($8.84 - 9.37\%$):} \\ 
                LC-MVP w/ mixedLKJ(10, 1.5) [$8.84$, $1.57$], \\
                LC-MVP w/ mixedLKJ(24, 4) [$9.37$, $2.50$]. \\
                \\
                \underline{Worse ($10.5 - 12.4$):} \\ 
                LC-MVP w/ LKJ(10, 1.5) [$10.5$, $4.88$], \\ 
                LC-MVP w/ LKJ(24, 4) [$10.9$, $5.22$], \\
                LC-MVP w/ TruncLKJ(24, 4) [$11.0$, $4.27$], \\
                LC-MVP w/ TruncLKJ(10, 1.5) [$11.1$, $5.35$], \\
                Latent trait w/ Gamma(1, 1), [$11.1$, $4.68$], \\
                Latent trait w/ Weibull(1.33), [$11.7$, $3.16$], \\
                CI [$12.2$, $7.25$], \\
                Latent trait w/ Weibull(1.45) [$12.4$, $5.63$].}
  & \makecell[l]{\underline{Best ($4.20 - 4.84$):} \\ 
                LC-MVP w/ mixedLKJ(10, 1.5) [$4.20$, $0.96$], \\
                LC-MVP w/ mixedLKJ(24, 4) [$4.84$, $2.11$]. \\
                \\
                \underline{Worse ($5.79 - 13.7$):} \\ 
                LC-MVP w/ TruncLKJ(10, 1.5) [$5.79$, $4.17$] , \\ 
                LC-MVP w/ LKJ(10, 1.5) [$5.86$, $2.64$], \\ 
                LC-MVP w/ LKJ(24, 4) [$6.09$, $3.45$], \\
                LC-MVP w/ TruncLKJ(24, 4) [$6.14$, $4.21$], \\
                CI [$9.34$, $8.59$], \\
                Latent trait w/ Gamma(1, 1) [$13.3$, $5.21$], \\
                Latent trait w/ Weibull(1.45) [$13.3$, $8.56$], \\
                Latent trait w/ Weibull(1.33) [$13.7$, $9.03$].} \\    
\midrule
5 & \makecell[l]{\underline{Best ($7.71 - 8.54$):} \\ 
               Latent trait w/ Weibull(1.45) [$7.71$, $2.62$], \\
               LC-MVP w/ mixedLKJ(10, 1.5) [$7.96$, $3.21$], \\
               LC-MVP w/ mixedLKJ(24, 4) [$8.06$, $2.61$], \\ 
               LC-MVP w/ LKJ(10, 1.5) [$8.07$, $2.93$], \\
               Latent trait w/ Weibull(1.33) [$8.26$, $2.11$], \\
               LC-MVP w/ LKJ(24, 4) [$8.47$, $3.28$], \\
               CI [$8.54$, $3.72$]. \\
               \\
               \underline{Worse ($9.27 - 10.7$):} \\  
               LC-MVP w/ TruncLKJ(24, 4) [$9.27$, $3.89$], \\
               LC-MVP w/ TruncLKJ(10, 1.5) [$9.57$, $4.59$], \\
               Gamma(1, 1) [$10.7$, $5.78$].} 
  & \makecell[l]{\underline{Best ($3.53 - 4.32$):} \\ 
                LC-MVP w/ mixedLKJ(10, 1.5) [$3.53$, $1.51$], \\
                LC-MVP w/ mixedLKJ(24, 4) [$3.63$, $1.92$],  \\
                LC-MVP w/ LKJ(24, 4) [$4.22$, $2.71$], \\ 
                LC-MVP w/ LKJ(10, 1.5) [$4.32$, $2.54$]. \\
                \\
                \underline{Worse ($4.76 - 14.5$):} \\ 
                LC-MVP w/ TruncLKJ(24, 4) [$4.76$, $3.09$], \\
                LC-MVP w/ TruncLKJ(10, 1.5) [$5.20$, $3.58$], \\
                CI [$5.58$, $4.68$], \\
                latent trait w/ Weibull(1.45) [$10.3$, $7.89$], \\
                latent trait w/ Weibull(1.33) [$12.3$, $11.6$], \\
                latent trait w/ Gamma(1, 1) [$14.5$, $12.7$].} \\ 
\bottomrule
\end{tabular}
\begin{tablenotes}
\footnotesize
\item Note:
For each prior/constraint, both the RMSE and bias are shown in square brackets (i.e.: "prior [RMSE, bias]").
Recall that, for DGM \#3: \\
true sensitivities ($\%$) are: (65, 55, 60, 65, 70);
true specificities ($\%$) are: (99, 95, 90, 90, 85). \\
Recall that, for DGM \#5: \\ 
true sensitivities ($\%$) are: (92.5, 86, 87, 91, 86);
true specificities ($\%$) are: (95,   81, 70, 67, 85).
\end{tablenotes}
\label{Table:summary_table_overall_both_models_DGMs_3_5}
\end{table}
\underline{\textbf{DGM \#3 (Heterogenous LC-MVP corr structure w/ low-moderate Se, high Sp), $N = 300$:}} \\ \newline
For DGM \#3 at $N = 300$, table \ref{Table:summary_table_overall_both_models_DGMs_3_5} (left-hand side)
clearly shows LC-MVP's mixedLKJ priors dominating, with an RMSE range of $8.84 - 9.37$,
and with the less informative mixedLKJ(10, 1.5) in the lead - 
in terms of both RMSE ($8.84$ vs. $9.37$) as well as bias ($1.57$ vs. $2.50$).
All other approaches perform statistically significantly worse than the LC-MVP model with the mixedLKJ(10, 1.5) prior:
LC-MVP's standard LKJ priors (RMSE: $10.5 - 10.9$, bias: $4.88 - 5.22$), 
LC-MVP using TruncLKJ priors (RMSE: $11.0 - 11.1$, bias: $4.27 - 5.35$), then the
latent trait model regardless of prior choice (RMSE: $11.1 - 12.4$, bias: $3.16 - 5.63$), 
and the CI model (RMSE/bias = $12.2/7.25$).

Hence, the hierarchy here is:
LC-MVP mixedLKJ $>>$ 
other LC-MVP $\approx$ latent trait $>$ 
CI.
These findings here suggest that flexible correlation modelling with appropriate constraints works best 
for heterogeneous structures, at least in this case (DGM \#3, $N = 300$).

\underline{\textbf{DGM \#3 (Heterogenous LC-MVP corr structure w/ low-moderate Se, high Sp), $N = 3000$:}} \\ \newline
At $N = 3000$ (see right-hand side of figure \ref{Table:summary_table_overall_both_models_DGMs_3_5}), 
the pattern is similar - but with more separation between the different model types.
We can see here that the CI model performs extremely badly (RMSE/bias = $9.34/8.59$), 
with extremely poor sensitivity coverage of just $29\%$ (not shown in figures/tables).
We can also see that LC-MVP model with mixedLKJ priors is again the only model/prior combination which makes it into
the best-performing group, with RMSE roughly halved compared to when $N = 300$ (RMSE: $4.20 - 4.84$, bias: $0.91 - 2.11$).
Similarly to the $N = 300$ case, it is again the less informative mixedLKJ(10, 1.5) in the lead (RMSE/bias = $4.20/0.91$).
Additionally, this is also combined with excellent coverage 
($98.2\% - 99.6\%$; see table \ref{Appendix_table:lc_mvp_dgm_3_5} in appendix \ref{appendix_A_detailed_results_tables}).

Furthermore, even though they are in the "worse" group - 
since they're statistically significantly worse than the leading mixedLKJ(10, 1.5) prior - 
the LC-MVP with LKJ priors actually does relatively well here, and substantially better than the CI model 
(RMSE range for LKJ: $5.86 - 6.09$ vs. $9.34$ for CI model, with bias range of: $2.64 - 3.45$ vs. $8.59$ for CI model).
Similarly, both of the TruncLKJ priors also perform substantially better than the CI model
(RMSE range for TruncLKJ: $5.79 - 6.14$ vs. $9.34$ for CI model, with bias range of: $4.17 - 4.21$ vs. $8.59$ for CI model).

Hence, overall, all six of the prior/constraint combinations for the LC-MVP model here so substantially - 
and statistically significantly - better than the CI model.
On the other hand, we can see that the latent trait catastrophically deteriorates to being even worse than its $N = 300$ performance ($11.1 - 11.7$),
especially in terms of RMSE, which is our main measure here
(RMSE range for latent trait: $13.3 - 13.7$ vs. $9.34$ for CI model, bias range: $5.21 - 9.03$ vs. $8.59$ for CI model).
This is combined with catastrophic coverage failure 
($53\% - 69\%$ coverage; see table \ref{Appendix_table:latent_trait_dgm_3_5} in appendix \ref{appendix_A_detailed_results_tables}).
This performance gap here occurs despite the latent trait model being more parsimonious,
perhaps because the true correlation structure is too heterogeneous for the latent trait's restrictive correlation structure.

\underline{\textbf{DGM \#5 (Heterogenous LC-MVP corr structure w/ high Se), $N = 300$:}} \\ \newline
For DGM \#5 at $N = 300$ (see bottom-left of table \ref{Table:summary_table_overall_both_models_DGMs_3_5}), 
seven priors including CI tie in the best group (RMSE range: $7.71 - 8.54$).
More specifically, the two models actually performed quite similarly - with the LC-MVP model
(with best four priors - which were all priors except for the two TruncLKJ priors) achieving an RMSE of $7.96 - 8.47$ (bias: $2.61 - 3.28$)
The worse-performming TruncLKJ priors however obtained an RMSE range of $9.27 - 9.57$ (with bias range: $3.89 - 4.59$).
versus latent trait with Weibull priors achieving an RMSE range of $7.71 - 8.26$ (with bias range: $2.11 - 2.62$).

The fact that the LC-MVP and latent trait models perform similarly here DGM \#5 (at $N = 300$) - but not for DGM \#3 -
is because DGM \#5 has much higher true sensitivity values, which likely makes correlation estimation less important
in this particular case; we discuss this concept in more detail in section \ref{section_ceiling_effect_paradox}.
However, note that when we increase the sample size ($N = 3000$), this pattern we observe for DGM \#5 changes - 
at this larger sample size, there is now a very clear difference between the LC-MVP model and the latent trait models;
we discuss this in more detail in this section below.

\underline{\textbf{DGM \#5 (Heterogenous LC-MVP corr structure w/ high Se), $N = 3000$:}} \\ \newline
At $N = 3000$, the pattern is very different to what it was at $N = 300$ - LC-MVP now dominates -
its best four priors (mixedLKJ and standard LKJ) achieve an RMSE range of $3.53 - 4.32$ (bias: $1.51 - 2.72$), 
and are the only four model/prior combinations which make it into the best-performing group.
However, both TruncLKJ priors are in the worse-perofmring grpup, although they do not do too badly
(RMSE: $4.76 - 5.20$, bias: $3.09 - 3.58$).
Notably, all LC-MVP prior/constraint combinations do better than the CI model (RMSE range: $3.53 - 5.20$ vs. $5.58$ for CI),
which also obtains much worse bias ($5.58$ vs. $1.51 - 3.58$ for LC-MVP).
In contrast to the LC-MVP, all latent trait priors fail catastrophically
(RMSE: $10.3 - 14.5$ vs. $5.58$ for CI, bias: $7.89 - 12.7$ vs. $4.68$ for CI),
with coverage below $50\%$ 
(see figure \ref{Appendix_figure:Sim_study_Coverage_DGM_5_} in appendix \ref{appendix_C_coverage_plots}).

In this case, the latent trait model's rigid structure - 
trying to estimate 10 correlations (per latent class) via just 5 $b^{[d]}_{T}$ parameters - 
causes performance to deteriorate with increased sample size for both DGMs here, 
since the true correlation structure for these two DGMs is highly heterogeneous and too difficult to estimate for the latent trait model,
and/or estimating it comes at the expense of distorting the actual test accuracy estimates (which rely on the "b" parameters -
unlike the LC-MVP model where accuracy estimates do not explicitly rely on any correlation-related parameters),
despite it being more parsimonious than the LC-MVP model.
These results demonstrate that when the true correlation structure conflicts with model constraints, increasing sample size 
paradoxically worsens performance by more precisely estimating the wrong structure, leading to systematic bias that 
destroys coverage (see figures \ref{Appendix_figure:Sim_study_Coverage_DGM_3_} and \ref{Appendix_figure:Sim_study_Coverage_DGM_5_} 
in appendix \ref{appendix_C_coverage_plots} for coverage plots for DGM \#3 and DGM \#5, respectively; 
for both the latent trait model and LC-MVP model), 
despite obtaining narrower intervals
(see figures \ref{Appendix_figure:Sim_study_Width_DGM_2_} and \ref{Appendix_figure:Sim_study_Width_DGM_4_} 
in appendix \ref{appendix_D_interval_width_plots} for interval width plots for DGM \#3 and DGM \#5, respectively; 
for both the latent trait model and LC-MVP models).

\subsubsection{ Correlation analysis (DGM's \#3 \& \#5) }
\label{section_results_DGMs_3_5_correlation_analysis} 
\begin{table}[H]
\centering
\caption{
Correlation RMSE and bias for LC-MVP model in diseased group; 
for DGMs \#3 and \#5
}
\scriptsize
\begin{tabular}{cclcccc}
\toprule
\textbf{DGM} & \textbf{N} & \textbf{Prior} & \textbf{RMSE(Se)} & \textbf{Total Corr} & \textbf{Test 1 pairs} & \textbf{Tests 2-5 pairs} \\
             &            &                & \textbf{(\%)}     & {RMSE/Bias}         & {RMSE/Bias}          & {RMSE/Bias} \\
\midrule
3 & 300 & LKJ(10, 1.5)      & 8.49 & 2.41/-1.50 & 0.73/-0.15 & 1.69/-1.35 \\
3 & 300 & LKJ(24, 4)        & 8.78 & 2.36/-1.74 & 0.59/-0.17 & 1.77/-1.57 \\
3 & 300 & TruncLKJ(10, 1.5) & 9.03 & 1.64/+0.79 & 0.89/+0.81 & 0.75/-0.02 \\
3 & 300 & TruncLKJ(24, 4)   & 8.82 & 1.44/+0.08 & 0.64/+0.56 & 0.80/-0.48 \\
3 & 300 & mixedLKJ(10, 1.5) & 6.99 & 1.49/-0.35 & 0.70/-0.11 & 0.79/-0.24 \\
3 & 300 & mixedLKJ(24, 4)   & 7.47 & 1.43/-0.75 & 0.56/-0.13 & 0.87/-0.61 \\
\midrule
3 & 3000 & LKJ(10, 1.5)      & 4.72 & 1.12/+0.01 & 0.53/+0.31 & 0.59/-0.30 \\
3 & 3000 & LKJ(24, 4)        & 4.77 & 1.17/-0.17 & 0.49/+0.31 & 0.68/-0.47 \\
3 & 3000 & TruncLKJ(10, 1.5) & 4.86 & 1.00/+0.55 & 0.59/+0.51 & 0.40/+0.04 \\
3 & 3000 & TruncLKJ(24, 4)   & 4.99 & 0.96/+0.27 & 0.53/+0.45 & 0.43/-0.18 \\
3 & 3000 & mixedLKJ(10, 1.5) & 3.45 & 0.87/-0.06 & 0.48/+0.01 & 0.39/-0.08 \\
3 & 3000 & mixedLKJ(24, 4)   & 3.92 & 0.91/-0.10 & 0.44/+0.15 & 0.47/-0.25 \\
\midrule
\midrule
5 & 300 & LKJ(10, 1.5)      & 5.40 & 2.35/-1.60 & 0.47/+0.04 & 1.89/-1.64 \\
5 & 300 & LKJ(24, 4)        & 5.68 & 2.35/-1.94 & 0.37/-0.08 & 1.98/-1.85 \\
5 & 300 & TruncLKJ(10, 1.5) & 6.94 & 1.97/+1.16 & 1.16/+1.14 & 0.81/+0.01 \\
5 & 300 & TruncLKJ(24, 4)   & 6.56 & 1.64/+0.17 & 0.76/+0.73 & 0.89/-0.56 \\
5 & 300 & mixedLKJ(10, 1.5) & 5.50 & 1.25/-0.06 & 0.40/+0.09 & 0.85/-0.15 \\
5 & 300 & mixedLKJ(24, 4)   & 5.45 & 1.25/-0.73 & 0.31/-0.09 & 0.93/-0.64 \\
\midrule
5 & 3000 & LKJ(10, 1.5)      & 3.19 & 1.95/-1.05 & 0.52/+0.09 & 1.43/-1.13 \\
5 & 3000 & LKJ(24, 4)        & 3.16 & 1.85/-1.18 & 0.40/+0.07 & 1.45/-1.25 \\
5 & 3000 & TruncLKJ(10, 1.5) & 4.04 & 1.86/+1.07 & 1.22/+1.18 & 0.64/-0.12 \\
5 & 3000 & TruncLKJ(24, 4)   & 3.61 & 1.61/+0.37 & 0.88/+0.84 & 0.74/-0.47 \\
5 & 3000 & mixedLKJ(10, 1.5) & 2.52 & 1.16/+0.05 & 0.52/+0.09 & 0.64/-0.04 \\
5 & 3000 & mixedLKJ(24, 4)   & 2.57 & 1.13/-0.36 & 0.42/+0.06 & 0.71/-0.42 \\
\midrule
\bottomrule
\end{tabular}
\begin{tablenotes}
\footnotesize
\item Note: Values shown as RMSE/bias. Negative bias indicates underestimation of correlations.
\end{tablenotes}
\label{Table:corr_rmse_LC_MVP_DGMs_3_5}
\end{table}
\begin{table}[H]
\centering
\caption{
Correlation RMSE and bias for latent trait model in diseased group; 
for DGMs \#3 and \#5
}
\scriptsize
\begin{tabular}{cclcccc}
\toprule
\textbf{DGM} & \textbf{N} & \textbf{Prior} & \textbf{RMSE(Se)} & \textbf{Total Corr} & \textbf{Test 1 pairs} & \textbf{Tests 2-5 pairs} \\
             &            &                & \textbf{(\%)}     & {RMSE/Bias}         & {RMSE/Bias}          & {RMSE/Bias} \\
\midrule
3 & 300 & gamma(1, 1) & 8.90 & 1.84/+0.46 & 0.74/+0.58 & 1.11/-0.13 \\
3 & 300 & Weibull($\frac{1.59}{1.45}, \frac{0.468}{0.881}$) & 9.99 & 1.57/-1.25 & 0.16/+0.11 & 1.40/-1.36 \\
3 & 300 & Weibull($\frac{1.52}{1.33}, \frac{0.633}{1.25}$) & 9.29 & 1.49/-0.27 & 0.48/+0.43 & 1.01/-0.69 \\
\midrule
3 & 3000 & gamma(1, 1) & 10.86 & 2.13/+0.31 & 0.82/+0.59 & 1.31/-0.29 \\
3 & 3000 & Weibull($\frac{1.59}{1.45}, \frac{0.468}{0.881}$) & 11.03 & 1.77/-0.52 & 0.56/+0.36 & 1.21/-0.88 \\
3 & 3000 & Weibull($\frac{1.52}{1.33}, \frac{0.633}{1.25}$) & 11.19 & 1.92/0.00 & 0.73/+0.58 & 1.19/-0.59 \\
\midrule
\midrule
5 & 300 & gamma(1, 1) & 7.92 & 2.35/+1.23 & 1.24/+1.16 & 1.11/+0.07 \\
5 & 300 & Weibull($\frac{1.59}{1.45}, \frac{0.468}{0.881}$) & 4.97 & 1.65/-1.37 & 0.14/+0.12 & 1.51/-1.50 \\
5 & 300 & Weibull($\frac{1.52}{1.33}, \frac{0.633}{1.25}$) & 5.44 & 1.63/-0.38 & 0.53/+0.51 & 1.10/-0.89 \\
\midrule
5 & 3000 & gamma(1, 1) & 11.6 & 3.17/+1.51 & 1.78/+1.69 & 1.39/-0.18 \\
5 & 3000 & Weibull($\frac{1.59}{1.45}, \frac{0.468}{0.881}$) & 8.01 & 2.35/-0.01 & 1.10/+0.96 & 1.25/-0.96 \\
5 & 3000 & Weibull($\frac{1.52}{1.33}, \frac{0.633}{1.25}$) & 9.77 & 2.59/+0.80 & 1.40/+1.38 & 1.19/-0.58 \\
\midrule
\bottomrule
\end{tabular}
\begin{tablenotes}
\footnotesize
\item Note: Values shown as RMSE/bias. Positive bias indicates overestimation of correlations.
\end{tablenotes}
\label{Table:corr_rmse_latent_trait_DGMs_3_5}
\end{table}
\begin{table}[H]
\centering
\caption{DGMs 3 \& 5 (True LC-MVP Structure): Correlations between RMSE(Se) and correlation recovery}
\begin{tabular}{clccc}
\toprule
DGM & Sample Size/Model & 
\makecell[l]{ Cor(RMSE(Se), \\ RMSE(Cor) } & 
\makecell[l]{ Cor(RMSE(Se), \\ RMSE(test-1-Cor) } & 
\makecell[l]{ Cor(RMSE(Se), \\RMSE(tests-2-5-Cor) } \\
\midrule
\multirow{4}{*}{3} & N=300, LC-MVP & 0.41 & 0.35 & 0.31 \\
                   & N=300, Latent Trait & -0.62 & -0.99 & 0.82 \\
                   & N=3000, LC-MVP & 0.66 & 0.68 & 0.36 \\
                   & N=3000, Latent Trait & -0.59 & -0.36 & -0.94 \\
\midrule
\multirow{4}{*}{5} & N=300, LC-MVP & 0.09 & 0.95 & -0.48 \\
                   & N=300, Latent Trait & 0.99 & 0.98 & -0.61 \\
                   & N=3000, LC-MVP & 0.75 & 0.86 & -0.01 \\
                   & N=3000, Latent Trait & 0.98 & 1.00 & 0.69 \\
\bottomrule
\end{tabular}
\label{Table:corr_patterns_RMSE_both_models_DGMs_3_5}
\end{table}
We will now look at the relationship between correlation recovery/RMSE and RMSE(Se), using the results from tables
\ref{Table:corr_patterns_RMSE_both_models_DGMs_3_5},
\ref{Table:corr_rmse_LC_MVP_DGMs_3_5} and \ref{Table:corr_rmse_latent_trait_DGMs_3_5}).
The correlation patterns in table \ref{Table:corr_patterns_RMSE_both_models_DGMs_3_5} require careful interpretation 
due to several limitations. 
Firstly, with only six LC-MVP priors and three latent trait priors, 
these correlations are susceptible to substantial MCSE noise. 
Secondly, our RMSE(Se) metric averages across all five tests - because presenting test-specific RMSE values 
would be impractical and overwhelming for readers 
(requiring five times as many tables and figures, or much more complex tables/figures, or a combination of both) - 
this averaging could mask important test-specific patterns where certain priors might excel for some tests but fail for others.
However, despite these limitations, some patterns emerge.

\underline{\textbf{DGM \#3 (Heterogenous LC-MVP corr structure w/ low-moderate Se, high Sp), $N = 300$:}} \\ \newline
For the latent trait model (for DGM \#3, $N = 300$),
we observe a strong positive correlation with tests 2-5 recovery (r = $0.82$; table \ref{Table:corr_patterns_RMSE_both_models_DGMs_3_5}), 
but strongly negative with the reference test (r = $-0.99$). 
With only three priors, this likely reflects a forced trade-off - 
priors that successfully recover the tests 2-5 correlations tend to sacrifice test-1 recovery, 
and for this heterogeneous structure and/or model, tests 2-5 recovery appears more critical for overall accuracy.
For the LC-MVP model, the moderate correlations (r = $0.31 - 0.41$) could reflect several factors: 
insufficient variation in RMSE(Se) across priors to detect strong patterns, 
the averaging across tests obscuring clearer test-specific relationships, or simply MCSE with limited priors.
Nevertheless, the best-performing mixedLKJ priors achieve both reasonable accuracy 
(RMSE(Se) = $6.99 - 7.47$ vs. $8.90 - 9.99$ for latent trait)
and correlation recovery (RMSE(Corr) = $1.43 - 1.49$; see table \ref{Table:corr_rmse_LC_MVP_DGMs_3_5}),
outperforming latent trait on both metrics (RMSE(Corr) for latent trait: $1.49 - 1.84$).

We also notice something very unusual: the LKJ priors active very bad correlation RMSE compared to (for instance) TruncLKJ
($2.36 - 2.41$ for LKJ vs. $1.44 - 1.64$ for TruncLKJ),
yet their RMSE(Se) is actually better than both TruncLKJ as well as the latent trait model
(RMSE(Se) for LKJ: $8.49 - 8.78$ vs. $8.82 - 9.03$ for TruncLKJ vs. $8.90 - 9.99$ for latent trait).
This paradox could be due to the following: notice that if we just focus on test-1 correlation recovery, the LKJ priors actually
do better than TruncLKJ, and are competetive with latent trait
(RMSE(test-1-pairs) for LKJ: $0.59 - 0.73$ vs. $0.64 - 0.89$ for TruncLKJ vs $0.16 - 0.74$ for latent trait).
Hence, it could be that test-1 correlations matter more for the LC-MVP model for this DGM,
but there just isn't enough information to reflect this in the correlations in table \ref{Table:corr_patterns_RMSE_both_models_DGMs_3_5},
hence giving us a Cor(RMSE(Se), RMSE(test-1-Cor)) of only $0.35$.

\underline{\textbf{DGM \#3 (Heterogenous LC-MVP corr structure w/ low-moderate Se, high Sp), $N = 3000$:}} \\ \newline
When looking at the correlation patterns (table \ref{Table:corr_patterns_RMSE_both_models_DGMs_3_5}), as usual these require careful 
interpretation, especially the ones for the latent trait which are all negative (between $-0.36$ to $-0.94$).
This is likely because of the extremely narrow range of RMSE(Se) we obtained for this model ($10.86 - 11.19$), 
which will easily obscure any patterns/correlations due to lack of variation. 
On the other hand, for the LC-MVP model, which had more variation for RMSE(Se) (range of: $3.45 - 4.99$ across all six priors), 
we obtained correlations of $0.66$ for Cor(RMSE(Se), RMSE(Cor)), 
$0.68$ for Cor(RMSE(Se), RMSE(test 1 Cor)), and
$0.36$ for Cor(RMSE(Se), RMSE(test 2-5 Cor)).
This suggests that for the LC-MVP model here, the test-1 correlation recovery is likely more important for predicting RMSE(Se).

Looking at the actual correlation RMSE values 
(table \ref{Table:corr_rmse_LC_MVP_DGMs_3_5} and table \ref{Table:corr_rmse_latent_trait_DGMs_3_5} 
for the LC-MVP and latent trait, respectively), 
we can see that the best-performing mixedLKJ priors (RMSE(Se): $3.45 - 3.92$ vs. $10.9 - 11.2$ for latent trait)
achieve an RMSE(Corr) of $0.87 - 0.91$ (vs. $1.77 - 2.13$ for latent trait),
whilst the middle-performing standard LKJ priors (RMSE(Se):  $4.72 - 4.77$, vs. $10.9 - 11.2$ for latent trait) 
have an RMSE(Corr) of $1.12 - 1.17$ (vs. $1.77 - 2.13$ for latent trait),
and the worst-performing TruncLKJ priors (RMSE(Se): $4.86 - 4.99$)
have an RMSE(Corr) of: $0.96 - 1.00$ (vs. $1.77 - 2.13$ for latent trait).
These findings are less counter-intuitive than the $N = 300$ case for DGM - 
as the LC-MVP model consistently obtains around 2-fold less RMSE(Corr) compared to the latent trait model - 
regardless of which prior we use, which is very consistent with its $\sim 2-3$ fold improvement in RMSE(Se)
($3.45 - 4.99$ for LC-MVP vs. $10.9 - 11.2$ for latent trait).

\underline{\textbf{DGM \#5 (Heterogenous LC-MVP corr structure w/ high Se), $N = 300$:}} \\ \newline
The correlation patterns for DGM \#5, $N = 300$ (see table \ref{Table:corr_patterns_RMSE_both_models_DGMs_3_5})
show more clear patterns compared to DGM \#3. 
More specifically, we can see that for the LC-MVP model, the correlation between RMSE(Se) 
(which has a range of $5.40 - 5.68$ for the best four priors, and $6.56 - 6.94$ for the two worst-performing priors, which were the 
TruncLKJ priors) and the RMSE of the reference test correlations is extremely strong at $0.95$, suggesting that -
for this model/DGM/N combination - that estimating reference test accuracies well predicted mean RMSE(Se).
However, neither Cor(RMSE(Se), RMSE(Cor)) ($0.09$) nor Cor(RMSE(Se), RMSE(test 2-5 Cor)) ($-0.48$) predicted the mean RMSE(Se) in this case.
A similar pattern holds for the latent trait model 
(RMSE(Se): $4.97 - 5.44$ for two best-performing priors and $4.97 - 7.92$ for all three priors) - 
where we obtained a correlation of $0.98$ for Cor(RMSE(Se), RMSE(test 1 Cor)), and also $0.99$ for Cor(RMSE(Se), RMSE(Cor)),
but $-0.61$ for Cor(RMSE(Se), RMSE(test 2-5 Cor)).

Now, looking at the actual correlation RMSE values 
(see table \ref{Table:corr_rmse_LC_MVP_DGMs_3_5} for LC-MVP and table \ref{Table:corr_rmse_latent_trait_DGMs_3_5} for latent trait), 
we can see that the best-performing priors for the latent trait model (both Weibull priors) obtain virtually identical RMSE(Cor) 
($1.63 - 1.65$) and they obtain a range of $0.14 - 0.53$ specifically for the RMSE of the reference test correlations, 
with the more highly informative Weibull prior obtaining just $0.14$ - which was the best-performing prior in terms of RMSE(Se).
On the other hand, the worst-performing prior (Gamma(1, 1)) here obtains a much worse RMSE(Cor) of $2.35$, and it obtains 
$0.82$ specifically for the RMSE of the reference test correlations.

For the LC-MVP model, we can see (from table \ref{Table:summary_table_overall_both_models_DGMs_3_5}) 
that the best four priors obtained an extremely narrow RMSE(Se) range of $5.40 - 5.68$
(vs. $6.56 - 6.94$ for the TruncLKJ priors - which were the only two priors in the worst-performing group).
However, when we look at the RMSE of the correlations, 
we can clearly see why we only obtained a Cor(RMSE(Se), RMSE(Cor)) of only $0.09$.
To be more specific, from table \ref{Table:corr_rmse_LC_MVP_DGMs_3_5}, we can see quite clearly that both LKJ priors obtained much worse
RMSE(Cor) (both equal to $2.35$), compared to either the worst-performing TruncLKJ priors 
(range: $1.64 - 1.97$) or the better-performing mixedLKJ priors (both obtained $1.25$),
yet the RMSE(Se) of the LKJ priors was actually better than TruncLKJ (RMSE(Se): $5.40 - 5.68$ vs. $6.56 - 6.94$ for TruncLKJ).
On the other hand, focusing on the RMSE of the test 1 correlations only, the pattern becomes much clearer, and we can see
why we obtained Cor(RMSE(Se), RMSE(test-1-Cor)) of $0.95$;
more specifically, 
the LKJ priors obtain RMSE(Se) of $5.40 - 5.68$ and RMSE(test-1-Cor) of $0.37 - 0.47$,
mixedLKJ priors obtain RMSE(Se) of $5.45 - 5.50$ and RMSE(test-1-Cor) of $0.31 - 0.40$,
and truncLKJ priors obtain RMSE(Se) of $6.56 - 6.94$ and RMSE(test-1-Cor) of $0.76 - 1.16$,
Furthermore, recall that for the latent trait model, for the two best-performing priors (Weibull priors) obtained
an RMSE(Se) of $4.97 - 5.44$ - 
which is comparable to LC-MVP with LKJ or mixedLKJ priors ($5.40 - 5.68$) - 
and they obtained RMSE(test-1-Cor) of $0.14 - 0.53$.

\underline{\textbf{DGM \#5 (Heterogenous LC-MVP corr structure w/ high Se), $N = 3000$:}} \\ \newline
When we look at the correlation patterns (table \ref{Table:corr_patterns_RMSE_both_models_DGMs_3_5}), 
we can see that for latent trait, there's near-perfect correlation between RMSE(Se) and correlation recovery
(Cor(RMSE(Se), RMSE(Cor)) = $0.98$, Cor(RMSE(Se), RMSE(test-1-Cor)) = $1.00$, Cor(RMSE(Se), RMSE(tests-2-5-Cor)) = $0.69$).
For the LC-MVP model, the correlation patterns show Cor(RMSE(Se), RMSE(Cor)) = $0.75$ and particularly strong Cor(RMSE(Se), RMSE(test-1-Cor)) = $0.86$,
indicating that test-1 correlation recovery is more important here.


Looking at actual correlation RMSE values (Tables \ref{Table:corr_rmse_LC_MVP_DGMs_3_5} and \ref{Table:corr_rmse_latent_trait_DGMs_3_5}),
LC-MVP's best priors (mixedLKJ; RMSE(Se) of $2.52 - 2.57$) achieve RMSE(Cor) of $1.13 - 1.16$ (with RMSE(test-1-Cor) of $0.42 - 0.52$),
then the LKJ priors achieve (RMSE(Se) of: $3.16 - 3.19$) achieve RMSE(Cor) of $1.85 - 1.95$ (with RMSE(test-1-Cor) of $0.40 - 0.52$),
and the TruncLKJ priors (RMSE(Se) of: $3.61 - 4.04$) achieve RMSE(Cor) of $1.61 - 1.86$ (with RMSE(test-1-Cor) of $0.88 - 1.22$).
On the other hand, for the latent trait model, the overall RMSE(Se) range across all 3 priors was: $8.01 - 11.6$, 
and the RMSE(Cor) was $2.35 - 3.17$, with an RMSE(test-1-Cor) of $1.10 - 1.78$
These results make sense: the latent trait model's RMSE(Se) is around 2-3 fold worse than the LC-MVP's, and it's 
poorer correlation recovery (especially test-1 correlation recovery) reflects this, since it's notably worse.

However, this doesn't fully capture or explain the poorer performance of the latent trait model, 
since correlation recovery is around $\sim 1.5-2$ fold worse, whereas the RMSE(Se) is $\sim 2-3$ fold worse compared to the LC-MVP model.
This could be explained by the latent trait's constrained parameterization creating asymmetric trade-offs.
The model can achieve partial correlation recovery (only $\sim 1.5-2$ fold worse than LC-MVP) by finding 
$b$ parameters that approximate the heterogeneous, non-latent-trait correlation pattern.
However, these "compromise" $b$ values - which don't match the true generating structure - 
are then pushed through the non-linear transformation $\Phi(a_t/\sqrt{1 + b^{[d]}_t})$ for accuracy estimation.
This transformation amplifies the misspecification, particularly at high sensitivity values where small 
changes in the arguments produce larger changes in the resulting probability.
Essentially, the latent trait achieves its moderate correlation recovery by selecting $b$ values that are 
catastrophic for accuracy estimation - it can partially match one target (correlations) but at 
disproportionate cost to the other (accuracies).
In contrast, LC-MVP's independent parameterization means correlation errors don't directly "contaminate"
accuracy estimates, allowing it to maintain better performance even with imperfect correlation recovery.

\underline{\textbf{Comparison between DGM \#3 and \#5}} \\ \newline
Another important thing to note here is that - 
despite worse correlation recovery (for $N = 3000$) - 
both models achieve better RMSE(Se) for DGM \#5 than DGM \#3,
for instance, for $N = 3000$, for LC-MVP we get a range of: $2.52 - 4.04$ for DGM \#5 vs $3.45 - 4.99$ for DGM \#3, 
and for latent trait $8.01 - 11.6$ for DGM \#5 vs $10.86 - 11.19$ for DGM \#3).
This is consistent with the ceiling effect (see section \ref{section_ceiling_effect_paradox}):
with high sensitivities ($86 - 92.5\%$), correlation misspecification matters less for accuracy estimation -
allowing both models to achieve better accuracy - despite worse correlation recovery.
Also, the latent trait model still potentially suffers from its structural constraints 
(b parameters having to determine both accuracy and correlation),
but the ceiling effect may partially mask this incompatibility compared to DGM \#3.

This superior accuracy recovery for DGM \#5 occurs because the higher sensitivity values ($86\% - 92.5\%$, with test 1 at $92.5\%$) - 
combined with the mildly-informative prior for the reference test 
($95\%$ prior interval of $(0.63, 0.99)$ with a prior median of $0.90$) -
create an overwhelming signal that "most diseased individuals test positive", 
allowing the data to more easily dominate the likelihood, even with $N = 300$.
In contrast, DGM \#3's lower sensitivities ($55\% - 70\%$, with test 1 at $65\%$) 
and less informative prior ($95\%$ prior interval of $(0.43, 0.90)$) 
provide a weaker signal with more variability in outcomes.
This means that overall, the correlation RMSE does not matter as much for DGM \#5 compared to DGM \#3, 
which explains why - 
despite the RMSE of the correlations being very similar to DGM \#3 (or worse in the $N = 3000$ case) -
DGM still \#5 achieves better accuracy (than DGM \#3), despite these similar or worse correlation estimates.

Looking at the correlation bias patterns for these heterogeneous structures does not seem to provide additional mechanistic insight.
For instance, for DGM \#3 at $N = 300$, the LC-MVP model with LKJ priors maintains good sensitivity estimation (RMSE(Se): $8.49 - 8.78$) - 
despite substantial negative correlation bias ($-1.50$ to $-1.74$) -
which suggests that correlation recovery is neither necessary nor sufficient for this flexible model 
(at least when using standard LKJ priors).
These necessary \& sufficient conditions (for both models) are discussed in more detail in section \ref{section_results_DGMs_2_4_between_model_comparison_and_corr_analysis},
where this idea helps explain the patterns seen there more.

\subsection{DGM's \#2 and \#4 (latent trait structure DGM's)}
\label{section_results_DGMs_2_4}
\begin{figure}[H]
      \centering
    \includegraphics[width=16cm]{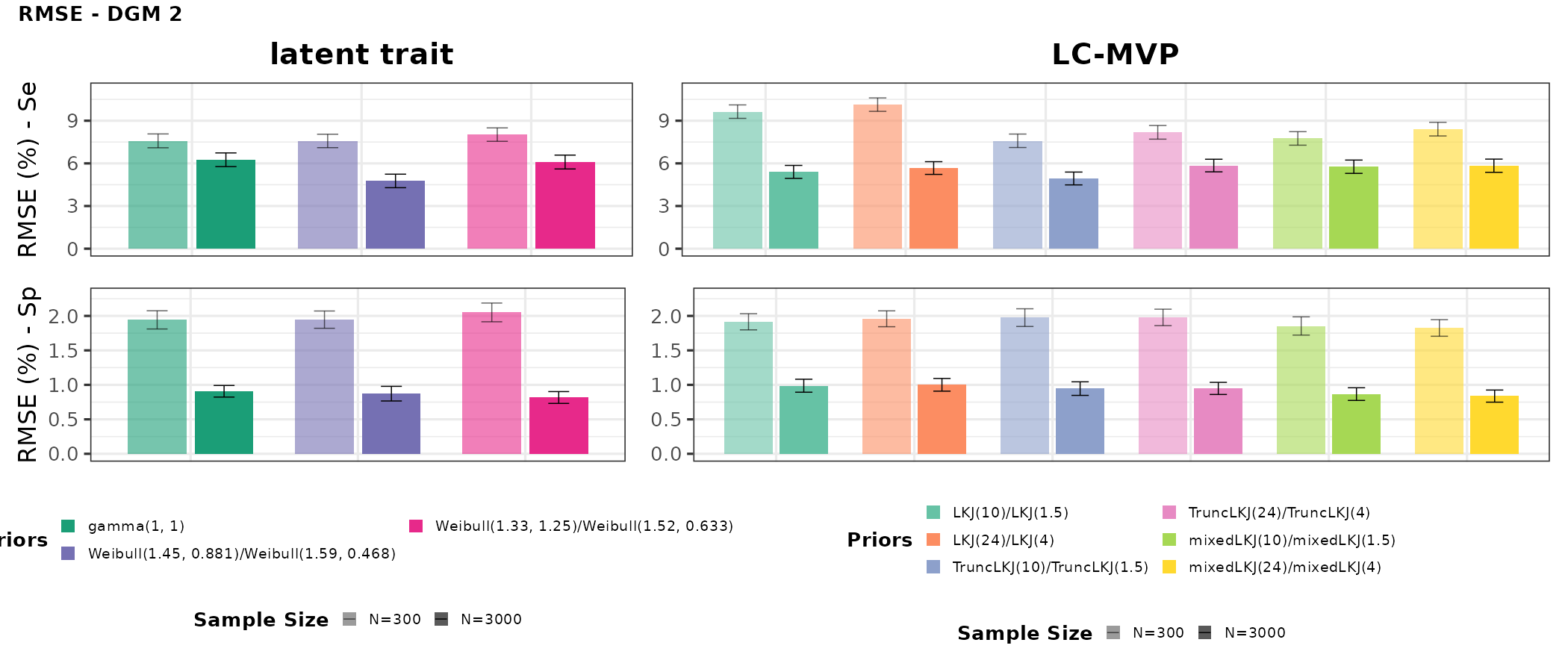}
    \caption{ Simulation study results (Se and Sp) for DGM \#2 - RMSE }
    \label{Figure:Sim_study_RMSE_DGM_2_}
\end{figure}
\begin{figure}[H]
      \centering
    \includegraphics[width=16cm]{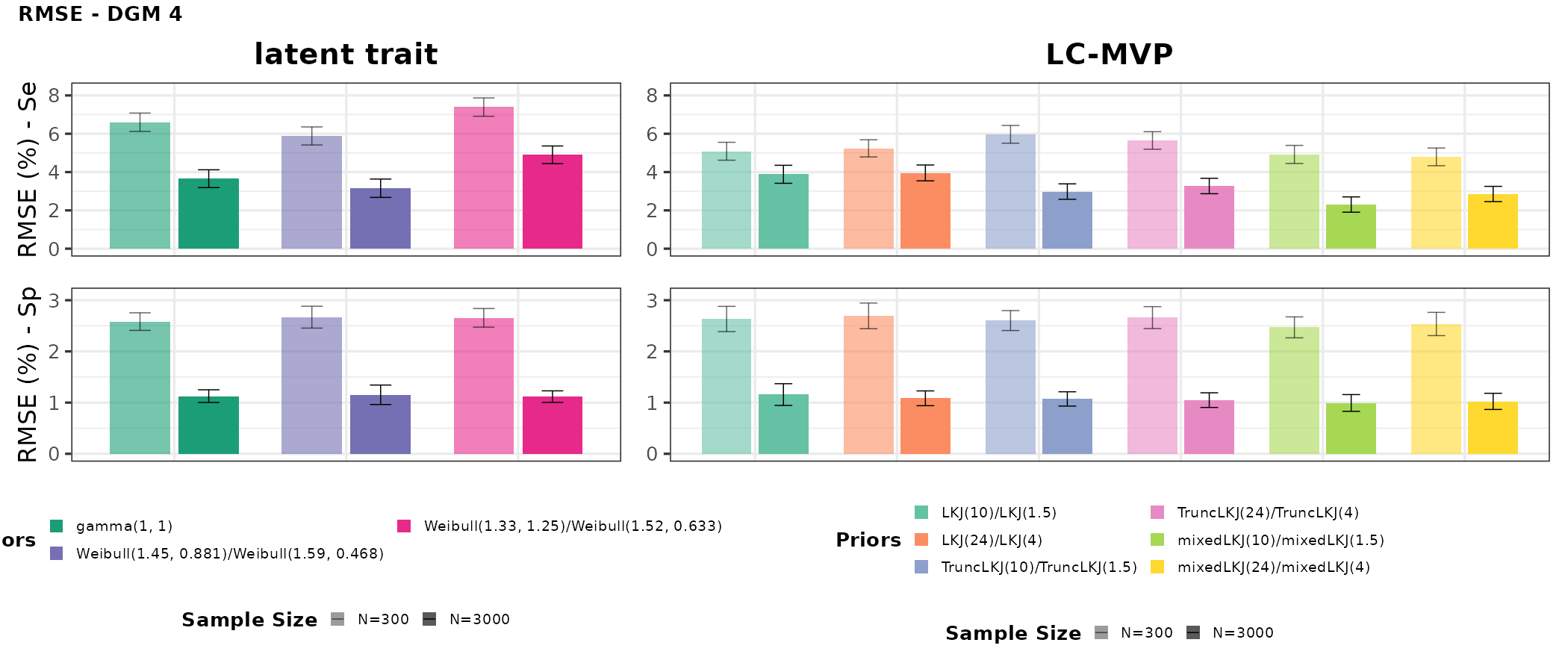}
    \caption{ Simulation study results (Se and Sp) for DGM \#4 - RMSE }
    \label{Figure:Sim_study_RMSE_DGM_4_}
\end{figure}

\subsubsection{ Model \& prior performance (DGM's \#2 \& \#4) }
\label{section_results_DGMs_2_4_model_and_prior_performance} 
\begin{table}[H]
\centering
\caption{
LC-MVP prior performance for DGMs \#2 \& \#4 (Latent Trait structure)
(statistically equivalent groups at 95\% CI; ranked by RMSE)
}
\footnotesize
\setlength{\tabcolsep}{4pt}
\begin{tabular}{c|l|l}
\toprule
\textbf{DGM} & \textbf{N=300} & \textbf{N=3000} \\
\midrule
\multicolumn{3}{c}{\textbf{By Overall RMSE (Se + Sp)}} \\
\midrule
2 & \makecell[l]{\underline{Best ($9.52 - 10.2$):} \\ 
               Latent trait w/ Weibull(1.45) [$9.52$, $2.78$], \\
               Latent trait w/ Gamma(1, 1) [$9.53$, $2.37$], \\
               LC-MVP w/ TruncLKJ(10, 1.5) [$9.56$, $2.53$], \\
               LC-MVP w/ mixedLKJ(10, 1.5) [$9.61$, $4.15$], \\
               Latent trait w/ Weibull(1.33) [$10.1$, $4.11$], \\
               LC-MVP w/ TruncLKJ(24, 4) [$10.2$, $3.47$], \\
               LC-MVP w/ mixedLKJ(24, 4) [$10.2$, $4.69$]. \\
               \\
               \underline{Worse ($11.6 - 13.8$):} \\
               LC-MVP w/ LKJ(10, 1.5) [$11.6$, $6.47$], \\
               LC-MVP w/ LKJ(24, 4) [$12.1$, $6.99$], \\
               CI[$13.8$, $9.41$].}
  & \makecell[l]{\underline{Best ($5.64 - 6.39$):} \\
               Latent trait w/ Weibull(1.45) [$5.64$, $2.25$], \\
               LC-MVP w/ TruncLKJ(10, 1.5) [$5.88$, $3.89$], \\
               LC-MVP w/ LKJ(10, 1.5) [$6.39$, $3.44$]. \\   
               \\
               \underline{Worse ($6.63 - 12.4$:} \\
               LC-MVP w/ mixedLKJ(10, 1.5) [$6.63$, $4.78$], \\
               LC-MVP w/ mixedLKJ(24, 4) [$6.67$, $4.90$], \\
               LC-MVP w/ LKJ(24, 4) [$6.67$, $4.35$], \\
               LC-MVP w/ TruncLKJ(24, 4) [$6.80$, $5.05$]. \\
               Latent trait w/ Weibull(1.33) [$6.91$, $4.14$], \\
               Latent trait w/ Gamma(1, 1) [$7.16$, $3.26$], \\
               CI [$12.4$, $11.6$].}  \\
\midrule
4 & \makecell[l]{\underline{Best ($7.33 - 8.31$):} \\ 
               LC-MVP w/ mixedLKJ(24, 4) [$7.33$, $1.64$], \\ 
               LC-MVP w/ mixedLKJ(10, 1.5) [$7.39$, $1.91$], \\
               LC-MVP w/ LKJ(10, 1.5) [$7.72$, $2.61$], \\
               LC-MVP w/ LKJ(24, 4) [$7.93$, $3.10$], \\
               CI [$8.18$, $3.92$], \\
               LC-MVP w/ TruncLKJ(24, 4) [$8.31$, $2.54$]. \\
               \\
               \underline{Worse ($8.55 - 10.1$):} \\
               Latent trait w/ Weibull(1.45) [$8.55$, $2.91$], \\
               LC-MVP w/ TruncLKJ(10, 1.5) [$8.57$, $3.11$], \\
               Latent trait w/ Gamma(1, 1) [$9.18$, $4.27$], \\
               Latent trait w/ Weibull(1.33) [$10.1$, $5.47$].}
  & \makecell[l]{\underline{Best ($3.30 - 4.05$):} \\ 
                LC-MVP w/ mixedLKJ(10, 1.5) [$3.30$, $1.39$], \\
                LC-MVP w/ mixedLKJ(24, 4) [$3.38$, $2.47$], \\ 
                LC-MVP w/ TruncLKJ(10, 1.5) [$4.05$, $2.51$]. \\
                \\
                \underline{Worse ($4.31 - 6.02$):} \\ 
                Latent trait w/ Weibull(1.45) [$4.31$, $1.93$], \\
                LC-MVP w/ TruncLKJ(24, 4) [$4.32$, $3.11$], \\
                Latent trait w/ Gamma(1, 1) [$4.78$, $1.91$], \\
                LC-MVP w/ LKJ(24, 4) [$5.03$, $3.88$], \\  
                LC-MVP w/ LKJ(10, 1.5) [$5.04$, $3.79$], \\
                CI [$5.76$, $5.14$], \\
                Latent trait w/ Weibull(1.33) [$6.02$, $3.98$].} \\ 
\bottomrule
\end{tabular}
\begin{tablenotes}
\footnotesize
\item Note:
Recall that, for DGM \#2: \\
true sensitivities ($\%$) are: (65, 55, 60, 65, 70);
true specificities ($\%$) are: (99, 95, 90, 90, 85). \\
Recall that, for DGM \#4: \\ 
true sensitivities ($\%$) are: (92.5, 86, 87, 91, 86);
true specificities ($\%$) are: (95,   81, 70, 67, 85).
\end{tablenotes}
\label{Table:summary_table_overall_all_DGMs_2_4}
\end{table}
\underline{\textbf{DGM \#2 (Latent trait corr structure w/ low-moderate Se, high Sp), $N = 300$:}} \\ \newline
Table \ref{Table:summary_table_overall_all_DGMs_2_4} (top, left-hand side) reveals that at $N = 300$, 
seven model/prior combinations tie in the best group (RMSE: $9.52 - 10.2$, bias: $2.37 - 4.69$): all three latent trait priors 
and LC-MVP's constrained priors (both TruncLKJ and mixedLKJ).
On the other hand, the LC-MVP model with standard LKJ performs worse (RMSE: $11.6 - 12.1$, bias: $6.47 - 6.99$), 
whilst the CI model fails catastrophically (RMSE/bias = $13.8/9.41$).
This hierarchy - latent trait $\approx$ LC-MVP constrained $>>$ LC-MVP standard LKJ $>>$ CI - 
confirms that matching or approximating the true positive correlation structure is essential here.

\underline{\textbf{DGM \#2 (Latent trait corr structure w/ low-moderate Se, high Sp), $N = 3000$:}} \\ \newline
At $N = 3000$ (table \ref{Table:summary_table_overall_all_DGMs_2_4}, top right-hand side), 
for DGM \#2, only three priors now remain in the best-performing group:
with latent trait's more highly informative Weibull prior leading at $5.64$ 
(also with statistically significantly lower bias than any other model/prior combination: $2.25$), 
followed by LC-MVP's TruncLKJ(10, 1.5) at $5.88$ (bias: $3.89$), then LKJ(10, 1.5) at $6.39$ (bias: $3.44$).
The remaining four priors for the LC-MVP perform slightly worse (RMSE: $6.63 - 6.80$), and the remaining two priors
for the latent trait perform worse (RMSE: $6.91 - 7.16$).
The CI model by far performs the worst here, almost as badly as it did for $N = 300$ (RMSE: $12.4$) 
with sensitivity coverage collapsing to $15\%$.

\underline{\textbf{DGM \#4 (Latent trait corr structure w/ high Se + moderate-high Sp), $N = 300$:}} \\ \newline
Now we are looking at DGM \#4 for $N = 300$ (see table \ref{Table:summary_table_overall_all_DGMs_2_4}, bottom left-hand side). 
Despite using the exact same true correlation structure as DGM \#2, we obtain a much different pattern here.
Namely, 5 out of 6 LC-MVP prior/constraint configurations make it into the best-performing group, 
with the more informative mixedLKJ(24, 4) in the league (RMSE/bias: $7.33/1.64$), 
followed by the less informative mixedLKJ(10, 1.5) (RMSE/bias = $7.39/1.91$), 
which is then followed by both LKJ priors, which perform almost identically (RMSE: $7.72 - 7.93$, bias: $2.61 - 3.10$),
followed by the CI model in 5th place (RMSE/bias = $8.18/3.92$), 
and finally, the LC-MVP with TruncLKJ(24, 4) prior comes in at last place within the best-performing group
(RMSE/bias = $8.31/2.54$).
On the other hand, all three of the latent trait priors perform worse than the CI model 
(RMSE for latent trait: $8.55 - 10.1$ vs. $8.18$ for CI), 
and statistically significantly worse than the LC-MVP with the mixedLKJ(24, 4) prior 
(RMSE: $8.55 - 10.1$ vs. $7.33$ for mixedLKJ(24, 4)).

It is quite interesting here that, for the LC-MVP model - unlike DGM \#2 at $N = 300$ -
the standard LKJ priors performed well here (RMSE range: $7.72 - 7.93$).
This is likely because of the ceiling effect 
(which we discuss in section \ref{section_ceiling_effect_paradox}), and because we have wider credible intervals here at $N = 300$,
unlike $N = 3000$, which we will discuss in more detail next when we discuss this DGM for the larger $N = 3000$ sample size.
It is even more interesting here that the CI model actually performs quite well, 
despite there being substantial conditional dependence for this DGM. 
This must be due to the relatively higher sensitivities here ($86\% - 92.5\%$ vs. $55\% - 70\%$ for DGM \#2), 
creating the aforementioned "ceiling effect" which means that correlation structure matters less.
We will discuss how correlation structure matters less here in more detail (and with explicit evidence) in section \ref{section_results_DGMs_2_4_between_model_comparison_and_corr_analysis} below.

\underline{\textbf{DGM \#4 (Latent trait corr structure w/ high Se + moderate-high Sp), $N = 3000$:}} \\ \newline
At $N = 3000$ (bottom, right-hand side of table \ref{Table:summary_table_overall_all_DGMs_2_4}),
the pattern changes somewhat - whilst both mixedLKJ priors are still in the league (RMSE: $3.30 - 3.38$, bias: $1.39 - 2.47$),
but now it's the leff informative mixedLKJ(10, 1.5) in the league (RMSE/bias: $3.30/1.39$), 
and the TruncLKJ(10, 1.5) now does much better compared to the $N = 300$ case, coming in at third place 
(RMSE/bias: $4.05/2.51$).
This is now the only three model/prior combinations that make up the best-performing group.

In other words, the other three LC-MVP priors, as well as all three latent trait model priors, as well as the CI model,
are now all in the worst-performing group, all being statistically significantly worse than the best-performing mixedLKJ(10, 1.5) prior.
To be more specific, the RMSE range for the worse-performing group is $4.31 - 6.02$ (vs. $3.30 - 4.05$ for best),
with the three worst-performing LC-MVP priors still performing better than the CI model (RMSE: $4.32 - 5.04$ vs. $5.76$ for CI).
Regarding the latent trait model, its best prior (the more informative Weibull prior) actually performed quite well, 
and is the least bad of the "worse" group (RMSE/bias: $4.31/1.93$), 
with its bias actually being better than 2 out of 3 of the best-performing LC-MVP priors.
Then, the Gamma(1, 1) prior performed only slightly worse (RMSE/bias: $4.78/1.91$ - but still within MCSE of the Weibull(1.45) prior).
However, the less informative Weibull prior performed substantially and statistically significantly worse than either of the other 
two latent trait priors, being the only model/prior combination which actually had an RMSE even worse than the CI model
(RMSE/bias of Weibull(1.33) = $6.02/3.98$, vs. $5.76/5.14$ for CI model).

\underline{\textbf{Latent trait model's performance for DGM \#4:}} \\ \newline
The latent trait model's performance on its own generated data reveals an important limitation.
At $N = 300$, all three latent trait priors perform worse than even the CI model 
(RMSE: $8.55 - 10.1$ vs $8.18$ for CI), which is particularly surprising.
At $N = 3000$, whilst the model's best prior (Weibull 1.45) achieves reasonable performance (RMSE: $4.31$) - 
nearly within MCSE of the third-best configuration (TruncLKJ(10,1.5): $4.05$) - 
it still cannot match LC-MVP's top configurations ($3.30 - 3.88$) despite this being data generated from its own structure.
This underperformance on its own generated data - where we would expect the generating model to excel - 
highlights how high sensitivities ($86\% - 92.5\%$) create challenges even for the true model.

If we look at the correlation recovery results 
(tables \ref{Table:corr_rmse_LC_MVP_DGMs_2_4} and \ref{Table:corr_rmse_latent_trait_DGMs_2_4}),
the latent trait actually achieves very good correlation RMSE ($1.33 - 1.71$) -
paradoxically much better than the better-performing (in terms of RMSE(Se)) LC-MVP with standard LKJ priors ($3.05 - 3.26$) -
and comparable to the other LC-MVP priors 
(note that we will discuss the correlation recovery/RMSE results in more detail in section 
\ref{section_results_DGMs_2_4_between_model_comparison_and_corr_analysis} below).
This confirms the model is more or less correctly recovering the correlation structure - 
which is unsurprising since the data was generated from it.
However, in this case, this correlation accuracy comes at the direct expense of sensitivity/specificity estimation.
This is likely due to how the ceiling effect (see section \ref{section_ceiling_effect_paradox})
interacts with the "inverse problem" of the latent trait model - 
which is the term we are using to describe 
the fact that the latent trait model is essentially trying to recover (for dimension 5) 
10 correlations from just 5 $b$ parameters. 
Please see section \ref{section_latent_trait_inverse_problem} for more details of this inverse problem.

For DGM \#2 with moderate sensitivities ($55\% - 70\%$), there's substantial variation in test outcomes 
among the diseased - some test positive, others negative - providing relatively rich information for solving 
the inverse problem (see section \ref{section_latent_trait_inverse_problem} of recovering 10 correlations from just 5 $b$ parameters.
But for DGM \#4 with high sensitivities ($86\% - 92.5\%$), most diseased individuals test positive 
on most tests, creating far less variation despite the same $N = 300$ ($\sim 60$ diseased people).
Hence, this reduced variation means each diseased individual provides much less information about the underlying 
correlation structure, making the already challenging inverse problem more difficult 
(despite the model being more parsimonious than the LC-MVP model), 
compared to just directly estimating the correlations directly like the LC-MVP model does.


On the other hand, the LC-MVP's direct parameterization - despite being less parsimonious -
avoids this inverse problem entirely, whilst the ceiling effect masks its poorer correlation recovery performance
(when using standard LKJ priors that is) - 
explaining why it excels even when it's correlation recovery is worse than the latent trait model.
This reveals a crucial interaction: high sensitivities that create ceiling effects 
not only reduce the importance of correlation modelling, but they may also make inverse problems 
in constrained models - like the latent trait model - 
dramatically harder to solve due to the data being less informative about the correlations.


\subsubsection{ Correlation analysis (DGM's \#2 \& \#4) }
\label{section_results_DGMs_2_4_between_model_comparison_and_corr_analysis} 
\begin{table}[H]
\centering
\caption{
Correlation RMSE and bias for LC-MVP model in diseased group; 
for DGMs \#2 and \#4
}
\scriptsize
\begin{tabular}{cclcccc}
\toprule
\textbf{DGM} & \textbf{N} & \textbf{Prior} & \textbf{RMSE(Se)} & \textbf{Total Corr} & \textbf{Test 1 pairs} & \textbf{Tests 2-5 pairs} \\
             &            &                & \textbf{(\%)}     & {RMSE/Bias}         & {RMSE/Bias}          & {RMSE/Bias} \\
\midrule
2 & 300 & LKJ(10, 1.5)      & 9.64 & 3.01/-2.39 & 0.92/-0.61 & 2.08/-1.78 \\
2 & 300 & LKJ(24, 4)        & 10.1 & 3.12/-2.76 & 0.89/-0.70 & 2.23/-2.06 \\
2 & 300 & TruncLKJ(10, 1.5) & 7.59 & 1.34/-0.20 & 0.49/+0.29 & 0.85/-0.50 \\
2 & 300 & TruncLKJ(24, 4)   & 8.18 & 1.52/-1.02 & 0.35/+0.01 & 1.17/-1.03 \\
2 & 300 & mixedLKJ(10, 1.5) & 7.75 & 1.89/-1.30 & 0.88/-0.58 & 1.00/-0.72 \\
2 & 300 & mixedLKJ(24, 4)   & 8.41 & 2.14/-1.83 & 0.85/-0.67 & 1.28/-1.16 \\
\midrule
2 & 3000 & LKJ(10, 1.5)      & 5.40 & 1.34/-0.66 & 0.54/-0.10 & 0.80/-0.56 \\
2 & 3000 & LKJ(24, 4)        & 5.67 & 1.48/-1.02 & 0.49/-0.18 & 0.99/-0.83 \\
2 & 3000 & TruncLKJ(10, 1.5) & 4.94 & 1.06/-0.41 & 0.36/+0.05 & 0.70/-0.47 \\
2 & 3000 & TruncLKJ(24, 4)   & 5.85 & 1.28/-0.88 & 0.34/-0.08 & 0.94/-0.80 \\
2 & 3000 & mixedLKJ(10, 1.5) & 5.76 & 1.37/-0.93 & 0.66/-0.41 & 0.71/-0.51 \\
2 & 3000 & mixedLKJ(24, 4)   & 5.83 & 1.49/-1.17 & 0.58/-0.39 & 0.91/-0.78 \\
\midrule
\midrule
4 & 300 & LKJ(10, 1.5)      & 5.09 & 3.05/-2.69 & 0.74/-0.57 & 2.31/-2.11 \\
4 & 300 & LKJ(24, 4)        & 5.24 & 3.26/-3.08 & 0.80/-0.71 & 2.46/-2.37 \\
4 & 300 & TruncLKJ(10, 1.5) & 5.97 & 1.30/+0.06 & 0.59/+0.54 & 0.71/-0.48 \\
4 & 300 & TruncLKJ(24, 4)   & 5.65 & 1.40/-0.97 & 0.25/+0.12 & 1.15/-1.08 \\
4 & 300 & mixedLKJ(10, 1.5) & 4.92 & 1.52/-1.15 & 0.65/-0.52 & 0.86/-0.63 \\
4 & 300 & mixedLKJ(24, 4)   & 4.79 & 2.00/-1.87 & 0.76/-0.71 & 1.24/-1.16 \\
\midrule
4 & 3000 & LKJ(10, 1.5)      & 3.88 & 2.62/-2.20 & 0.67/-0.47 & 1.95/-1.72 \\
4 & 3000 & LKJ(24, 4)        & 3.95 & 2.68/-2.42 & 0.68/-0.58 & 2.00/-1.85 \\
4 & 3000 & TruncLKJ(10, 1.5) & 2.98 & 1.58/+0.01 & 0.71/+0.63 & 0.87/-0.62 \\
4 & 3000 & TruncLKJ(24, 4)   & 3.27 & 1.54/-0.82 & 0.37/+0.22 & 1.17/-1.04 \\
4 & 3000 & mixedLKJ(10, 1.5) & 2.30 & 1.37/-0.81 & 0.63/-0.33 & 0.74/-0.48 \\
4 & 3000 & mixedLKJ(24, 4)   & 2.86 & 1.73/-1.46 & 0.63/-0.49 & 1.10/-0.96 \\
\midrule
\bottomrule
\end{tabular}
\begin{tablenotes}
\footnotesize
\item Note: Values shown as RMSE/bias. Negative bias indicates underestimation of correlations.
\end{tablenotes}
\label{Table:corr_rmse_LC_MVP_DGMs_2_4}
\end{table}
\begin{table}[H]
\centering
\caption{
Correlation RMSE and bias for latent trait model in diseased group; 
for DGMs \#2 and \#4
}
\footnotesize
\begin{tabular}{cclcccc}
\toprule
\textbf{DGM} & \textbf{N} & \textbf{Prior} & \textbf{RMSE}     & \textbf{Total Corr}   & \textbf{Test 1 pairs}   & \textbf{Tests 2-5 pairs} \\
             &            &                & \textbf{Se(\%)}   & {RMSE/Bias}          & {RMSE/Bias}            & {RMSE/Bias} \\
\midrule
2 & 300 & gamma(1, 1) & 7.58 & 1.49/-0.35 & 0.50/+0.16 & 0.99/-0.52 \\
2 & 300 & Weibull($\frac{1.59}{1.45}, \frac{0.468}{0.881}$) & 7.58 & 1.33/-0.60 & 0.38/+0.14 & 0.96/-0.74 \\
2 & 300 & Weibull($\frac{1.52}{1.33}, \frac{0.633}{1.25}$) & 8.02 & 1.43/+0.43 & 0.69/+0.53 & 0.74/-0.10 \\
\midrule
2 & 3000 & gamma(1, 1) & 6.25 & 1.12/+0.29 & 0.52/+0.32 & 0.60/-0.03 \\
2 & 3000 & Weibull($\frac{1.59}{1.45}, \frac{0.468}{0.881}$) & 4.76 & 0.85/+0.10 & 0.38/+0.22 & 0.46/-0.12 \\
2 & 3000 & Weibull($\frac{1.52}{1.33}, \frac{0.633}{1.25}$) & 6.09 & 1.00/+0.48 & 0.53/+0.38 & 0.47/+0.10 \\
\midrule
\midrule
4 & 300 & gamma(1, 1) & 6.60 & 1.54/+0.16 & 0.69/+0.55 & 0.85/-0.39 \\
4 & 300 & Weibull($\frac{1.59}{1.45}, \frac{0.468}{0.881}$) & 5.89 & 1.33/-0.32 & 0.46/+0.39 & 0.86/-0.71 \\
4 & 300 & Weibull($\frac{1.52}{1.33}, \frac{0.633}{1.25}$) & 7.39 & 1.71/+1.21 & 1.04/+0.99 & 0.67/+0.22 \\
\midrule
4 & 3000 & gamma(1, 1) & 3.65 & 1.69/+0.28 & 0.81/+0.68 & 0.88/-0.40 \\
4 & 3000 & Weibull($\frac{1.59}{1.45}, \frac{0.468}{0.881}$) & 3.16 & 1.42/-0.01 & 0.60/+0.52 & 0.82/-0.53 \\
4 & 3000 & Weibull($\frac{1.52}{1.33}, \frac{0.633}{1.25}$) & 4.90 & 1.79/+1.15 & 1.10/+1.04 & 0.69/+0.11 \\
\midrule
\bottomrule
\end{tabular}
\begin{tablenotes}
\footnotesize
\item Note: Values shown as RMSE/bias. Positive bias indicates overestimation of correlations.
\end{tablenotes}
\label{Table:corr_rmse_latent_trait_DGMs_2_4}
\end{table}
\begin{table}[H]
\centering
\caption{DGMs 2 \& 4 (True Latent Trait Structure): Correlations between RMSE(Se) and correlation recovery}
\begin{tabular}{clccc}
\toprule
DGM & Sample Size/Model &  
\makecell[l]{ Cor(RMSE(Se), \\ RMSE(Cor) } & 
\makecell[l]{ Cor(RMSE(Se), \\ RMSE(test-1-Cor) } & 
\makecell[l]{ Cor(RMSE(Se), \\RMSE(tests-2-5-Cor) } \\
\midrule
\multirow{4}{*}{2} & N=300, LC-MVP & 0.95 & 0.55 & \textbf{1.00} \\
                   & N=300, Latent Trait & 0.14 & 0.92 & -0.99$^*$ \\
                   & N=3000, LC-MVP & 0.79 & 0.39 & 0.61 \\
                   & N=3000, Latent Trait & 0.94 & 0.99 & 0.63 \\
\midrule
\multirow{4}{*}{4} & N=300, LC-MVP & -0.40 & -0.58 & -0.30 \\
                   & N=300, Latent Trait & 1.00 & 1.00 & -0.90$^*$ \\
                   & N=3000, LC-MVP & 0.91 & 0.08 & 0.95 \\
                   & N=3000, Latent Trait & 0.87 & 0.99 & -0.83$^*$ \\
\bottomrule
\multicolumn{5}{l}{\footnotesize $^*$Negative correlations reflect constraint trade-offs with only 3 priors}
\end{tabular}
\label{Table:corr_patterns_RMSE_both_models_DGMs_2_4}
\end{table}
We will now look at the relationship between correlation recovery
(see tables \ref{Table:corr_patterns_RMSE_both_models_DGMs_2_4},
\ref{Table:corr_rmse_LC_MVP_DGMs_2_4} and \ref{Table:corr_rmse_latent_trait_DGMs_2_4})
and sensitivity estimation (RMSE(Se)).

\underline{\textbf{DGM \#2 (Latent trait corr structure w/ low-moderate Se, high Sp), $N = 300$:}} \\ \newline
For LC-MVP at $N = 300$, correlation recovery strongly predicts sensitivity estimation
(Cor(RMSE(Se), RMSE(Cor)) = $0.95$), with tests 2-5 correlations being particularly important
(Cor(RMSE(Se), RMSE(tests-2-5-Cor)) = $1.00$ vs. Cor(RMSE(Se), RMSE(test-1-Cor)) = $0.55$).
This pattern is also clearly reflected in the actual RMSE correlation values; for instance, standard LKJ priors - 
the only ones in the "worst" group (RMSE(Se): $9.64 - 10.1$) -
obtain poor correlation recovery (RMSE(Cor): $3.01 - 3.12$),
whereas the best-performing TruncLKJ and mixedLKJ priors (RMSE(Se): $7.59 - 8.41$)
achieve greatly superior correlation recovery (RMSE(Cor): $1.34 - 2.14$).

The bias patterns reveal the mechanism: LKJ's massive negative bias ($-2.39$ to $-2.76$) 
reflects systematic underestimation of the true positive correlations, 
whilst TruncLKJ ($-0.20$ to $-1.02$) and mixedLKJ ($-1.30$ to $-1.83$) 
show more moderate bias patterns that better align with the true structure.

For the latent trait model, the pattern differs: test-1 correlations strongly predict performance 
(as opposed to test-2-5 correlations)
(Cor(RMSE(Se), RMSE(test-1-Cor)) = $0.92$, Cor(RMSE(Se), RMSE(test-2-5-Cor) = $-0.99$);
however, the very the narrow RMSE(Se) range for the latent trait model ($7.58 - 8.02$)
greatly limits interpretation. 
Also, it is worth noting here that the latent trait model achieves similar performance to LC-MVP model with TruncLKJ priors
(RMSE(Se): $7.58 - 8.02$ vs. $7.59 - 8.18$ for TruncLKJ) -
with very similar correlation recovery/RMSE (RMSE(Cor): $1.33 - 1.49$ vs. $1.34 - 1.52$) -
which suggests that matching the true positive correlation structure is important for this DGM.

\underline{\textbf{ DGM \#2 (Latent trait corr structure w/ low-moderate Se, high Sp), $N = 3000$:}} \\ \newline
At $N = 3000$, latent trait shows clear patterns: unlike the $N = 300$ case, 
RMSE(Cor) strongly predicts RMSE(Se) (cor = $0.94$),
with test-1 correlations being extremely predictive (Cor(RMSE(Se), RMSE(test-1-Cor)) = $0.99$), 
and test-2-5 correlations more moderately predictive of sensitivity recovery (cor = $0.63$).
The best-performing Weibull(1.45) prior (RMSE(Se) = $4.76$) achieves the best correlation recovery
(RMSE(Cor) = $0.85$), notably better than the other priors ($1.00 - 1.12$),
which is consistent with this prior also having the best - and also statistically significantly better -
RMSE(Se) ($4.76$ vs. $6.09 - 6.26$ for the other two latent trait priors).

For LC-MVP, the patterns are less clear. While overall correlation recovery shows moderate correlation 
with RMSE(Se) (r = $0.79$), test-1 correlations show weak correlation (r = $0.39$) 
and tests-2-5 show moderate correlation (r = $0.61$).
However, all six LC-MVP priors perform similarly (RMSE(Se): $4.94 - 5.85$ - less than 1 percentage point range), 
making it difficult to detect meaningful patterns.
The correlation RMSE values are correspondingly quite tight, considering there's six different priors (RMSE(Cor): $1.06 - 1.49$).

Now comparing between models: LC-MVP achieves comparable or better RMSE(Se) ($4.94 - 5.85$) 
compared to latent trait ($4.76 - 6.25$), despite having worse overall correlation recovery 
($1.06 - 1.49$ for LC-MVP vs. $0.85 - 1.12$ for latent trait).
However - both models interestingly achieve similar test-1 correlation recovery 
($0.34 - 0.66$ for LC-MVP vs. $0.38 - 0.52$ for latent trait),
which suggests that this might be more important here for sensitivity recovery, despite
the Cor(RMSE(Se), RMSE(test-1-Cor)) being only $0.39$ for the LC-MVP model
(perhaps just not enough variation to pick up the correlation fully).

On the other hand, this discrepancy in Cor(RMSE(Se), RMSE(test-1-Cor)) could suggest different mechanisms 
(at least in this particular case):
perhaps the latent trait model requires more accurate test-1 correlation recovery 
for good performance on its own generated data (supported by the near-perfect correlation of $0.99$), 
whilst the LC-MVP model's performance could be less dependent on any specific correlation pattern 
(test-1 cor only $0.39$), and perhaps more dependent on the overall pattern (supported by Cor(RMSE(Se), RMSE(Cor)) = $0.79$).
This could be because LC-MVP's flexible parameterization - where accuracies and correlations are estimated independently - 
allows it to maintain more stable performance across different prior choices, and perhaps compensate for correlation misspecification 
in ways the constrained latent trait cannot - making it more sensitive to how well specific correlations are recovered.
The narrow performance range across all six LC-MVP priors ($4.94 - 5.85$) versus the wider range 
for just three latent trait priors ($4.76 - 6.25$) supports this interpretation - 
LC-MVP appears more robust to prior choice for this DGM at larger sample sizes.

\underline{\textbf{DGM \#4 (Latent trait corr structure w/ high Se + moderate-high Sp), $N = 300$:}} \\ \newline
At $N = 300$, the ceiling effect dominates. 
LC-MVP shows negative correlations between RMSE(Se) and correlation recovery (r = $-0.40$),
but with only 1.2 percentage points variation in RMSE(Se) ($4.79 - 5.97$), 
this is likely noise rather than true signal.
All LC-MVP priors perform similarly, including standard LKJ (RMSE(Se): $5.09 - 5.24$)
despite its poor correlation recovery (RMSE(Cor): $3.05 - 3.26$ vs. $1.30 - 2.00$ for others).

On the other hand, the latent trait model shows perfect correlation between RMSE(Se) and correlation recovery (r = $1.00$),
with wider (especially considering only three priors vs. six for LC-MVP) variation in RMSE(Se) ($5.89 - 7.39$) -
providing more meaningful signal to detect this correlation.
However, even the best latent trait prior (Weibull)1.45), with RMSE(Se) = $5.89$) performs worse than the LC-MVP when using standard LKJ priors (RMSE(Se): $5.09 - 5.24$ vs. $5.89$ for latent trait), despite the latent trait model obtaing around 2-fold better correlation recovery (RMSE(Cor): $1.33 - 1.71$ vs. LC-MVP's $3.05 - 3.26$ for LKJ).
This paradox very clearly reveals the ceiling effect (see section \ref{section_ceiling_effect_paradox}):
with sensitivities of $86 - 92.5\%$,
correlation structure does not matter as much - since diseased individuals test positive regardless.
However, this may not hold at other sample sizes - but it does in this case ($N = 300$).

\underline{\textbf{DGM \#4 (Latent trait corr structure w/ high Se + moderate-high Sp), $N = 3000$:}} \\ \newline
At $N = 3000$, the ceiling effect (which dominated at $N = 300$) vanishes -
and correlation recovery appears to become more critical for both models
(Cor(RMSE(Se), RMSE(Cor)): $0.91$ for LC-MVP, $0.87$ for latent trait).
Interestingly, the correlation patterns (table \ref{Table:corr_patterns_RMSE_both_models_DGMs_2_4}) 
suggest that the models may prioritize different correlations
(as usual though, these findings need to be interpreted with caution as this could be more noise rather than signal - 
especially when correlations are very low or negative).
More specifically, for the LC-MVP model, sensitivity recovery appears to be more dependent on tests 2-5 correlations
(r = $0.95$ vs. $0.08$ for test-1 correlations, with overall Cor(RMSE(Se), RMSE(Cor)) = $0.91$),
whilst the latent trait model prioritizes test-1 correlations
(r = $0.99$ vs. $-0.83$ for tests 2-5, and $0.87$ for overall Cor(RMSE(Se), RMSE(Cor))).

When we look at the actual RMSE of the correlations, we can see that the best LC-MVP priors 
(both mixedLKJ and TruncLKJ(10,1.5) priors - the only three model/prior combinations which made it into the "best" group) 
obtained an  RMSE(Se) of $2.30 - 2.98$, with an RMSE(Cor) of $1.37 - 1.73$.
On the other hand, LC-MVP with standard LKJ priors achieved substantially worse sensitivity recovery 
(RMSE(Se): $3.88 - 3.95$), 
with a correspondly much worse correlation recovery (RMSE(Cor): $2.62 - 2.68$).

For the latent trait model, we obtain good correlation recovery across all priors (RMSE(Cor): $1.42 - 1.79$),
yet only its best prior 
(Weibull(1.45); RMSE(Se) = $3.16$ vs. $3.65 - 4.90$ for the other two priors, with Gamma(1, 1) being in the middle) 
comes close to matching the LC-MVP model's performance.
This RMSE(Cor) range we obtained for the latent trait model ($1.42 - 1.79$) is also interesting because
it is much better than the LC-MVP model with the worst-performing standard LKJ priors 
(RMSE(Cor) for LC-MVP with LKJ priors: $2.62 - 2.68$), despite the fact that standard LKJ still achieves better RMSE(Se) 
than the latent trait's worst-performing prior 
(RMSE(Se) for LC-MVP with LKJ: $3.88 - 3.95$ vs. $4.90$ for latent trait's worse Weibull(1.33) prior).

The correlation bias analysis shows LC-MVP with LKJ maintains substantial negative bias ($-2.20$ to $-2.42$) - even here at $N = 3000$ -
yet still achieves competitive RMSE(Se). 
This persistent systematic bias - not just variance - 
demonstrates that LC-MVP's flexible parameterization can compensate for correlation misspecification
in ways the structurally constrained latent trait model cannot.

These findings here suggest that the ceiling effect (see section \ref{section_ceiling_effect_paradox}) 
that protects at $N = 300$ (for this DGM - DGM \#4) disappears with the tighter intervals at $N = 3000$.

The above patterns also suggest the following might be true: for the true generating model (latent trait),
good correlation recovery is a necessary condition - but not a sufficient condition - for predicting RMSE(Se).
We can see this very clearly if, for instance, we compare the latent trait model with it's worst-performing prior (Weibull(1.33)) -
which obtains an RMSE(Se) of $4.90$ and RMSE(Cor) of $1.79$ - 
to the LC-MVP with the mixedLKJ(24, 4) prior, which obtains virtually identical RMSE(Cor) ($1.73$),
yet a statistically significantly better (and also substantially better in magnitude) RMSE(Se) of $2.86$.
Another way to see this (for the latent trait) is by this observation: all three latent trait priors achieve reasonably good
correlation recovery ($1.42 - 1.79$) - but this is clearly not sufficient for good sensitivity estimation, as RMSE(Se) 
still varies widely ($3.16 - 4.90$).
In other words - this suggests that good correlation recovery is necessary (or may be) for the latent trait mode
(since all three priors have it), but not sufficient - since having it does not guarantee good RMSE(Se).

On the other hand, for the LC-MVP model, the LKJ priors have relatively terrible correlation recovery
(2.62 - 2.68), yet they still manage to get good RMSE(Se) here ($3.88 - 3.95$) - an overall better range than the 
latent trait model ($3.16 - 4.90$).
This proves that good correlation recovery is \textit{not} a necessary condition for the LC-MVP - 
in other words, the LC-MVP model can obtain good RMSE(Se) even with terrible correlation recovery - 
at least when using the unrestricted LKJ priors.

\underline{\textbf{Comparison between DGM \#2 and \#4}} \\ \newline
It is also important to note and discuss why the RMSE(Se) is much better for DGM \#4 compared to DGM \#2 - 
especially for $N = 3000$ - despite obtaining much worse correlation recovery.
More specifically, for DGM \#2 we obtain an RMSE(Se) range of $4.94 - 5.85$ for LC-MVP model
(and $4.76 - 6.25$ for latent trait), 
compared to a range of only $2.30 - 3.95$ for DGM \#4 (and $3.16 - 4.90$ for latent trait).
However, for DGM \#2 we obtain correlation recovery between $1.06 - 1.49$ for the LC-MVP model ($0.85 - 1.12$ for latent trait),
but for DGM \#4 we obtain correlation RMSE between $1.37 - 2.68$ for LC-MVP ($1.42 - 1.79$ for latent trait).
This apparent paradox is due to the ceiling effect 
(see section \ref{section_ceiling_effect_paradox} for more details on this),
and parallels the discrepancy we also observed when comparing DGMs \#3 and \#5 to one another 
(see section \ref{section_results_DGMs_3_5_correlation_analysis}).
Essentially, with the high sensitivities of DGM \#4 ($86 - 92.5\%$ vs. $55\% - 70\%$ for DGM \#2), 
correlation misspecification does not matter as much for accuracy estimation -
allowing both models to achieve better accuracy - despite worse correlation recovery.


The correlation bias patterns reveal why this paradox occurs: 
for DGM \#2 at $N = 3000$, all LC-MVP priors converge to similar moderate negative bias ($-0.41$ to $-1.17$) 
because the low-moderate sensitivity values ($55 - 70\%$) provide sufficient information for correlation recovery.
In contrast, for DGM \#4 at $N = 3000$, the high sensitivity values ($86 - 92.5\%$) create a ceiling effect 
(see section\ref{section_ceiling_effect_paradox}) that prevents this convergence - 
LKJ maintains extreme negative bias ($-2.20$ to $-2.42$) while other LC-MVP priors achieve much better bias ($-1.46$ to $+0.01$).
This differential bias convergence explains within-LC-MVP performance differences: 
for DGM \#4, priors with better correlation bias achieve better RMSE(Se). 
However, between models, LC-MVP's flexible parameterization allows it to maintain competitive performance despite poor correlation recovery, 
while the latent trait's constrained structure cannot fully compensate even with better correlation estimates. 
This fundamental structural difference - not correlation recovery quality - determines model performance.

Additionally, it is possible that the latent trait's constrained parameterisation - 
even when doing a good job at recovering correlations -
still distorts accuracy estimates through the non-linear transformation,
whilst LC-MVP's flexible parameterisation with appropriate constraints 
achieves better or comparable accuracy - despite sometimes worse correlation recovery - 
especially when using standard LKJ priors, where we often note good RMSE(Se) performance,
despite relatively poor RMSE(Cor) performance.

These findings also have potential implications for model checking - and suggest that simple methods like the 
"correlation residual plot" (Qu et al, 1996 \supercite{Qu1996}) may not be very useful for comparing different latent class models,
and more robust methods such as the LOO-IC (assuming sufficiently good pareto-k diagnostics),
or full K-fold cross validation might be needed,
or at least focusing on the full 2^T table residuals - like we do in our DTA-MA R shiny application - 
MetaBayesDTA (Cerullo et al, 2023\supercite{cerullo_shiny_app_MetaBayesDTA}).




\textbf{Key finding: Necessary \& sufficient conditions for correlation recovery}
\begin{itemize}
    \item For the latent trait model: Good correlation recovery appears necessary but not sufficient for optimal sensitivity estimation. 
    All priors achieve moderate RMSE(Cor) ($1.42 - 1.79$) due to structural constraints, 
    but varying bias patterns and implementation details lead to different RMSE(Se) outcomes.
    \item For the LC-MVP model: Good correlation recovery is neither necessary nor sufficient. 
    The model can achieve reasonable RMSE(Se) despite poor correlation recovery (RMSE(Cor): $2.62 - 2.68$) 
    and extreme negative bias ($-2.20$ to $-2.42$) through flexible parameterization that separates accuracy and correlation parameters.
    \item Critical role of ceiling effects: At $N = 3000$, DGM \#2 (low Se) allows bias convergence across priors, 
    whilst DGM \#4 (high Se) maintains disparate bias patterns due to ceiling effects 
    (i.e., higher Se makes the data not informative enough, even at $N = 3000$), explaining persistent performance differences.
\end{itemize}

\subsection{ Efficiency analysis}
\label{section_results_efficiency_analysis}
Computational efficiency varied dramatically across priors (see Appendix \ref{appendix_E_efficiency_analysis} for details). 
From table \ref{Appendix_table:efficiency_N300_full} (in appendix \ref{appendix_E_efficiency_analysis}), 
we can see that, for $N = 300$, the median time to reach a minimum of 1000 effective samples (ESS) ranged from $6.1$ seconds to $145.1$ seconds.
For the LC-MVP model, it was clearly the TruncLKJ(10, 1.5) prior which was the least efficient - having time(min(ESS) = 1000) between $43.2$ to $87.6$ seconds.
For the latent trait model, the most informative Weibull(1.45, 0.881) prior was the most efficient (time between $9.9$ to $21.3$ seconds), and
the Gamma(1, 1) prior was the least efficient (time between $80.2$ seconds to $145.1$ seconds).

For $N = 3000$ (see table \ref{Appendix_table:efficiency_N3000_full} in appendix \ref{appendix_E_efficiency_analysis}),
we can see that now time(min(ESS) = 1000) ranged from $78.3$ to $1229.8$ seconds.
For the LC-MVP model, just like the $N = 300$ case, it was again the TruncLKJ(10, 1.5) prior that was the least efficient - 
obtaining effciiency between $459.5$ seconds to $1229.8$ seconds.
For the latent trait model, it was again the more informative Weibull(1.45, 0.881) prior which was the most efficiency (between $78.3$ to $182.1$ seconds) 
and the Gamma(1, 1) prior again performed the worst with efficiency between $350.1$ to $955.8$ seconds.

Regarding the inefficiency of the LC-MVP model when using TruncLKJ priors, we believe this inefficiency may stem from:
(i) initial values of $0.001$ for raw correlation parameters potentially starting far from typical posterior values;
(ii) prior-data conflict when TruncLKJ forces positive correlations despite true near-zero values 
(e.g., $0.10$ in diseased, $0.05$ in non-diseased for reference test correlations); 
(iii) Efficiency issues with the method we used to implement TruncLKJ (Pinkney et al, 2024 \supercite{Sean_Pinkney_2024_shortnoteflexiblecholesky})
(however, it is important to note that the method we used is not necessarily the most up-to-date Pinkney et al method)
or: a combination of two or more of the above.

Overall, standard LKJ priors consistently provided the best computational efficiency for LC-MVP models, with mixedLKJ coming in second,
followed lastly by TruncLKJ (see above for why this may have been the case here). 
For the latent trait model, it is clearly the Weibull priors which performed best, especially the more informative Weibull(1.45, 0.881) prior; 
however, even the less-informative Weibull(1.33, 1.25) prior generally was much more efficient than the Gamma(1, 1) prior.
Furthermore, unlike the case discussed above (i.e., LC-MVP model using TruncLKJ priors), this issue with Gamma(1, 1) 
was very unlikely to be due to starting values. 
Hence, the poor performance of the Gamma(1, 1) prior here suggests that there may be fundamental incompatibility between the 
gamma(1,1) prior and typical posterior values, potentially contradicting its use in previous literature 
(e.g., Keddie et al, 2023 \supercite{Keddie2023}).

\section{Discussion}
\label{section_discussion}
\subsection{ Discussion of key findings}
\label{section_discussion_discussion_of_key_findings}
From section \ref{section_results}, the LC-MVP model demonstrated superior overall performance (on average) across the five DGMs,
particularly with mixedLKJ priors.
The ceiling effect (see section \ref{section_ceiling_effect_paradox}) emerged as an important factor - 
high sensitivities universally reduce the importance of correlation recovery for sensitivity estimation.

\subsubsection{ Discussion of key findings: DGM \#1}
\label{section_discussion_discussion_of_key_findings_DGM_1}
For conditional independence (DGM \#1), LC-MVP with standard LKJ priors performs best at our smaller sample size ($N = 300$), 
correctly estimating zero correlations with minimal bias. 
However, TruncLKJ and all latent trait priors force positive correlations where none exist,
resulting in both worse RMSE and substantially higher bias. 
The CI model also performs very well, as expected given it matches the true structure.
On the other hand, for $N = 3000$, when looking at RMSE (our main measure), we found that the latent trait model actually did quite well - 
with RMSE within MCSE of the best-performing CI model - and the only LC-MVP prior which made it into the best-performing
group was the LKJ(24, 4) prior; although the LKJ(10, 1.5) prior only just didn't make it into this group.
However, when we look at bias, the LC-MVP with LKJ priors did the best (besides the CI model) - 
essentially matching the CI model's performance.

\subsubsection{ Discussion of key findings: DGMs \#3 and \#5}
\label{section_discussion_discussion_of_key_findings_DGMs_3_5}
DGMs \#3 and \#5 used a heterogenous correlation structure, generated from the LC-MVP model.
The only difference between them is that DGM \#3 had lower sensitivities, but higher specificities, compared to DGM \#5.
For DGM \#3 at $N = 300$, we found that the only two priors which made it into the best-performing group
was the LC-MVP model with the mixedLKJ priors - all other model/prior combinations performed substantially worse -
including the latent trait model and the CI model. The overall hierarchy we found was approximately:
LC-MVP w/ mixedLKJ $>>$ other LC-MVP priors $\sim$ latent trait $>$ CI.

For $N = 3000$ (DGM \#3), we found that the RMSE for all LC-MVP priors approximately halved compared to $N = 300$,
and it also improved notably for the CI model; however, for the latent trait model, the RMSE actually got
around 20\% worse with larger sample size.
Furthermore, we found once again that the only two model/prior combinations in the best-performing group was 
the LC-MVP model with mixedLKJ priors, which obtained substantially better RMSE than all other model/prior combinations.
In addition, all latent trait priors actually performed much worse (beyond MCSE) than the CI model.
Notably, the latent trait achieved comparable correlation recovery to LC-MVP with standard LKJ 
(RMSE(Cor) $1.77 - 2.13$ vs $1.12 - 1.17$),
yet accuracy RMSE was over 2-fold worse ($10.86 - 11.19$ vs $4.72 - 4.77$) - 
demonstrating how the structural constraint can cause very poor accuracy estimates despite reasonable correlation recovery.
Hence here, the overall hierarchy we found was:
LC-MVP w/ mixedLKJ $>>$ other LC-MVP priors $>>$ CI $>>$ latent trait.

For DGM \#5 at $N = 300$, the results were less clear and more mixed, and there was essentially no clear pattern or heirarchy - 
with the latent trait with the more informative Weibull prior in the lead - but tying with six other model/prior configurations,
including: 
the latent trait model with the less informative Weibull priors, 
the LC-MVP with both mixedLKJ priors, 
the LC-MVP with both LKJ priors, 
and also the CI model.
On the other hand, the only three model/prior combinations in the worst-performing group was 
the LC-MVP model with both TruncLKJ priors, and the latent trait model with the Gamma(1, 1) prior, which performed the worst. 
Furthermore, the reason for the discrepancy between DGMs \#3 and \#5 here for $N = 300$ - 
in other words, why DGM \#3 has quite a clear hierarchy whereas why DGM \#5 does not - 
is because the higher sensitivities of DGM \#5 (relative to DGM \#3) make the data less informative,
especially for such a small sample ($N = 300 \implies \sim 60$ people in the diseased group),
which also correlation estimation less important.
We discuss this more in the results section 
(section \ref{section_results} - 
specifically sections \ref{section_ceiling_effect_paradox} and 
\ref{section_results_DGMs_3_5_correlation_analysis}).

For $N = 3000$ (DGM \#5),
with the RMSE approximately halving with the increased sample size (but getting worse for latent trait model, just as with DGM \#3).
Furthermore, the pattern here was much clearer (compared to $N = 3000$).
More specifically, we found that only four model/prior combinations now made it into the best-performing group, 
which were:
the LC-MVP model with both mixedLKJ priors (with mixedLKJ(10, 1.5) in the lead), 
followed by the LC-MVP with standard LKJ priors - which performed nearly as well as mixedLKJ  - unlike DGM \#3 at the same sample size.
On the other hand, the LC-MVP model with TruncLKJ priors performed statistically significantly worse than the leading mixedLKJ(10, 1.5) 
priors (although only slightly worse than the standard LKJ priors), which were then followed by the CI model, 
and then finally all three latent trait model priors performed worst and extremely badly - having on average double the RMSE and bias
of the CI model, and around 3-fold worse RMSE than the LC-MVP model with standard LKJ priors.
Hence here, unlike $N = 300$, we had a very clear hierarchy:
LC-MVP w/ mixedLKJ $>$ LC-MVP w/ LKJ $>$ LC-MVP w/ TruncLKJ $>$ CI $>>>$ latent trait.

\subsubsection{ Discussion of key findings: DGMs \#2 and \#4}
\label{section_discussion_discussion_of_key_findings_DGMs_2_4}
DGMs \#2 and \#4 used a more homogenous correlation structure (relative to \#3 and \#5), which was generated from the more restricted
latent trait model. As with DGMs \#3 and \#5, the only difference between them was that DGM \#2 has lower sensitivities, but 
higher specificities. 
For DGM \#2 at $N = 300$, we found that all three latent trait priors, as well as LC-MVP model with TruncLKJ and mixedLKJ priors,
all performed equally as well in terms of RMSE. 
However, the LC-MVP model with standard LKJ priors was notably worse (in terms of both RMSE and bias), 
with the CI model being the very worst model.
Hence, the hierarchy here was: 
latent trait $\approx$ LC-MVP constrained $>>$ LC-MVP w/ standard LKJ $>>$ CI

For $N = 3000$ (DGM \#2), there was only three model/prior combinations in the best-performing group, which were:
the latent trait model with the more informative Weibull prior, the LC-MVP model with TruncLKJ(10, 1.5) prior, 
and finally the LC-MVP model with LKJ(10, 1.5) prior.
The latent trait model with Weibull(1.45) was the best, especially when taking its bias into account,
which was statistically significantly lower than any other model/prior combination.
Furthermore, although they did not make it into the best-performing group,
all other LC-MVP priors (LKJ(24, 4), TruncLKJ(24, 4), and both mixedLKJ priors) -
as well as the other two latent trait priors - performed approximately the same
(RMSE: $6.63 - 7.16$) and only slight worse than the priors in the best-performing group (RMSE: $5.64 - 6.39$);
however, when looking at bias, these four LC-MVP priors - 
but not the two latent trait ones - did perform notably worse (around or over $1\%$ worse bias).
Finally, the CI model was by far the worst here, with RMSE and bias both around $2-3$ fold worse than other model/prior combinations.

For DGM \#4 at $N = 300$, the high sensitivities create a ceiling effect where most diseased individuals 
test positive regardless of correlation structure - making correlation misspecification less harmful.
This explains why five out of six LC-MVP configurations perform well, including standard LKJ priors 
which struggled for DGM \#2 - their poor correlation recovery doesn't hurt because diseased people 
test positive anyway.
Even the CI model performs reasonably well despite assuming independence, because the high sensitivities 
protect against this misspecification.
Surprisingly, all three latent trait priors perform worse than CI, despite this being their own generated data.
The hierarchy was approximately: 
LC-MVP w/ mixedLKJ $>$ LC-MVP w/ LKJ $\sim$ CI $>$ LC-MVP w/ TruncLKJ $\sim$ latent trait w/ Weibull(1.45) $>$ latent trait.

For $N = 3000$ (DGM \#4), only three configurations achieve best performance:
LC-MVP with mixedLKJ(10, 1.5) [RMSE: $3.30$], mixedLKJ(24, 4) [RMSE: $3.88$], and TruncLKJ(10, 1.5) [RMSE: $4.05$].
The latent trait model achieves good correlation recovery (RMSE $1.42$ for Weibull(1.45)),
yet all three latent trait priors fall into the worse-performing group 
(overall RMSE $4.31 - 6.02$ vs $3.30 - 4.05$ for best group),
alongside LC-MVP with standard LKJ priors and CI.
Notably, the best latent trait prior (Weibull 1.45: $4.31$) comes close to competitive performance - 
only $0.26$ percentage points behind TruncLKJ(10,1.5) - but cannot match LC-MVP's top configurations (mixedLKJ)
despite this being its own generated data.
As we discussed in section \ref{section_results_DGMs_2_4},
this may occur because high sensitivities ($86 - 92.5\%$) create an information-poor environment - 
most diseased test positive, making the inverse problem (see section \ref{section_latent_trait_inverse_problem})
of recovering precise $b$ parameters from limited variation extremely difficult.
Hence, whilst the model successfully recovers correlations, small errors in $b$ parameters get amplified through 
the non-linear transformation $\Phi(a_t/\sqrt{1 + b_t})$, degrading accuracy estimation.
On the other hand, the LC-MVP model avoids this inverse problem through direct parameterisation.
Even standard LKJ maintains reasonable performance (overall RMSE $5.04$) despite poor correlation recovery -
protected by the ceiling effect - though it too falls into the worse group.

Furthermore, it is important to note that we generally obtained much better RMSE for DGM \#4 compared to DGM \#2 - 
especially at the larger $N = 3000$ sample size - 
due to the ceiling effect (because of the higher sensitivities of DGM \#4), which make correlation estimation less important.
For example, for DGM \#2 at $N = 3000$, RMSE range across all models as $5.64 - 12.4$ (with CI model obtaining $12.4$), 
but for DGM \#4 (at $N = 3000$), the range was only: $3.30 - 6.02$, with the CI model obtaining only $5.76$ RMSE.

\subsubsection{Discussion of key findings: Necessary \& sufficient conditions}
\label{section_discussion_discussion_of_key_findings_necessary_and_sufficient_conditions}
The relationship between correlation recovery and performance 
(detailed in section \ref{section_results_DGMs_2_4_between_model_comparison_and_corr_analysis})
reveals a fundamental paradox for LC-MVP with standard LKJ priors.
For DGM \#2, standard LKJ shows poor correlation recovery at $N = 300$ 
(RMSE(Cor) $3.01 - 3.12$; table \ref{Table:corr_rmse_LC_MVP_DGMs_2_4}),
but improves substantially at $N = 3000$ (RMSE(Cor) $1.34 - 1.48$) because low sensitivities ($55 - 70\%$) 
provide sufficient variation in test outcomes to inform correlation estimation with more data.
For DGM \#4, standard LKJ maintains poor correlation recovery even at $N = 3000$ 
(RMSE(Cor) $3.05 - 3.26$ to $2.62 - 2.68$; table \ref{Table:corr_rmse_LC_MVP_DGMs_2_4}) 
because high sensitivities ($86 - 92.5\%$) create a ceiling effect (see section \ref{section_ceiling_effect_paradox}) - 
most diseased test positive regardless, providing little information about correlation structure even with more data.
Yet paradoxically, LC-MVP with standard LKJ achieves better performance on DGM \#4 than DGM \#2 at both sample sizes
(see table \ref{Table:summary_table_overall_all_DGMs_2_4}),
demonstrating that ceiling effects both impair correlation recovery yet also protect against its consequences.

For heterogeneous structures (DGMs \#3 and \#5; section \ref{section_results_DGMs_3_5_correlation_analysis}), 
standard LKJ shows poor correlation recovery at $N = 300$
(DGM \#3: RMSE(Cor) $2.36 - 2.41$; DGM \#5: RMSE(Cor) $2.35$; table \ref{Table:corr_rmse_LC_MVP_DGMs_3_5}),
yet still achieves competitive performance (table \ref{Table:summary_table_overall_both_models_DGMs_3_5}).
At $N = 3000$, correlation recovery improves for standard LKJ on both heterogeneous DGMs
(DGM \#3: RMSE(Cor) $1.12 - 1.17$; DGM \#5: RMSE(Cor) $1.85 - 1.95$; table \ref{Table:corr_rmse_LC_MVP_DGMs_3_5}),
with DGM \#3 showing more improvement due to lower sensitivities allowing better correlation learning,
while DGM \#5's high sensitivities limit improvement due to ceiling effects.
This demonstrates that for LC-MVP with standard LKJ priors, good correlation recovery is neither necessary nor sufficient.

Meanwhile, the latent trait model achieves good correlation recovery across all scenarios
(tables \ref{Table:corr_rmse_latent_trait_DGMs_2_4} and \ref{Table:corr_rmse_latent_trait_DGMs_3_5}),
yet cannot consistently outperform LC-MVP's best configurations, and generally does very badly on DGMs \#3 and \#5,
showing that good correlation recovery is not sufficient for optimal performance.

\subsubsection{ Discussion of key findings: Summary}
\label{section_discussion_discussion_of_key_findings_summary}
Our key findings establish that, 
whilst the latent trait model performs acceptably when its more restrictive assumptions are met
(e.g., coverage usually $>98\%$ for latent-trait-generated DGMs \#2 and \#4 - 
see figures \ref{Appendix_figure:Sim_study_Coverage_DGM_2_} 
and \ref{Appendix_figure:Sim_study_Coverage_DGM_4_} 
in appendix \ref{appendix_C_coverage_plots}
and/or table \ref{Appendix_table:table_latent_trait_dgm_2_4} in appendix \ref{appendix_A_detailed_results_tables}),
it fails catastrophically when these assumptions are violated, with coverage collapsing below $50\%$ for more general 
- but realistic - correlation structures at $N = 3000$.
In contrast, the LC-MVP model generally maintains more consistent performance across scenarios.

Overall, the latent trait model only showed acceptable performance when its assumptions held (DGMs \#2 and \#4),
and even then it was only better than the LC-MVP model for one out of four cases (DGM\#2 at $N = 300$), whereas for the other three cases, 
it was either only better in terms of bias (DGM \#2 at $N = 3000$ - but only $\sim 1\%$ better bias) or it was actually worse (for DGM \#4),
and also it often had more variation between prior choices compared to the LC-MVP model.

On the other hand, for the more heterogeneous correlation structures (DGMs \#3 and \#5), the latent trait model failed catastrophically -
with worsening performance as sample size increases - a hallmark of severe model misspecification.
Furthermore, our analysis reveals why the latent trait model fails for heterogeneous structures:
the b parameters must simultaneously determine both correlations and accuracy estimates.
For non-latent-trait structures, achieving reasonable correlation recovery may require b values that distort sensitivity estimation - 
which our findings here suggest, because in many cases, the latent trait model achieved good correlation RMSE (or even better than LC-MVP),
yet poorer accuracy estimates.

In contrast, LC-MVP's independent parameterization avoids this constraint, explaining its more robust performance across all scenarios.
Importantly, ceiling effects modulate these patterns - 
protecting against correlation misspecification at small samples for high-sensitivity DGMs -
but making correlation recovery critical at large samples when the ceiling effect dissipates.
The LC-MVP's flexibility to accommodate any correlation structure - 
as opposed to enforcing the structural constraints of the latent trait model -
makes it more robust for real-world applications where correlation patterns may not follow such a specific structure.

\subsection{ Potential practical \& clinical implications}
\label{section_discussion_practical_and_clinical_implications}
The latent trait model's mixed performance in our study reveals important insights that may have been somewhat overlooked in the
diagnostic accuracy literature since its introduction to this field\supercite{Qu1996, dendukuri2001}, perhaps because it has not  
previously been compared to a model as flexible as the LC-MVP\supercite{Xu2009, Xu2013} model.
Furthermore, the model's requirement for strictly positive factor loadings ($b_t > 0$) - 
which forces the within study correlations to also all be positive -
combined with estimating a full correlation matrix from fewer parameters,
provides identifiability but at a cost\supercite{Albert_dodd_2004_cautionary_note}.
This constrained correlation structure may be reasonable in psychometrics where multiple items intentionally measure a 
single construct\supercite{lord1968statistical, Embretson2013},
and our results confirm it can work for diagnostic tests when this assumption holds.
However, whilst it performs adequately when its assumptions are met,
the severe consequences of misspecification - 
combined with the potential difficulties of verifying these assumptions in practice\supercite{Pepe_Janes_2007_insights_LCMs} - 
make it a more risky choice.

On the other hand, the fact that the LC-MVP achieves comparable or better performance - even on latent trait data (e.g., for DGM \#4) -
combined with the fact that the LC-MVP even with TruncLKJ priors (which force positive correlations just like the latent trait model) 
outperforms the latent trait model, whilst maintaining flexibility for other structures,
suggests adopting the more general model as the default approach for diagnostic/screening test accuracy studies without a perfect gold standard.

For practical application, our results emphasize the importance of carefully considering correlation structures when evaluating 
diagnostic/screening tests without a perfect gold standard.
Whilst our BayesMVP R package\supercite{Cerullo_BayesMVP_2025} can fit both LC-MVP and latent trait models very efficiently,
the severe consequences of latent trait misspecification observed in our study - 
with coverage collapsing below $50\%$ for realistic heterogeneous structures - 
reinforce that the more flexible LC-MVP approach is generally safer.

Furthermore, whilst our mixedLKJ priors makes sense if reference test correlations are known or thought to be lower, 
and/or the sensitivity/specificity of the reference test is very high (but not perfect), 
correlation structures which are different and/or more complex than our mixedLKJ formulation may be needed, 
guided by clinical knowledge of the specific tests.
For instance, for COVID-19 diagnosis, PCR and rapid antigen tests might be strongly correlated in infected individuals 
(both detecting viral material), but essentially uncorrelated with antibody tests which measure immune response - 
a pattern requiring different correlation constraints for different test pairs.
One might also consider a structure which allows all correlations to vary freely 
(i.e. each  $\Omega_{i, j}^{[d-]} \in (-1.0, 1.0)$) in the non-diseased group - 
which does not push correlations to be greater than zero   - 
whilst imposing structured constraints in the diseased group based on biological mechanisms

Disease prevalence also critically affects which correlation structure matters most.
Our study assumed low prevalence ($0.20$), making sensitivity estimates - not specificity estimates -
the most sensitive to prior specification and correlation constraints.
For intermediate prevalence ($0.30-0.60$), both sensitivity and specificity may be equally affected, 
requiring careful specification of both correlation matrices.
For high prevalence settings ($> 0.75$), the non-diseased correlation structure becomes paramount - 
with $N^{[d-]} < N^{[d+]}$ - as specificity estimates become highly sensitive to prior/correlation constraints, 
potentially reversing the pattern observed in our more common low-prevalence scenarios.

The BayesMVP R package \supercite{Cerullo_BayesMVP_2025} we developed, 
leveraging Sean Pinkney's novel constrained correlation method 
\supercite{Sean_Pinkney_2024_shortnoteflexiblecholesky}, 
facilitates the aforementioned tailored approach by allowing practitioners to specify different 
LKJ parameters and hard constraints for each individual correlation element, as well as allowing different constraints and priors 
for each latent class independently.
Furthermore, our R package also even allows the use of element-wise beta priors (instead of LKJ priors), so different "soft" constraints
can be specified for each individual correlation element (these more nuanced priors were not explored in this paper).
With computational speeds $10-100+$ times faster than other state-of-the-art implementations 
(e.g., Mplus \supercite{mplus}, Stan\supercite{Carpenter2017}), 
analysts can feasibly conduct comprehensive sensitivity analyses across multiple correlation specifications - 
including full K-fold cross-validation on just a regular laptop even with very large sample sizes - 
allowing them to compare results across different priors and assess robustness to these assumptions.


Regarding model selection between different LC-MVP correlation constraint configurations, we recommend full K-fold cross-validation as the
gold standard when computational resources permit\supercite{Vehtari_2017_LOO_IC_published}. 
K-fold provides unbiased estimates of predictive performance by evaluating models on genuinely held-out data, 
avoiding the optimistic bias inherent in within-sample metrics. 
For computationally constrained settings, leave-one-out cross-validation via Pareto-smoothed importance sampling (LOO-IC) offers a reasonable alternative\supercite{Vehtari_2017_LOO_IC_published}, but only when all Pareto-$\hat{k}$ diagnostics remain below 0.7 - 
values exceeding this threshold indicate unreliable estimates - which is particularly common in hierarchical models 
with limited groups\supercite{vehtari_2024_arxiv_pareto_smoothed_imp_samp, Bürkner_2020_approx_loo}. 
For instance, it almost always fails spectacularly for meta-analysis and network-meta-analysis models such as those in our recently developed 
MetaOrdDTA R package \supercite{Cerullo_2025_MetaOrdDTA, Cerullo_2025_ord_MA_NMA_MetaOrdDTA_preprint}.

The BayesMVP R package we developed\supercite{Cerullo_BayesMVP_2025} implements parallel K-fold cross-validation (forthcoming release) 
to reduce computational burden. 
We strongly discourage the use of DIC for latent class models, as it assumes fixed parameters and - unlike the LOO-IC - 
it fails "silently" (no warnings), e.g. for hierarchical structures\supercite{Plummer_2008_DIC_limitations, Gelman_2014_WAIC}.
Similarly, we advise against correlation residual plots \supercite{Qu1996} which - 
although historically used in the imperfect-gold-standard test accuracy literature - 
lack theoretical justification for model comparison and, like the DIC, 
often fails to distinguish between competing models effectively \supercite{Pepe_Janes_2007_insights_LCMs}.
Furthermore, if incorporating covariates - 
which is possible with our BayesMVP R package implementation of the LC-MVP) -
statistical significance of parameters should generally not be used for model selection, 
as it conflates hypothesis testing with predictive performance\supercite{Piironen_2017_model_selection}, 
which is what researchers and clinicians are ultimately interested in.

\subsection{ Limitations \& future work}
\label{section_discussion_limitations_and_future_work}
\subsubsection{Study limitations}
Our study has several limitations that merit consideration. We fixed disease prevalence at $20\%$, representing typical screening scenarios, 
but correlation structure sensitivity likely varies across the prevalence spectrum. 
At high prevalence ($>75\%$), the smaller non-diseased group would make specificity estimates more sensitive to correlation constraints, 
potentially reversing the patterns we observed. 
Similarly, we examined only two sample sizes - $N = 300$ and $N = 3000$ - 
which correspond to (on average) only $\sim 60$ (for $N = 300$) and $\sim 600$ (for $N = 3000$) people in the diseased group.
Intermediate sizes might reveal non-linear transitions in prior sensitivity, and larger samples would likely reveal generally
better sensitivity parameter recovery (except perhaps for the latent trait for DGMs \#3 and \#5).

Furthermore, our choice to use five tests limits generalisability to smaller (3-4 tests) or larger (6+) 
test batteries, where correlation dimensionality effects may differ substantially. 
We also did not systematically examine negative correlations — 
although these are likely rare in diagnostic testing - 
but they could potentially occur when tests measure competing biological processes. 
Finally, we assumed complete data, whereas real test accuracy datasets sometimes involve missing test results, which could 
interact with correlation assumptions and influence parameter recovery patterns across models and priors/constraints.

\subsubsection{Future methodological research}
For future work, additional simulation studies could also compare alternative flexible correlation approaches, 
such as copulas, against the LC-MVP framework. 
Furthermore, other authors have warned that model misspecification for latent class models may often be impossible to detect,
and potentially more robust model checking methods, such as LOO-IC (assuming adequate perto-K values) and/or full K-fold cross-validation -
which is generally regarded as being the gold standard method and generally far superior to the DIC - have not, to our knowledge,
been formally validated in any simulation study specifically for latent class models such as the LC-MVP models and/or latent trait models, 
especially within the context of test accuracy. 
Future simulation studies should look at comparing these model checking methods to methods such as the correlation residual plot 
proposed by Qu et al, 1996 \supercite{Qu1996} and to our "2^T contingency table residual table/plot" method which we 
proposed and implemented into our R shiny app - MetaBayesDTA (Cerullo et al, 2023\supercite{cerullo_shiny_app_MetaBayesDTA}).

\subsubsection{Extensions in development}
Future work should also look at extending the methodology to more complex scenarios. 
For instance, the incorporation of ordinal and mixed ordinal-binary test data into the LC-MVP framework presents immediate challenges, 
as cutpoint parameters add another layer of complexity that likely exacerbates correlation specification issues. 
Particularly important is the extension to individual participant data (IPD) meta-analysis and network meta-analysis settings,
which would enable re-evaluation of major diagnostic/screening accuracy studies without assuming perfect gold standards.
For instance, the influential PHQ-9 and PHQ-2 JAMA study by Levis et al \supercite{levis_JAMA_2020}
used the MINI as a reference standard in a significant proportion of included studies - 
yet the MINI itself has imperfect accuracy.
Our framework could provide more accurate estimates by properly accounting for this reference standard imperfection.

These functionalities are currently being developed for the BayesMVP R package \supercite{Cerullo_BayesMVP_2025},
using manual gradients as well as other optimisations for computational efficiency.
Additional extensions could also include longitudinal testing scenarios, where correlations evolve over time,
polytomous latent classes for disease subtypes or severity levels,
and integration with health economic decision frameworks for optimal test selection, 
which consider both diagnostic/screening performance and cost-effectiveness.

\subsection{ Conclusions}
\label{section_discussion_conclusions}
This comprehensive simulation study demonstrates that the LC-MVP model with flexible correlation structures provides 
a more robust framework for diagnostic/screening test evaluation without a perfect gold standard than the more commonly-used latent trait model.
Whilst the latent trait model performs acceptably when its restrictive assumptions are met (e.g., DGMs \#2 and \#4),
our findings suggest that it fails catastrophically for realistic heterogeneous correlation structures (e.g., DGMs \#3 and \#5), 
with coverage often collapsing below $50\%$ at $N = 3000$.

In contrast, LC-MVP maintains more consistent performance across diverse scenarios.
This holds true even if one only looks at the LC-MVP model with TruncLKJ priors - 
which more closely mimics the latent trait model (as they force all correlations to be positive) - 
yet is still more flexible than the latent trait model is, due to estimating the full correlation matrix.

Furthermore, the fact that model checking methods for latent class models have not been properly evaluated 
(as we discuss in section \ref{section_discussion_limitations_and_future_work}),
and are known to not necessarily be reliable \supercite{Pepe_Janes_2007_insights_LCMs}
(especially the simple correlation residual plot\supercite{Qu1996}),
makes a priori model choice even more important, 
and hence leaning towards a more flexible model like the LC-MVP makes even more sense given this context.

The critical advantage of LC-MVP lies in its independent parameterisation and flexibility - 
accuracies and correlations are estimated separately, 
and correlation structures can be tailored to specific clinical knowledge and/or biological mechanisms,
especially combined with using the method proposed by Pinkney et al, 2024\supercite{Sean_Pinkney_2024_shortnoteflexiblecholesky},
which is implemented into our BayesMVP R package\supercite{Cerullo_BayesMVP_2025}.
This flexibility of the LC-MVP model is particularly valuable given that diagnostic/screening tests in practice may not necessarily 
satisfy the restrictive correlation structure of the latent trait model,
and they may sometimes not even satisfy the assumption of universally positive correlations (like those we considered in this paper), 
as different tests sometimes measure very distinct biological processes or disease manifestations.

In general, we would strongly recommend that analysts carefully consider correlation structures
based on the biological mechanisms and/or clinical knowledge underlying the specific tests being evaluated.
We also strongly recommend incorporating prior information directly on the sensitivities and specificities of the tests,
especially reference test(s) in which clinicians and researchers might be more confident about their accuracy in general 
(and in specific populations).

Overall, given the severe consequences of model misspecification observed in our study - 
as well as the potential difficulties of detecting such misspecification in practice - 
choosing the more flexible LC-MVP model (especially with appropriately customized correlation constraints)
represents the safer, more defensible choice for researchers evaluating diagnostic/screening tests without a perfect gold standard.

\newpage
\appendix 
\setcounter{figure}{0}
\setcounter{table}{0}
\renewcommand{\thefigure}{A.\arabic{figure}}
\renewcommand{\thetable}{A.\arabic{table}}
\section{Appendix A - Detailed results tables}
\label{appendix_A_detailed_results_tables}
\subsection{Detailed results tables for LC-MVP}
\label{appendix_A_detailed_results_tables_LC_MVP}
\subsubsection{Detailed results tables; LC-MVP: DGM \#1}
\label{appendix_A_detailed_results_tables_LC_MVP_DGM_1}
\begin{table}[H]
\centering
\caption{
LC-MVP model performance on DGM \#1 simulated data (Conditional Independence).
}
\small
\setlength{\tabcolsep}{3pt}
\begin{tabular}{cc|c|cc|cc|cc|cc}
\toprule
\multirow{2}{*}{\textbf{$N$}} & 
\multirow{2}{*}{\textbf{Prior}} &
\multirow{2}{*}{\textbf{$N_{\text{sim}}$}} & 
\multicolumn{8}{c}{\textbf{Performance Metrics}} \\
\cmidrule(lr){4-11}
& & & \multicolumn{2}{c|}{\textbf{RMSE (\%)}} & 
\multicolumn{2}{c|}{\textbf{$|$Bias$|$ (\%)}} &
\multicolumn{2}{c|}{\textbf{Cvg.}} & 
\multicolumn{2}{c}{\textbf{Width}} \\
\cmidrule(lr){4-5} \cmidrule(lr){6-7} \cmidrule(lr){8-9} \cmidrule(lr){10-11}
& & & \textbf{Se} & \textbf{Sp} & \textbf{Se} & \textbf{Sp} & \textbf{Se} & \textbf{Sp} & \textbf{Se} & \textbf{Sp} \\
\midrule
\midrule
300 & \footnotesize{LKJ(10, 1.5)} & 427 & 6.85 {\tiny(0.237)} & 1.86 {\tiny(0.068)} & 0.38 {\tiny(0.331)} & 0.58 {\tiny(0.085)} & 98.1 & 97.0 & 32.8 & 8.32 \\
300 & \footnotesize{LKJ(24, 4)} & 436 & 6.88 {\tiny(0.237)} & 1.83 {\tiny(0.066)} & 0.29 {\tiny(0.329)} & 0.37 {\tiny(0.086)} & 97.7 & 96.9 & 30.8 & 7.94 \\
300 & \footnotesize{TruncLKJ(10, 1.5)} & 602 & 11.6 {\tiny(0.246)} & 2.34 {\tiny(0.068)} & 9.58 {\tiny(0.268)} & 1.26 {\tiny(0.080)} & 90.2 & 97.3 & 37.9 & 10.3 \\
300 & \footnotesize{TruncLKJ(24, 4)} & 590 & 10.1 {\tiny(0.244)} & 1.99 {\tiny(0.059)} & 7.52 {\tiny(0.277)} & 0.63 {\tiny(0.078)} & 90.8 & 97.3 & 33.5 & 8.91 \\
300 & \footnotesize{mixedLKJ(10, 1.5)} & 520 & 7.99 {\tiny(0.245)} & 2.31 {\tiny(0.069)} & 4.21 {\tiny(0.297)} & 1.45 {\tiny(0.078)} & 97.0 & 95.6 & 36.8 & 8.82 \\
300 & \footnotesize{mixedLKJ(24, 4)} & 488 & 7.55 {\tiny(0.246)} & 2.06 {\tiny(0.065)} & 3.26 {\tiny(0.307)} & 1.01 {\tiny(0.080)} & 96.0 & 95.8 & 32.8 & 8.07 \\
\midrule
\midrule
3000 & \footnotesize{LKJ(10, 1.5)} & 124 & 3.31 {\tiny(0.197)} & 0.76 {\tiny(0.046)} & 0.28 {\tiny(0.297)} & 0.08 {\tiny(0.068)} & 98.1 & 98.9 & 15.9 & 3.68 \\
3000 & \footnotesize{LKJ(24, 4)} & 89 & 3.04 {\tiny(0.215)} & 0.70 {\tiny(0.049)} & 0.20 {\tiny(0.323)} & 0.09 {\tiny(0.073)} & 97.5 & 98.7 & 14.0 & 3.24 \\
3000 & \footnotesize{TruncLKJ(10, 1.5)} & 122 & 4.60 {\tiny(0.211)} & 0.69 {\tiny(0.041)} & 3.79 {\tiny(0.234)} & 0.26 {\tiny(0.057)} & 83.6 & 97.7 & 13.1 & 3.11 \\
3000 & \footnotesize{TruncLKJ(24, 4)} & 117 & 4.36 {\tiny(0.215)} & 0.70 {\tiny(0.042)} & 3.47 {\tiny(0.241)} & 0.21 {\tiny(0.060)} & 82.2 & 97.3 & 12.4 & 2.90 \\
3000 & \footnotesize{mixedLKJ(10, 1.5)} & 97 & 3.75 {\tiny(0.218)} & 1.17 {\tiny(0.055)} & 1.91 {\tiny(0.279)} & 1.01 {\tiny(0.060)} & 94.8 & 83.3 & 14.0 & 3.10 \\
3000 & \footnotesize{mixedLKJ(24, 4)} & 95 & 3.33 {\tiny(0.216)} & 0.99 {\tiny(0.056)} & 1.74 {\tiny(0.270)} & 0.79 {\tiny(0.061)} & 94.5 & 82.9 & 13.0 & 2.87 \\
\midrule
\bottomrule
\end{tabular}
\begin{tablenotes}
\footnotesize
\item Note: 
All performance metrics are means across the 5 tests; 
Values in parentheses are the Monte Carlo Standard Errors (MCSE).
\item RMSE = Root Mean Square Error; Cvg. = Coverage; Width = mean 95\% interval width.
\end{tablenotes}
\label{Appendix_table:lc_mvp_dgm_1}
\end{table}
\subsubsection{Detailed results tables; LC-MVP: DGMs \#3 \& \#5}
\label{appendix_A_detailed_results_tables_LC_MVP_DGMs_3_5}
\begin{table}[H]
\centering
\caption{LC\_MVP model performance on DGMs with LC-MVP structure (DGMs \#3, \#5).}
\small
\setlength{\tabcolsep}{2pt}
\begin{tabular}{c|cc|c|cc|cc|cc|cc}
\toprule
\multirow{2}{*}{\textbf{\scriptsize{DGM}}} &
\multirow{2}{*}{\textbf{$N$}} & 
\multirow{2}{*}{\textbf{Prior}} &
\multirow{2}{*}{\textbf{$N_{\text{sim}}$}} & 
\multicolumn{8}{c}{\textbf{Performance Metrics}} \\
\cmidrule(lr){5-12}
& & & & \multicolumn{2}{c|}{\textbf{RMSE (\%)}} & 
\multicolumn{2}{c|}{\textbf{$|$Bias$|$ (\%)}} &
\multicolumn{2}{c|}{\textbf{Cvg.}} & 
\multicolumn{2}{c}{\textbf{Width}} \\
\cmidrule(lr){5-6} \cmidrule(lr){7-8} \cmidrule(lr){9-10} \cmidrule(lr){11-12}
& & & & \textbf{Se} & \textbf{Sp} & \textbf{Se} & \textbf{Sp} & \textbf{Se} & \textbf{Sp} & \textbf{Se} & \textbf{Sp} \\
\midrule
\midrule
3 & 300 & \footnotesize{LKJ(10, 1.5)} & 
526 & 8.49 {\tiny(0.239)} & 2.00 {\tiny(0.063)} & 4.18 {\tiny(0.316)} & 0.71 {\tiny(0.080)} & 97.8 & 97.3 & 35.6 & 8.66 \\
3 & 300 & \footnotesize{LKJ(24, 4)} & 
561 & 8.78 {\tiny(0.241)} & 2.09 {\tiny(0.062)} & 4.34 {\tiny(0.314)} & 0.89 {\tiny(0.078)} & 95.6 & 94.8 & 32.8 & 8.14 \\
3 & 300 & \footnotesize{TruncLKJ(10, 1.5)} & 
534 & 9.03 {\tiny(0.247)} & 2.03 {\tiny(0.066)} & 4.91 {\tiny(0.311)} & 0.45 {\tiny(0.084)} & 97.0 & 98.4 & 38.7 & 9.79 \\
3 & 300 & \footnotesize{TruncLKJ(24, 4)} &
547 & 8.82 {\tiny(0.245)} & 2.13 {\tiny(0.066)} & 3.49 {\tiny(0.320)} & 0.78 {\tiny(0.084)} & 95.0 & 96.5 & 34.9 & 8.80 \\
3 & 300 & \footnotesize{mixedLKJ(10, 1.5)} & 
407 & 6.99 {\tiny(0.241)} & 1.85 {\tiny(0.073)} & 1.30 {\tiny(0.340)} & 0.27 {\tiny(0.089)} & 99.2 & 98.0 & 36.8 & 8.80 \\
3 & 300 & \footnotesize{mixedLKJ(24, 4)} & 
439 & 7.47 {\tiny(0.242)} & 1.90 {\tiny(0.070)} & 2.03 {\tiny(0.341)} & 0.47 {\tiny(0.088)} & 97.9 & 96.7 & 33.4 & 8.13 \\
\midrule
3 & 3000 & \footnotesize{LKJ(10, 1.5)} & 
163 & 4.72 {\tiny(0.231)} & 1.14 {\tiny(0.064)} & 1.92 {\tiny(0.306)} & 0.71 {\tiny(0.070)} & 99.3 & 97.4 & 27.2 & 5.04 \\
3 & 3000 & \footnotesize{LKJ(24, 4)} & 
159 & 4.77 {\tiny(0.232)} & 1.32 {\tiny(0.064)} & 2.45 {\tiny(0.288)} & 1.00 {\tiny(0.066)} & 96.5 & 89.4 & 22.0 & 4.20 \\
3 & 3000 & \footnotesize{TruncLKJ(10, 1.5)} & 
142 & 4.86 {\tiny(0.217)} & 0.93 {\tiny(0.056)} & 3.64 {\tiny(0.247)} & 0.52 {\tiny(0.063)} & 93.0 & 97.2 & 21.6 & 4.28 \\
3 & 3000 & \footnotesize{TruncLKJ(24, 4)} & 
157 & 4.99 {\tiny(0.220)} & 1.15 {\tiny(0.058)} & 3.36 {\tiny(0.256)} & 0.84 {\tiny(0.061)} & 88.0 & 90.2 & 18.5 & 3.73 \\
3 & 3000 & \footnotesize{mixedLKJ(10, 1.5)} & 
97 & 3.45 {\tiny(0.236)} & 0.75 {\tiny(0.057)} & 0.80 {\tiny(0.340)} & 0.17 {\tiny(0.073)} & 99.6 & 99.6 & 23.3 & 4.30 \\
3 & 3000 & \footnotesize{mixedLKJ(24, 4)} & 
147 & 3.92 {\tiny(0.224)} & 0.91 {\tiny(0.054)} & 1.61 {\tiny(0.275)} & 0.50 {\tiny(0.061)} & 98.2 & 95.9 & 19.4 & 3.73 \\
\midrule
\midrule
5 & 300 & \footnotesize{LKJ(10, 1.5)} & 
247 & 5.40 {\tiny(0.229)} & 2.67 {\tiny(0.121)} & 2.02 {\tiny(0.296)} & 0.91 {\tiny(0.159)} & 98.3 & 96.0 & 24.5 & 11.4 \\
5 & 300 & \footnotesize{LKJ(24, 4)} & 
247 & 5.68 {\tiny(0.240)} & 2.79 {\tiny(0.127)} & 2.25 {\tiny(0.309)} & 1.03 {\tiny(0.165)} & 96.5 & 94.9 & 22.5 & 10.9 \\
5 & 300 & \footnotesize{TruncLKJ(10, 1.5)} & 
389 & 6.94 {\tiny(0.246)} & 2.63 {\tiny(0.096)} & 4.21 {\tiny(0.268)} & 0.39 {\tiny(0.129)} & 97.5 & 97.6 & 29.3 & 12.1 \\
5 & 300 & \footnotesize{TruncLKJ(24, 4)} & 
354 & 6.56 {\tiny(0.245)} & 2.71 {\tiny(0.103)} & 3.24 {\tiny(0.283)} & 0.65 {\tiny(0.139)} & 95.8 & 96.2 & 25.5 & 11.4 \\
5 & 300 & \footnotesize{mixedLKJ(10, 1.5)} & 
276 & 5.50 {\tiny(0.245)} & 2.47 {\tiny(0.106)} & 2.93 {\tiny(0.276)} & 0.28 {\tiny(0.147)} & 98.0 & 97.2 & 25.7 & 11.2 \\
5 & 300 & \footnotesize{mixedLKJ(24, 4)} & 
252 & 5.45 {\tiny(0.243)} & 2.60 {\tiny(0.116)} & 2.10 {\tiny(0.306)} & 0.50 {\tiny(0.160)} & 97.5 & 96.0 & 23.1 & 10.9 \\
\midrule
5 & 3000 & \footnotesize{LKJ(10, 1.5)} & 
78 & 3.19 {\tiny(0.214)} & 1.13 {\tiny(0.082)} & 2.11 {\tiny(0.253)} & 0.43 {\tiny(0.112)} & 99.2 & 96.9 & 16.7 & 5.28 \\
5 & 3000 & \footnotesize{LKJ(24, 4)} & 
74 & 3.16 {\tiny(0.212)} & 1.06 {\tiny(0.075)} & 2.17 {\tiny(0.245)} & 0.54 {\tiny(0.102)} & 96.8 & 97.0 & 13.8 & 4.71 \\
5 & 3000 & \footnotesize{TruncLKJ(10, 1.5)} & 
81 & 4.04 {\tiny(0.235)} & 1.15 {\tiny(0.074)} & 2.90 {\tiny(0.266)} & 0.68 {\tiny(0.098)} & 93.3 & 96.3 & 19.2 & 5.49 \\
5 & 3000 & \footnotesize{TruncLKJ(24, 4)} & 
75 & 3.61 {\tiny(0.236)} & 1.15 {\tiny(0.075)} & 2.38 {\tiny(0.275)} & 0.71 {\tiny(0.099)} & 90.9 & 93.1 & 14.6 & 4.67 \\
5 & 3000 & \footnotesize{mixedLKJ(10, 1.5)} & 
70 & 2.52 {\tiny(0.208)} & 1.01 {\tiny(0.079)} & 1.23 {\tiny(0.250)} & 0.28 {\tiny(0.115)} & 99.1 & 98.6 & 16.7 & 5.02 \\
5 & 3000 & \footnotesize{mixedLKJ(24, 4)} & 
65 & 2.57 {\tiny(0.208)} & 1.06 {\tiny(0.083)} & 1.53 {\tiny(0.250)} & 0.39 {\tiny(0.122)} & 98.2 & 96.3 & 13.2 & 4.47 \\
\midrule
\bottomrule
\end{tabular}
\begin{tablenotes}
\footnotesize
\item Note: 
All performance metrics are means across the 5 tests; 
Values in parentheses are the Monte Carlo Standard Errors (MCSE).
\item RMSE = Root Mean Square Error; Cvg. = Coverage; Width = mean 95\% interval width.
\end{tablenotes}
\label{Appendix_table:lc_mvp_dgm_3_5}
\end{table}
\subsubsection{Detailed results tables; LC-MVP: DGMs \#2 \& \#4}
\label{appendix_A_detailed_results_tables_LC_MVP_DGMs_2_4}
\begin{table}[H]
\centering
\caption{LC\_MVP model performance on DGMs with Latent Trait structure (DGMs \#2, \#4).}
\small
\setlength{\tabcolsep}{2pt}
\begin{tabular}{c|cc|c|cc|cc|cc|cc}
\toprule
\multirow{2}{*}{\textbf{\scriptsize{DGM}}} &
\multirow{2}{*}{\textbf{$N$}} & 
\multirow{2}{*}{\textbf{Prior}} &
\multirow{2}{*}{\textbf{$N_{\text{sim}}$}} & 
\multicolumn{8}{c}{\textbf{Performance Metrics}} \\
\cmidrule(lr){5-12}
& & & & \multicolumn{2}{c|}{\textbf{RMSE (\%)}} & 
\multicolumn{2}{c|}{\textbf{$|$Bias$|$ (\%)}} &
\multicolumn{2}{c|}{\textbf{Cvg.}} & 
\multicolumn{2}{c}{\textbf{Width}} \\
\cmidrule(lr){5-6} \cmidrule(lr){7-8} \cmidrule(lr){9-10} \cmidrule(lr){11-12}
& & & & \textbf{Se} & \textbf{Sp} & \textbf{Se} & \textbf{Sp} & \textbf{Se} & \textbf{Sp} & \textbf{Se} & \textbf{Sp} \\
\midrule
\midrule
2 & 300 & \footnotesize{LKJ(10, 1.5)} & 567 & 9.64 {\tiny(0.242)} & 1.91 {\tiny(0.060)} & 5.94 {\tiny(0.298)} & 0.53 {\tiny(0.075)} & 95.9 & 96.9 & 35.8 & 8.43 \\
2 & 300 & \footnotesize{LKJ(24, 4)} & 599 & 10.1 {\tiny(0.241)} & 1.96 {\tiny(0.059)} & 6.42 {\tiny(0.297)} & 0.57 {\tiny(0.074)} & 91.7 & 94.9 & 32.8 & 7.92 \\
2 & 300 & \footnotesize{TruncLKJ(10, 1.5)} & 522 & 7.59 {\tiny(0.241)} & 1.98 {\tiny(0.066)} & 2.11 {\tiny(0.313)} & 0.42 {\tiny(0.082)} & 99.3 & 98.3 & 39.2 & 9.26 \\
2 & 300 & \footnotesize{TruncLKJ(24, 4)} & 576 & 8.18 {\tiny(0.245)} & 1.98 {\tiny(0.061)} & 2.96 {\tiny(0.311)} & 0.51 {\tiny(0.078)} & 97.7 & 96.8 & 35.2 & 8.42 \\
2 & 300 & \footnotesize{mixedLKJ(10, 1.5)} & 479 & 7.75 {\tiny(0.243)} & 1.85 {\tiny(0.068)} & 3.70 {\tiny(0.307)} & 0.45 {\tiny(0.081)} & 98.5 & 97.4 & 36.6 & 8.36 \\
2 & 300 & \footnotesize{mixedLKJ(24, 4)} & 537 & 8.41 {\tiny(0.244)} & 1.83 {\tiny(0.061)} & 4.28 {\tiny(0.304)} & 0.40 {\tiny(0.076)} & 96.4 & 96.5 & 33.1 & 7.80 \\
\midrule
2 & 3000 & \footnotesize{LKJ(10, 1.5)} & 200 & 5.40 {\tiny(0.232)} & 0.99 {\tiny(0.048)} & 3.07 {\tiny(0.303)} & 0.37 {\tiny(0.062)} & 99.4 & 98.4 & 29.9 & 5.02 \\
2 & 3000 & \footnotesize{LKJ(24, 4)} & 194 & 5.67 {\tiny(0.228)} & 1.00 {\tiny(0.047)} & 3.82 {\tiny(0.287)} & 0.53 {\tiny(0.058)} & 96.6 & 94.6 & 23.1 & 4.09 \\
2 & 3000 & \footnotesize{TruncLKJ(10, 1.5)} & 140 & 4.94 {\tiny(0.230)} & 0.95 {\tiny(0.050)} & 3.43 {\tiny(0.291)} & 0.46 {\tiny(0.063)} & 99.1 & 97.1 & 23.9 & 4.17 \\
2 & 3000 & \footnotesize{TruncLKJ(24, 4)} & 165 & 5.85 {\tiny(0.226)} & 0.95 {\tiny(0.045)} & 4.56 {\tiny(0.268)} & 0.49 {\tiny(0.056)} & 93.8 & 92.7 & 19.7 & 3.59 \\
2 & 3000 & \footnotesize{mixedLKJ(10, 1.5)} & 172 & 5.76 {\tiny(0.239)} & 0.87 {\tiny(0.047)} & 4.41 {\tiny(0.276)} & 0.37 {\tiny(0.056)} & 97.9 & 97.8 & 24.5 & 4.16 \\
2 & 3000 & \footnotesize{mixedLKJ(24, 4)} & 166 & 5.83 {\tiny(0.237)} & 0.84 {\tiny(0.045)} & 4.50 {\tiny(0.275)} & 0.40 {\tiny(0.055)} & 91.8 & 96.4 & 19.7 & 3.54 \\
\midrule
\midrule
4 & 300 & \footnotesize{LKJ(10, 1.5)} & 213 & 5.09 {\tiny(0.239)} & 2.63 {\tiny(0.126)} & 1.79 {\tiny(0.305)} & 0.82 {\tiny(0.171)} & 98.3 & 95.2 & 23.4 & 11.1 \\
4 & 300 & \footnotesize{LKJ(24, 4)} & 214 & 5.24 {\tiny(0.229)} & 2.69 {\tiny(0.128)} & 2.17 {\tiny(0.309)} & 0.93 {\tiny(0.173)} & 96.7 & 94.9 & 21.2 & 10.7 \\
4 & 300 & \footnotesize{TruncLKJ(10, 1.5)} & 352 & 5.97 {\tiny(0.236)} & 2.60 {\tiny(0.099)} & 2.75 {\tiny(0.267)} & 0.35 {\tiny(0.135)} & 98.4 & 97.3 & 27.8 & 11.6 \\
4 & 300 & \footnotesize{TruncLKJ(24, 4)} & 298 & 5.65 {\tiny(0.234)} & 2.66 {\tiny(0.109)} & 2.00 {\tiny(0.289)} & 0.54 {\tiny(0.150)} & 97.2 & 95.8 & 23.9 & 11.0 \\
4 & 300 & \footnotesize{mixedLKJ(10, 1.5)} & 265 & 4.92 {\tiny(0.242)} & 2.47 {\tiny(0.104)} & 1.74 {\tiny(0.279)} & 0.17 {\tiny(0.151)} & 98.1 & 97.2 & 24.5 & 11.0 \\
4 & 300 & \footnotesize{mixedLKJ(24, 4)} & 231 & 4.79 {\tiny(0.236)} & 2.54 {\tiny(0.116)} & 1.20 {\tiny(0.301)} & 0.45 {\tiny(0.164)} & 97.7 & 96.1 & 21.9 & 10.6 \\
\midrule
4 & 3000 & \footnotesize{LKJ(10, 1.5)} & 42 & 3.88 {\tiny(0.240)} & 1.16 {\tiny(0.108)} & 3.27 {\tiny(0.288)} & 0.52 {\tiny(0.139)} & 100 & 97.1 & 15.2 & 4.93 \\
4 & 3000 & \footnotesize{LKJ(24, 4)} & 71 & 3.95 {\tiny(0.211)} & 1.08 {\tiny(0.074)} & 3.36 {\tiny(0.221)} & 0.52 {\tiny(0.101)} & 90.7 & 93.5 & 11.8 & 4.31 \\
4 & 3000 & \footnotesize{TruncLKJ(10, 1.5)} & 68 & 2.98 {\tiny(0.205)} & 1.07 {\tiny(0.071)} & 1.99 {\tiny(0.237)} & 0.52 {\tiny(0.100)} & 97.6 & 95.0 & 16.9 & 4.89 \\
4 & 3000 & \footnotesize{TruncLKJ(24, 4)} & 68 & 3.27 {\tiny(0.206)} & 1.05 {\tiny(0.073)} & 2.58 {\tiny(0.233)} & 0.53 {\tiny(0.097)} & 95.0 & 91.2 & 12.3 & 4.24 \\
4 & 3000 & \footnotesize{mixedLKJ(10, 1.5)} & 64 & 2.30 {\tiny(0.204)} & 0.99 {\tiny(0.084)} & 1.15 {\tiny(0.238)} & 0.24 {\tiny(0.118)} & 100 & 98.1 & 15.6 & 4.76 \\
4 & 3000 & \footnotesize{mixedLKJ(24, 4)} & 64 & 2.86 {\tiny(0.203)} & 1.02 {\tiny(0.081)} & 2.13 {\tiny(0.222)} & 0.33 {\tiny(0.115)} & 96.6 & 96.6 & 11.9 & 4.21 \\
\midrule
\bottomrule
\end{tabular}
\begin{tablenotes}
\footnotesize
\item Note: 
All performance metrics are means across the 5 tests; 
Values in parentheses are the Monte Carlo Standard Errors (MCSE).
\item RMSE = Root Mean Square Error; Cvg. = Coverage; Width = mean 95\% interval width.
\end{tablenotes}
\label{Appendix_table:lc_mvp_dgm_2_4}
\end{table}

\subsection{Detailed results tables for latent trait}
\label{appendix_A_detailed_results_tables_latent_trait}
\subsubsection{Detailed results tables; latent trait: DGM \#1}
\label{appendix_A_detailed_results_tables_latent_trait_DGM_1}
\begin{table}[H]
\centering
\caption{Latent trait model performance on DGM \#1 (conditional independence).}
\small
\setlength{\tabcolsep}{2pt}
\begin{tabular}{c|cc|c|cc|cc|cc|cc}
\toprule
\multirow{2}{*}{\textbf{\scriptsize{DGM}}} &
\multirow{2}{*}{\textbf{$N$}} & 
\multirow{2}{*}{\textbf{Prior}} &
\multirow{2}{*}{\textbf{$N_{\text{sim}}$}} & 
\multicolumn{8}{c}{\textbf{Performance Metrics}} \\
\cmidrule(lr){5-12}
& & & & \multicolumn{2}{c|}{\textbf{RMSE (\%)}} & 
\multicolumn{2}{c|}{\textbf{$|$Bias$|$ (\%)}} &
\multicolumn{2}{c|}{\textbf{Cvg.}} & 
\multicolumn{2}{c}{\textbf{Width}} \\
\cmidrule(lr){5-6} \cmidrule(lr){7-8} \cmidrule(lr){9-10} \cmidrule(lr){11-12}
& & & & \textbf{Se} & \textbf{Sp} & \textbf{Se} & \textbf{Sp} & \textbf{Se} & \textbf{Sp} & \textbf{Se} & \textbf{Sp} \\
\midrule
\midrule
1 & 300 & \footnotesize{Gamma(1, 1)} & 
603 & 9.44 {\tiny(0.245)} & 2.23 {\tiny(0.067)} & 6.33 {\tiny(0.285)} & 1.23 {\tiny(0.076)} & 96.5 & 97.9 & 41.0 & 10.8 \\
1 & 300 & \footnotesize{Weibull$^{[\frac{d-}{d+}]}(\frac{1.59}{1.45}, \frac{0.468}{0.881})$} & 
583 & 9.64 {\tiny(0.246)} & 1.98 {\tiny(0.060)} & 6.92 {\tiny(0.278)} & 0.63 {\tiny(0.077)} & 93.1 & 97.6 & 35.4 & 9.20 \\
1 & 300 & \footnotesize{Weibull$^{[\frac{d-}{d+}]}(\frac{1.52}{1.33}, \frac{0.633}{1.25})$} &
628 & 10.9 {\tiny(0.243)} & 2.22 {\tiny(0.064)} & 8.65 {\tiny(0.265)} & 1.11 {\tiny(0.077)} & 93.4 & 98.2 & 39.9 & 10.6 \\
\midrule
1 & 3000 & \footnotesize{Gamma(1, 1)} & 
53 & 2.74 {\tiny(0.237)} & 0.58 {\tiny(0.049)} & 1.25 {\tiny(0.332)} & 0.13 {\tiny(0.078)} & 96.2 & 98.9 & 12.1 & 2.88 \\
1 & 3000 & \footnotesize{Weibull$^{[\frac{d-}{d+}]}(\frac{1.59}{1.45}, \frac{0.468}{0.881})$} &
64 & 3.15 {\tiny(0.245)} & 0.64 {\tiny(0.059)} & 2.03 {\tiny(0.300)} & 0.11 {\tiny(0.079)} & 94.4 & 96.6 & 12.5 & 2.86 \\
1 & 3000 & \footnotesize{Weibull$^{[\frac{d-}{d+}]}(\frac{1.52}{1.33}, \frac{0.633}{1.25})$}  & 
64 & 3.09 {\tiny(0.227)} & 0.67 {\tiny(0.053)} & 2.11 {\tiny(0.280)} & 0.21 {\tiny(0.080)} & 95.9 & 98.4 & 13.1 & 2.98 \\
\midrule
\bottomrule
\end{tabular}
\begin{tablenotes}
\footnotesize
\item Note: 
All performance metrics are means across the 5 tests; 
Values in parentheses are the Monte Carlo Standard Errors (MCSE).
\item RMSE = Root Mean Square Error; Cvg. = Coverage; Width = mean 95\% interval width.
\end{tablenotes}
\label{Appendix_table:latent_trait_dgm_1}
\end{table}
\subsubsection{Detailed results tables; latent trait: DGMs \#3 \& \#5}
\label{appendix_A_detailed_results_tables_latent_trait_DGM_1}
\begin{table}[H]
\centering
\caption{Latent trait model performance on DGMs with LC-MVP structure (DGMs \#3, \#5).}
\small
\setlength{\tabcolsep}{2pt}
\begin{tabular}{c|cc|c|cc|cc|cc|cc}
\toprule
\multirow{2}{*}{\textbf{\scriptsize{DGM}}} &
\multirow{2}{*}{\textbf{$N$}} & 
\multirow{2}{*}{\textbf{Prior}} &
\multirow{2}{*}{\textbf{$N_{\text{sim}}$}} & 
\multicolumn{8}{c}{\textbf{Performance Metrics}} \\
\cmidrule(lr){5-12}
& & & & \multicolumn{2}{c|}{\textbf{RMSE (\%)}} & 
\multicolumn{2}{c|}{\textbf{$|$Bias$|$ (\%)}} &
\multicolumn{2}{c|}{\textbf{Cvg.}} & 
\multicolumn{2}{c}{\textbf{Width}} \\
\cmidrule(lr){5-6} \cmidrule(lr){7-8} \cmidrule(lr){9-10} \cmidrule(lr){11-12}
& & & & \textbf{Se} & \textbf{Sp} & \textbf{Se} & \textbf{Sp} & \textbf{Se} & \textbf{Sp} & \textbf{Se} & \textbf{Sp} \\
\midrule
\midrule
3 & 300 & \footnotesize{Gamma(1, 1)} & 
618 & 8.90 {\tiny(0.248)} & 2.19 {\tiny(0.065)} & 3.89 {\tiny(0.308)} & 0.79 {\tiny(0.080)} & 99.1 & 99.3 & 46.2 & 11.4 \\
3 & 300 & \footnotesize{Weibull$^{[\frac{d-}{d+}]}(\frac{1.59}{1.45}, \frac{0.468}{0.881})$} &
763 & 9.99 {\tiny(0.244)} & 2.40 {\tiny(0.065)} & 4.56 {\tiny(0.318)} & 1.07 {\tiny(0.074)} & 96.1 & 94.7 & 39.3 & 9.80 \\
3 & 300 & \footnotesize{Weibull$^{[\frac{d-}{d+}]}(\frac{1.52}{1.33}, \frac{0.633}{1.25})$} & 
688 & 9.29 {\tiny(0.247)} & 2.44 {\tiny(0.076)} & 2.06 {\tiny(0.337)} & 1.10 {\tiny(0.080)} & 98.8 & 97.8 & 44.6 & 11.0 \\
\midrule
3 & 3000 & \footnotesize{Gamma(1, 1)} & 
522 & 10.9 {\tiny(0.248)} & 2.42 {\tiny(0.047)} & 4.20 {\tiny(0.411)} & 1.01 {\tiny(0.082)} & 69.1 & 47.4 & 24.1 & 4.24 \\
3 & 3000 & \footnotesize{Weibull$^{[\frac{d-}{d+}]}(\frac{1.59}{1.45}, \frac{0.468}{0.881})$} & 
558 & 11.0 {\tiny(0.246)} & 2.31 {\tiny(0.040)} & 7.40 {\tiny(0.336)} & 1.16 {\tiny(0.070)} & 53.8 & 48.4 & 17.5 & 3.73 \\
3 & 3000 & \footnotesize{Weibull$^{[\frac{d-}{d+}]}(\frac{1.52}{1.33}, \frac{0.633}{1.25})$} & 
447 & 11.2 {\tiny(0.249)} & 2.49 {\tiny(0.045)} & 7.39 {\tiny(0.363)} & 1.64 {\tiny(0.076)} & 53.3 & 44.4 & 20.0 & 4.02 \\
\midrule
\midrule
5 & 300 & \footnotesize{Gamma(1, 1)} & 
512 & 7.92 {\tiny(0.245)} & 2.78 {\tiny(0.089)} & 5.28 {\tiny(0.249)} & 0.49 {\tiny(0.118)} & 97.1 & 97.9 & 34.5 & 13.1 \\
5 & 300 & \footnotesize{Weibull$^{[\frac{d-}{d+}]}(\frac{1.59}{1.45}, \frac{0.468}{0.881})$} & 
189 & 4.97 {\tiny(0.243)} & 2.73 {\tiny(0.139)} & 1.93 {\tiny(0.329)} & 0.68 {\tiny(0.184)} & 98.3 & 95.4 & 22.1 & 11.2 \\
5 & 300 & \footnotesize{Weibull$^{[\frac{d-}{d+}]}(\frac{1.52}{1.33}, \frac{0.633}{1.25})$} & 
276 & 5.44 {\tiny(0.247)} & 2.82 {\tiny(0.118)} & 1.39 {\tiny(0.300)} & 0.71 {\tiny(0.159)} & 99.3 & 96.1 & 26.4 & 11.8 \\
\midrule
5 & 3000 & \footnotesize{Gamma(1, 1)} & 
227 & 11.6 {\tiny(0.242)} & 2.89 {\tiny(0.070)} & 10.3 {\tiny(0.262)} & 2.44 {\tiny(0.089)} & 42.1 & 46.1 & 21.1 & 5.20 \\
5 & 3000 & \footnotesize{Weibull$^{[\frac{d-}{d+}]}(\frac{1.59}{1.45}, \frac{0.468}{0.881})$} & 
165 & 8.01 {\tiny(0.236)} & 2.24 {\tiny(0.085)} & 6.29 {\tiny(0.371)} & 1.59 {\tiny(0.111)} & 49.7 & 57.9 & 11.9 & 4.25 \\
5 & 3000 & \footnotesize{Weibull$^{[\frac{d-}{d+}]}(\frac{1.52}{1.33}, \frac{0.633}{1.25})$} & 
92 & 9.77 {\tiny(0.230)} & 2.50 {\tiny(0.092)} & 9.39 {\tiny(0.246)} & 2.20 {\tiny(0.105)} & 42.4 & 48.7 & 14.0 & 4.38 \\
\midrule
\bottomrule
\end{tabular}
\begin{tablenotes}
\footnotesize
\item Note: 
All performance metrics are means across the 5 tests; 
Values in parentheses are the Monte Carlo Standard Errors (MCSE).
\item RMSE = Root Mean Square Error; Cvg. = Coverage; Width = mean 95\% interval width.
\end{tablenotes}
\label{Appendix_table:latent_trait_dgm_3_5}
\end{table}
\subsubsection{Detailed results tables; latent trait: DGMs \#2 \& \#4}
\label{appendix_A_detailed_results_tables_latent_trait_DGM_1}   
\begin{table}[H]
\centering
\caption{Latent trait model performance on DGMs with latent trait structure (DGMs \#2, \#4).}
\small
\setlength{\tabcolsep}{2pt}
\begin{tabular}{c|cc|c|cc|cc|cc|cc}
\toprule
\multirow{2}{*}{\textbf{\scriptsize{DGM}}} &
\multirow{2}{*}{\textbf{$N$}} & 
\multirow{2}{*}{\textbf{Prior}} &
\multirow{2}{*}{\textbf{$N_{\text{sim}}$}} & 
\multicolumn{8}{c}{\textbf{Performance Metrics}} \\
\cmidrule(lr){5-12}
& & & & \multicolumn{2}{c|}{\textbf{RMSE (\%)}} & 
\multicolumn{2}{c|}{\textbf{$|$Bias$|$ (\%)}} &
\multicolumn{2}{c|}{\textbf{Cvg.}} & 
\multicolumn{2}{c}{\textbf{Width}} \\
\cmidrule(lr){5-6} \cmidrule(lr){7-8} \cmidrule(lr){9-10} \cmidrule(lr){11-12}
& & & & \textbf{Se} & \textbf{Sp} & \textbf{Se} & \textbf{Sp} & \textbf{Se} & \textbf{Sp} & \textbf{Se} & \textbf{Sp} \\
\midrule
\midrule
2 & 300 & \footnotesize{Gamma(1, 1)} & 
476 & 7.58 {\tiny(0.248)} & 1.94 {\tiny(0.067)} & 1.72 {\tiny(0.326)} & 0.58 {\tiny(0.083)} & 99.7 & 99.0 & 45.2 & 10.4 \\
2 & 300 & \footnotesize{Weibull(1.45, 0.881)} & 
510 & 7.58 {\tiny(0.241)} & 1.95 {\tiny(0.064)} & 0.97 {\tiny(0.309)} & 0.07 {\tiny(0.083)} & 99.2 & 97.9 & 39.5 & 9.07 \\
2 & 300 & \footnotesize{Weibull(1.33, 1.25)} & 
508 & 8.02 {\tiny(0.241)} & 2.05 {\tiny(0.069)} & 3.74 {\tiny(0.301)} & 0.37 {\tiny(0.088)} & 99.6 & 98.4 & 44.2 & 10.1 \\
\midrule
2 & 3000 & \footnotesize{Gamma(1, 1)} & 
262 & 6.25 {\tiny(0.245)} & 0.91 {\tiny(0.043)} & 2.98 {\tiny(0.329)} & 0.01 {\tiny(0.051)} & 99.5 & 98.9 & 35.7 & 5.06 \\
2 & 3000 & \footnotesize{Weibull(1.45, 0.881)} & 
157 & 4.76 {\tiny(0.243)} & 0.87 {\tiny(0.054)} & 1.88 {\tiny(0.334)} & 0.16 {\tiny(0.062)} & 99.6 & 97.8 & 29.8 & 4.22 \\
2 & 3000 & \footnotesize{Weibull(1.33, 1.25)} & 
201 & 6.09 {\tiny(0.248)} & 0.82 {\tiny(0.044)} & 3.95 {\tiny(0.323)} & 0.02 {\tiny(0.055)} & 99.6 & 99.4 & 33.5 & 4.67 \\
\midrule
\midrule
4 & 300 & \footnotesize{Gamma(1, 1)} & 
408 & 6.60 {\tiny(0.245)} & 2.58 {\tiny(0.088)} & 3.89 {\tiny(0.252)} & 0.18 {\tiny(0.124)} & 98.6 & 98.8 & 32.8 & 12.5 \\
4 & 300 & \footnotesize{Weibull(1.45, 0.881)} & 
320 & 5.89 {\tiny(0.240)} & 2.67 {\tiny(0.109)} & 2.33 {\tiny(0.283)} & 0.23 {\tiny(0.145)} & 98.4 & 97.1 & 27.7 & 11.5 \\
4 & 300 & \footnotesize{Weibull(1.33, 1.25)} & 
445 & 7.39 {\tiny(0.244)} & 2.66 {\tiny(0.093)} & 5.11 {\tiny(0.245)} & 0.31 {\tiny(0.123)} & 97.9 & 98.2 & 33.9 & 12.5 \\
\midrule
4 & 3000 & \footnotesize{Gamma(1, 1)} & 
125 & 3.65 {\tiny(0.237)} & 1.13 {\tiny(0.063)} & 1.57 {\tiny(0.264)} & 0.10 {\tiny(0.091)} & 99.2 & 98.2 & 27.0 & 5.65 \\
4 & 3000 & \footnotesize{Weibull(1.45, 0.881)} & 
68 & 3.16 {\tiny(0.244)} & 1.15 {\tiny(0.097)} & 0.23 {\tiny(0.311)} & 0.09 {\tiny(0.123)} & 99.4 & 94.4 & 19.5 & 4.84 \\
4 & 3000 & \footnotesize{Weibull(1.33, 1.25)} & 
152 & 4.90 {\tiny(0.235)} & 1.12 {\tiny(0.058)} & 3.67 {\tiny(0.250)} & 0.07 {\tiny(0.083)} & 98.8 & 98.3 & 28.1 & 5.74 \\
\midrule
\bottomrule
\end{tabular}
\begin{tablenotes}
\footnotesize
\item Note: 
All performance metrics are means across the 5 tests; 
Values in parentheses are the Monte Carlo Standard Errors (MCSE).
\item RMSE = Root Mean Square Error; Cvg. = Coverage; Width = mean 95\% interval width.
\end{tablenotes}
\label{Appendix_table:table_latent_trait_dgm_2_4}
\end{table}
\newpage
\setcounter{figure}{0}
\setcounter{table}{0}
\renewcommand{\thefigure}{B.\arabic{figure}}
\renewcommand{\thetable}{B.\arabic{table}}
\section{Appendix B - Bias plots}
\label{appendix_B_bias_plots}
\subsection{Bias plots; DGM \#1}
\label{appendix_B_bias_plots_DGM_1}
\begin{figure}[H]
      \centering
    \includegraphics[width=16cm]{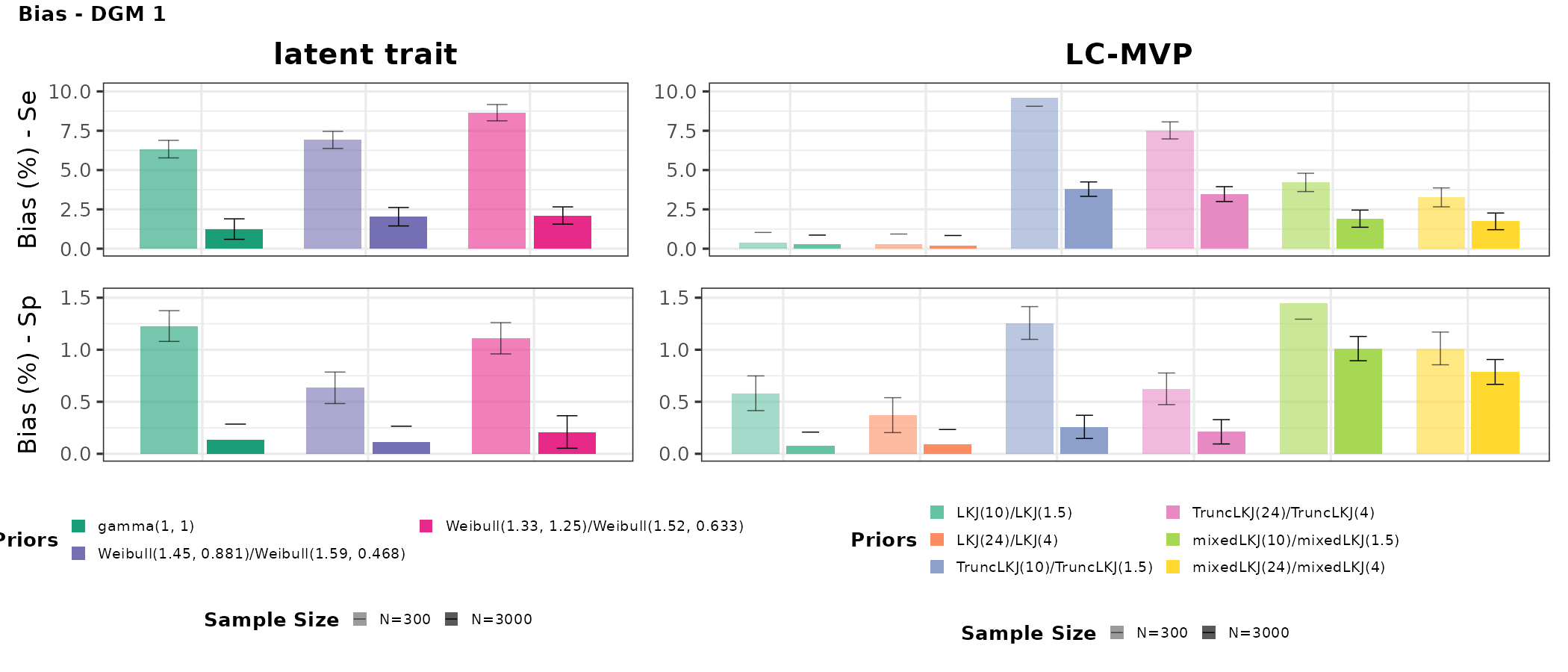}
    \caption{ Simulation study results (Se and Sp) for DGM 1 - Bias }
    \label{Appendix_figure:Sim_study_Bias_DGM_1_}
\end{figure}
\subsection{Bias plots; DGM \#2}
\label{appendix_B_bias_plots_DGM_2}
\begin{figure}[H]
      \centering
    \includegraphics[width=16cm]{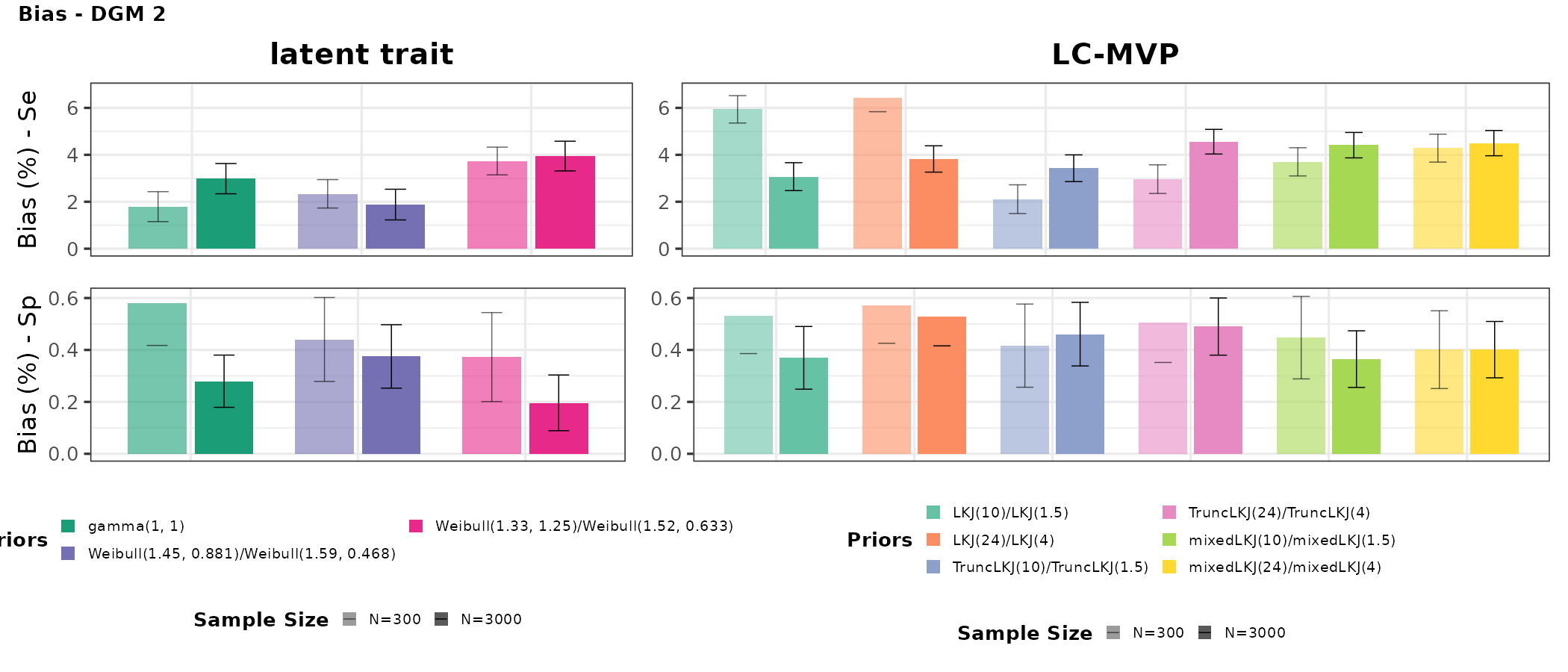}
    \caption{ Simulation study results (Se and Sp) for DGM \#2 - Bias }
    \label{Appendix_figure:Sim_study_Bias_DGM_2_}
\end{figure}
\subsection{Bias plots; DGM \#3}
\label{appendix_B_bias_plots_DGM_3}
\begin{figure}[H]
      \centering
    \includegraphics[width=16cm]{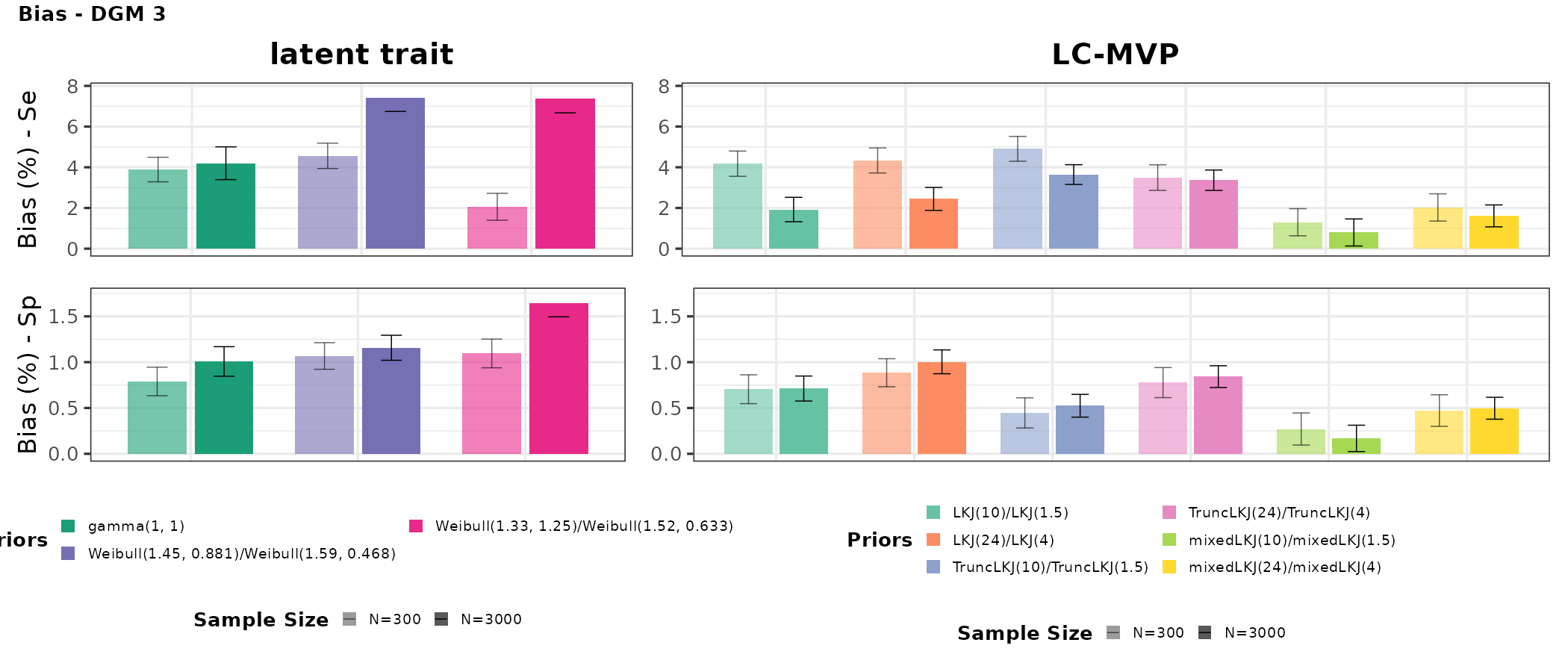}
    \caption{ Simulation study results (Se and Sp) for DGM \#3 - Bias }
    \label{Appendix_figure:Sim_study_Bias_DGM_3_}
\end{figure}
\subsection{Bias plots; DGM \#4}
\label{appendix_B_bias_plots_DGM_4}
\begin{figure}[H]
      \centering
    \includegraphics[width=16cm]{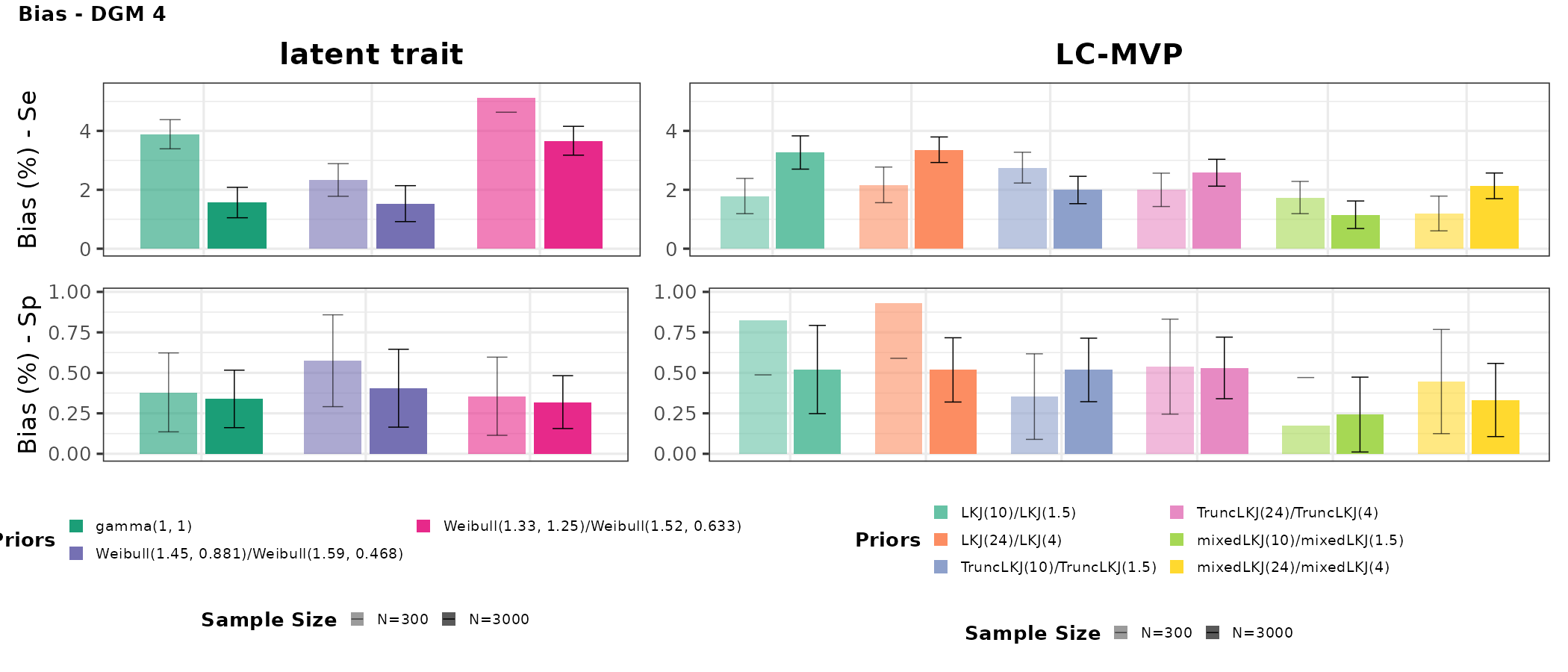}
    \caption{ Simulation study results (Se and Sp) for DGM \#4 - Bias }
    \label{Appendix_figure:Sim_study_Bias_DGM_4_}
\end{figure}
\subsection{Bias plots; DGM \#5}
\label{appendix_B_bias_plots_DGM_5}
\begin{figure}[H]
      \centering
    \includegraphics[width=16cm]{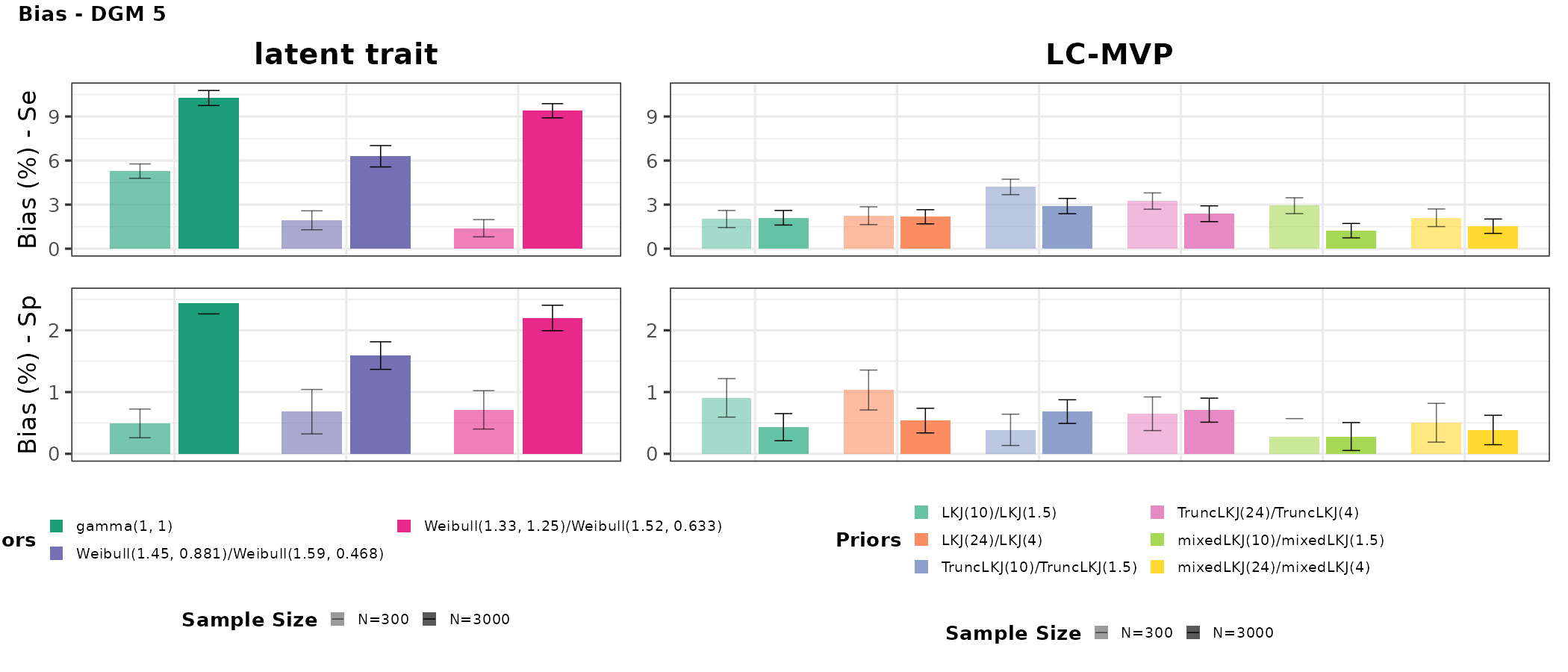}
    \caption{ Simulation study results (Se and Sp) for DGM \#5 - Bias }
    \label{Appendix_figure:Sim_study_Bias_DGM_5_}
\end{figure}
\newpage
\setcounter{figure}{0}
\setcounter{table}{0}
\renewcommand{\thefigure}{C.\arabic{figure}}
\renewcommand{\thetable}{C.\arabic{table}}
\section{Appendix C - Coverage plots}
\label{appendix_C_coverage_plots}
\subsection{Coverage plots; DGM \#1}
\label{appendix_C_coverage_plots_plots_DGM_1}
\begin{figure}[H]
      \centering
    \includegraphics[width=16cm]{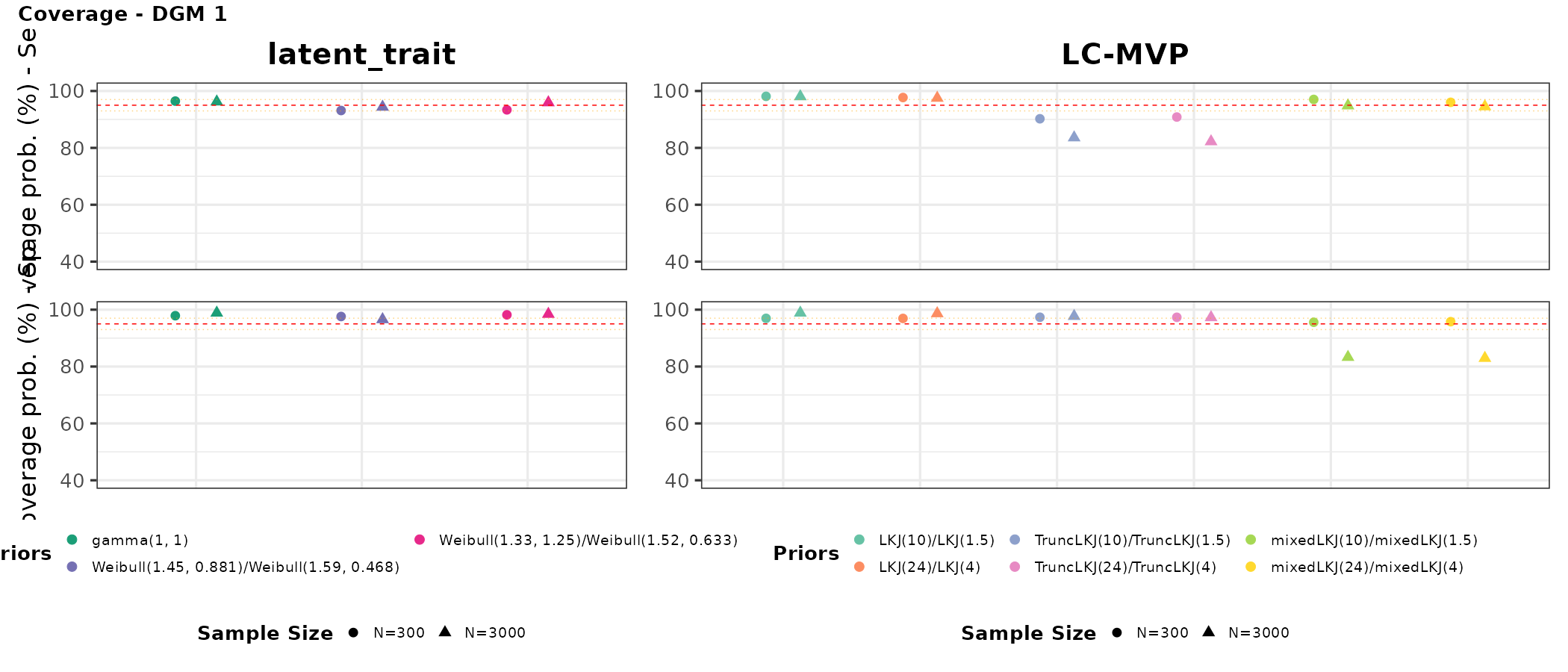}
    \caption{ Simulation study results (Se and Sp) for DGM \#1 - Coverage }
    \label{Appendix_figure:Sim_study_Coverage_DGM_1_}
\end{figure}
\subsection{Coverage plots; DGM \#2}
\label{appendix_C_coverage_plots_plots_DGM_2}
\begin{figure}[H]
      \centering
    \includegraphics[width=16cm]{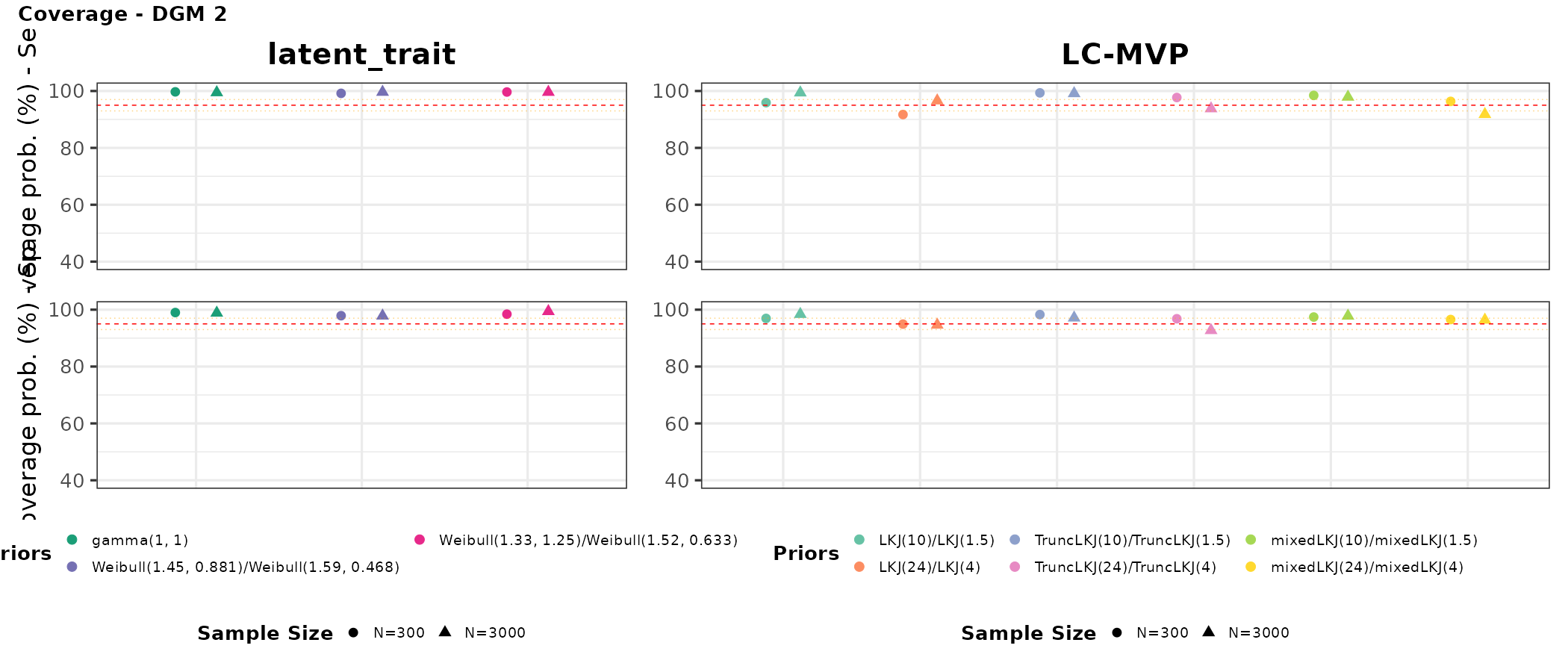}
    \caption{ Simulation study results (Se and Sp) for DGM \#2 - Coverage }
    \label{Appendix_figure:Sim_study_Coverage_DGM_2_}
\end{figure}
\subsection{Coverage plots; DGM \#3}
\label{appendix_C_coverage_plots_plots_DGM_3}
\begin{figure}[H]
      \centering
    \includegraphics[width=16cm]{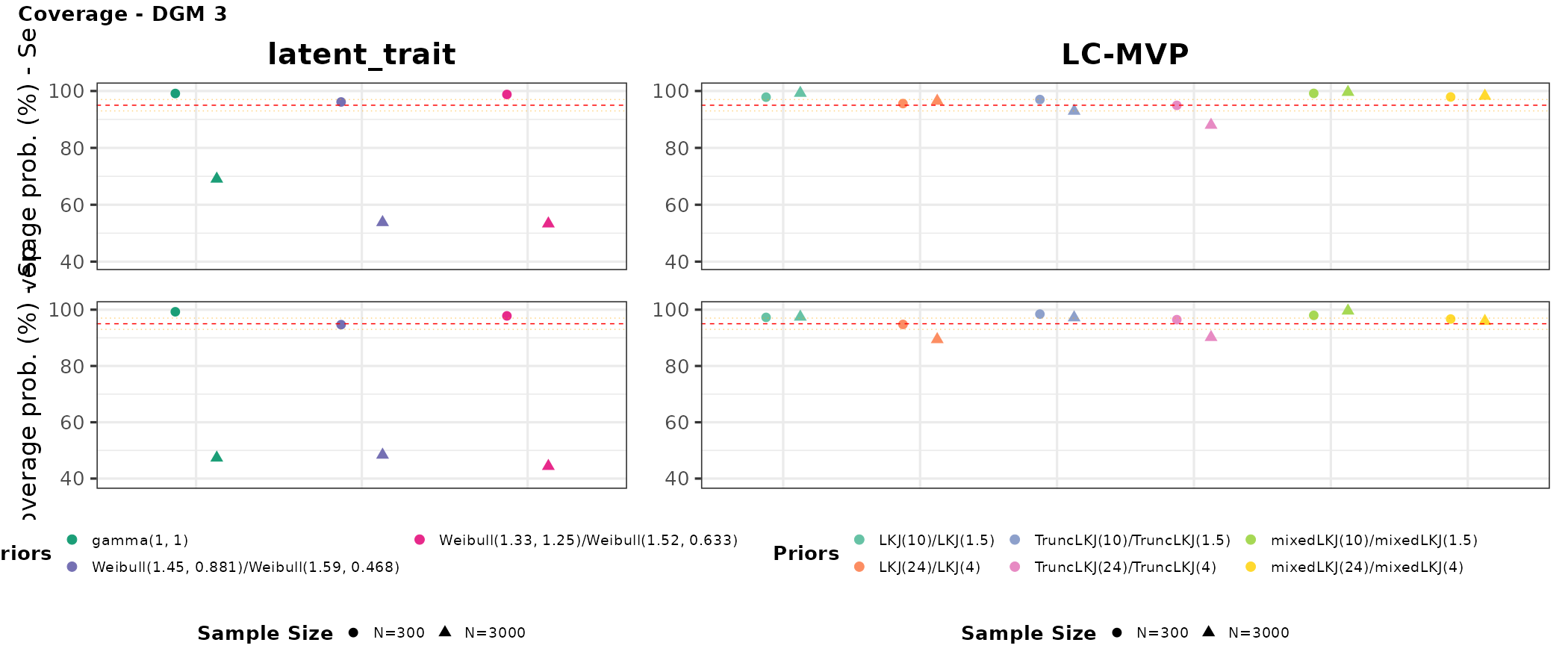}
    \caption{ Simulation study results (Se and Sp) for DGM \#3 - Coverage }
    \label{Appendix_figure:Sim_study_Coverage_DGM_3_}
\end{figure}
\subsection{Coverage plots; DGM \#4}
\label{appendix_C_coverage_plots_plots_DGM_4}
\begin{figure}[H]
      \centering
    \includegraphics[width=16cm]{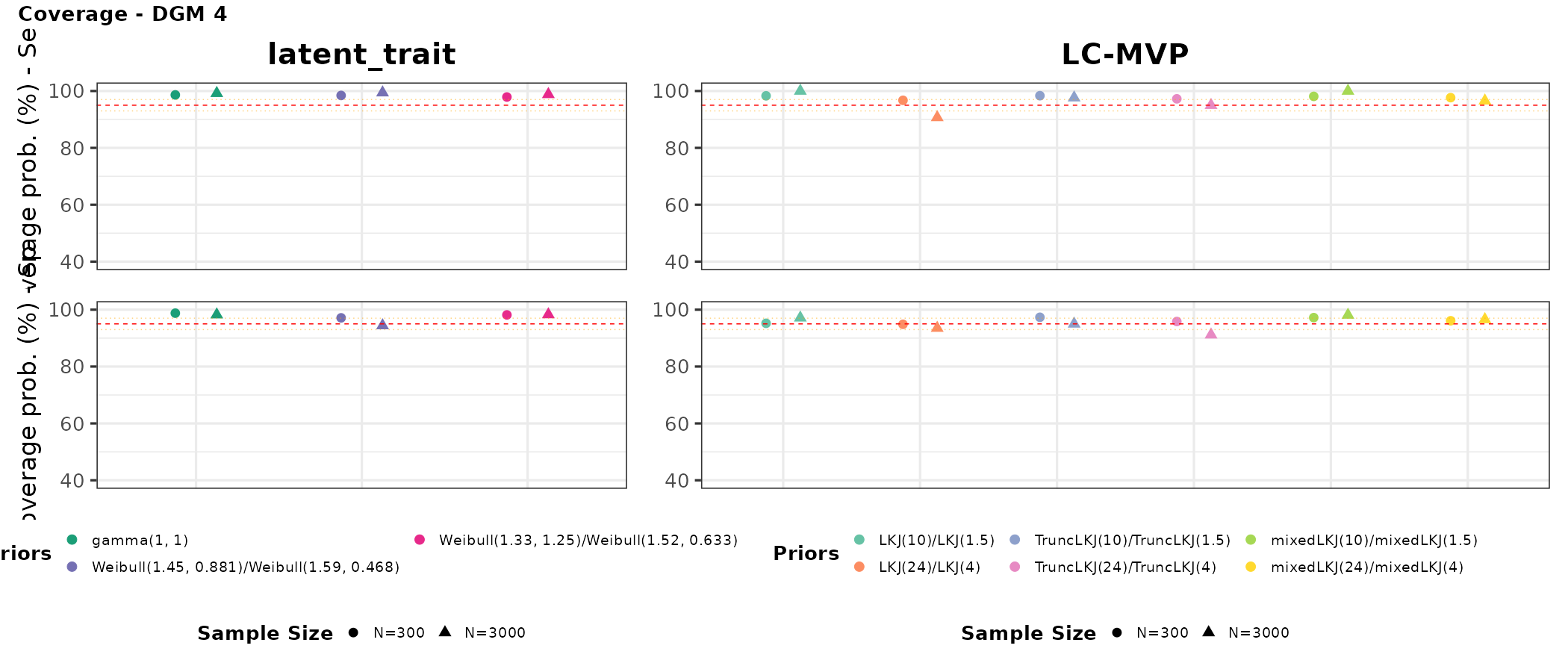}
    \caption{ Simulation study results (Se and Sp) for DGM \#4 - Coverage }
    \label{Appendix_figure:Sim_study_Coverage_DGM_4_}
\end{figure}
\subsection{Coverage plots; DGM \#5}
\label{appendix_C_coverage_plots_plots_DGM_5}
\begin{figure}[H]
      \centering
    \includegraphics[width=16cm]{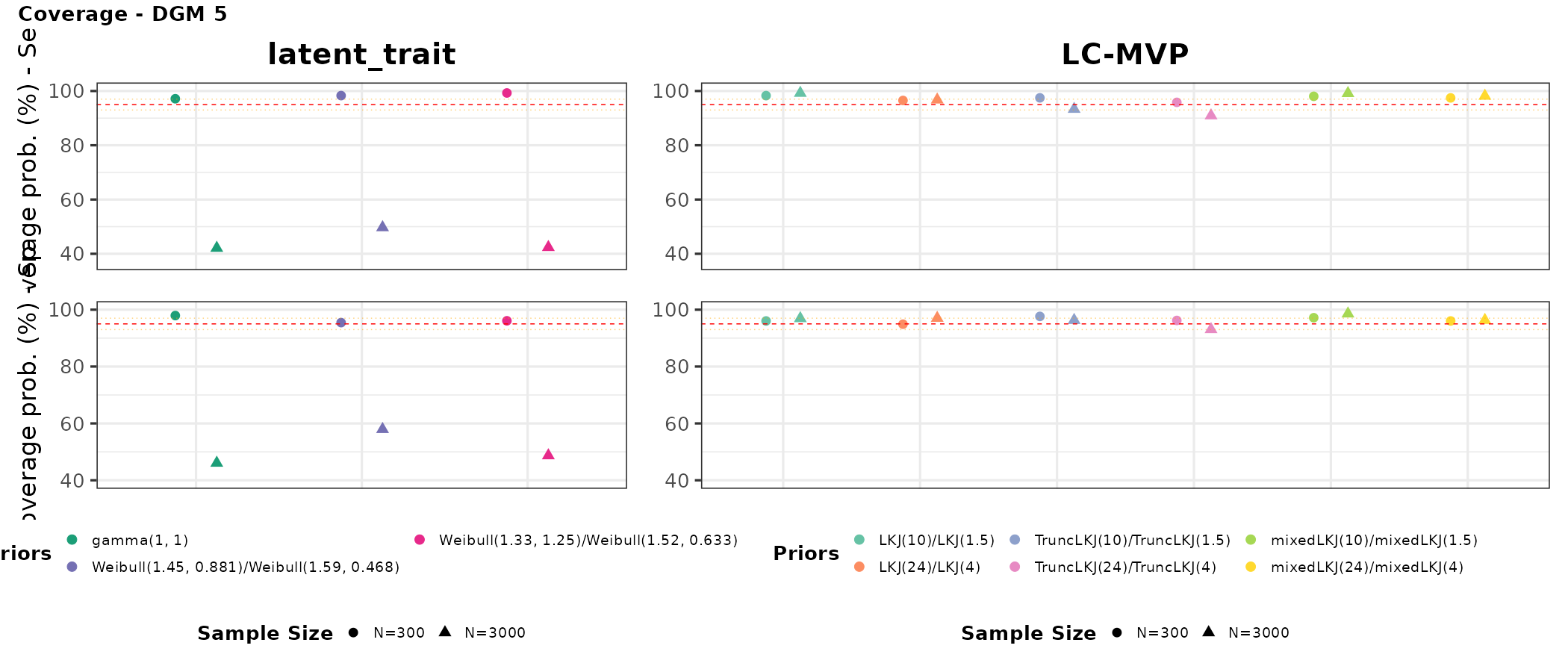}
    \caption{ Simulation study results (Se and Sp) for DGM \#5 - Coverage }
    \label{Appendix_figure:Sim_study_Coverage_DGM_5_}
\end{figure}
\newpage
\setcounter{figure}{0}
\setcounter{table}{0}
\renewcommand{\thefigure}{D.\arabic{figure}}
\renewcommand{\thetable}{D.\arabic{table}}
\section{Appendix D - Interval width plots}
\label{appendix_D_interval_width_plots}
\subsection{Interval width plots; DGM \#1}
\label{appendix_D_interval_width_plots_DGM_1}
\begin{figure}[H]
      \centering
    \includegraphics[width=16cm]{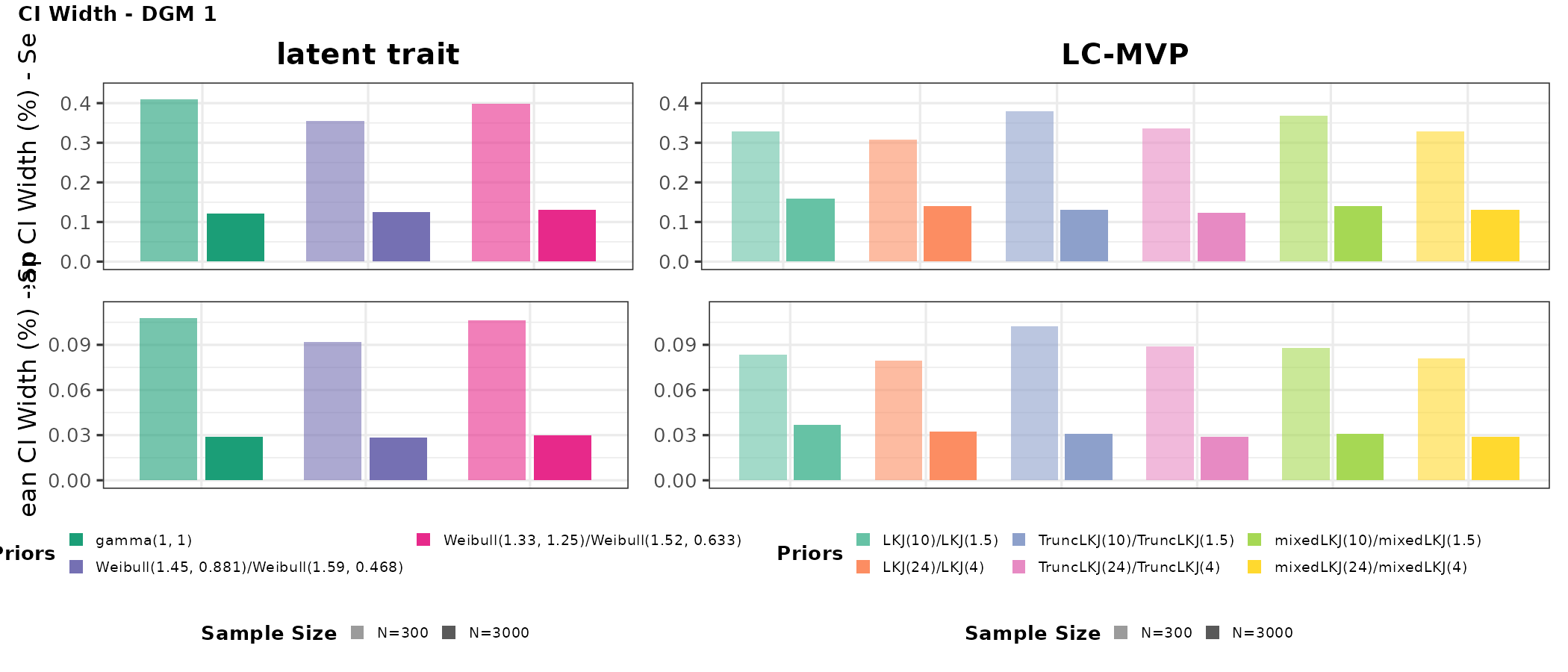}
    \caption{ Simulation study results (Se and Sp) for DGM \#1 - Interval width }
    \label{Appendix_figure:Sim_study_Width_DGM_1_}
\end{figure}
\subsection{Interval width plots; DGM \#2}
\label{appendix_D_interval_width_plots_DGM_2}
\begin{figure}[H]
      \centering
    \includegraphics[width=16cm]{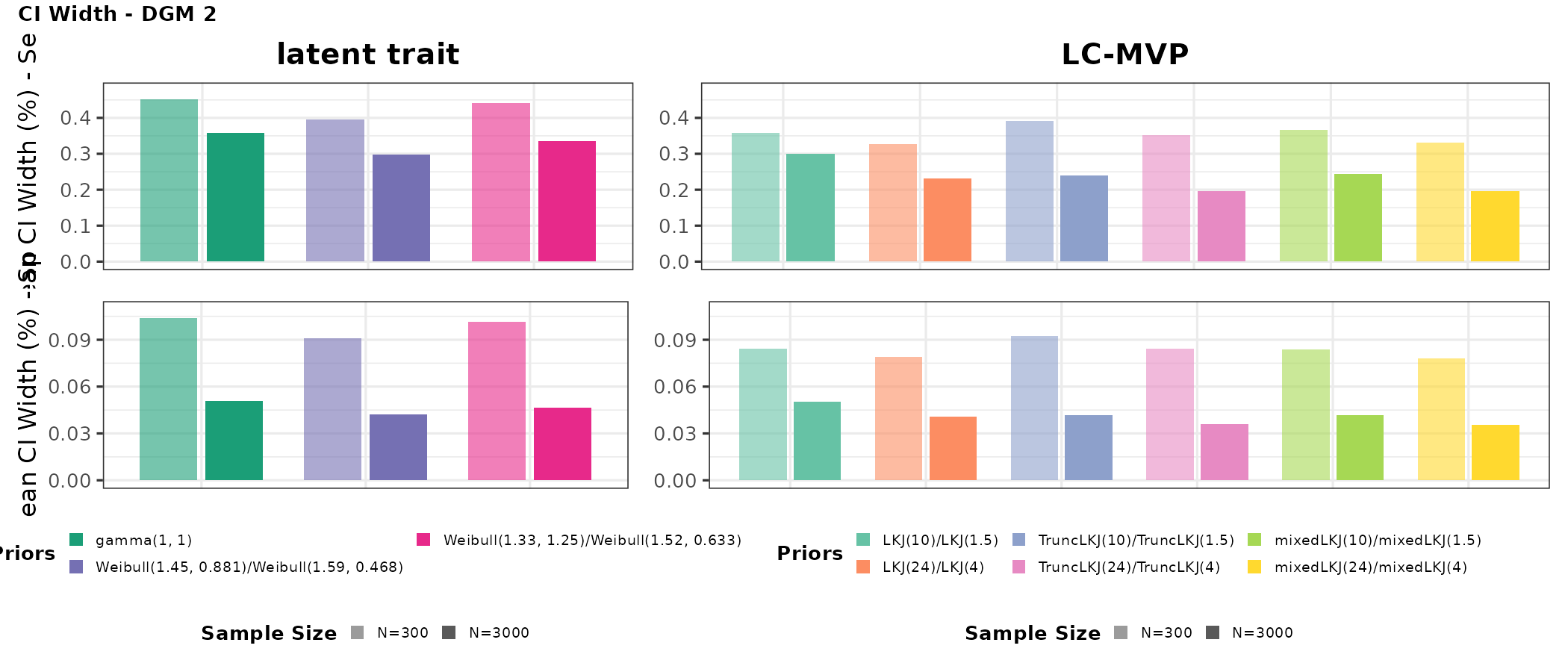}
    \caption{ Simulation study results (Se and Sp) for DGM \#2 - Interval width }
    \label{Appendix_figure:Sim_study_Width_DGM_2_}
\end{figure}
\subsection{Interval width plots; DGM \#3}
\label{appendix_D_interval_width_plots_DGM_3}
\begin{figure}[H]
      \centering
    \includegraphics[width=16cm]{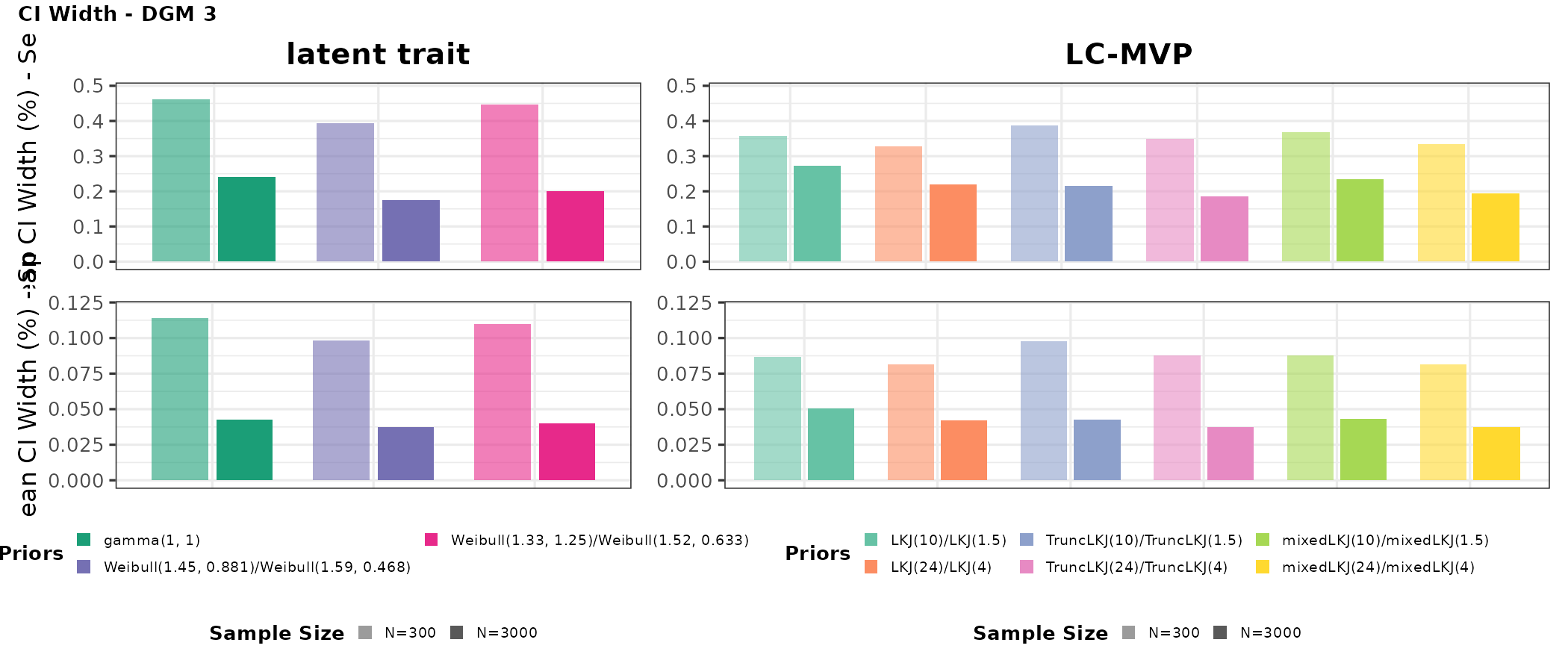}
    \caption{ Simulation study results (Se and Sp) for DGM \#3 - Interval width }
    \label{Appendix_figure:Sim_study_Width_DGM_3_}
\end{figure}
\subsection{Interval width plots; DGM \#4}
\label{appendix_D_interval_width_plots_DGM_4}
\begin{figure}[H]
      \centering
    \includegraphics[width=16cm]{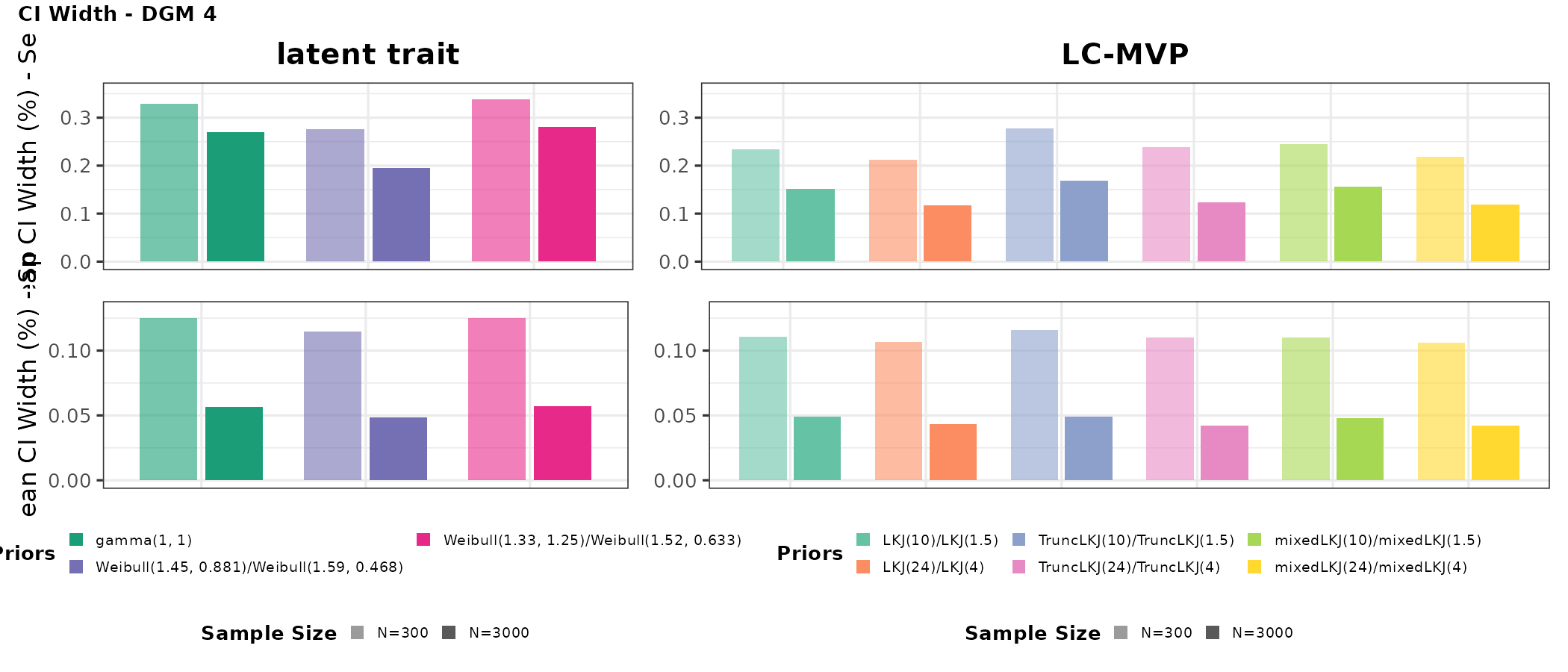}
    \caption{ Simulation study results (Se and Sp) for DGM \#4 - Interval width }
    \label{Appendix_figure:Sim_study_Width_DGM_4_}
\end{figure}
\subsection{Interval width plots; DGM \#5}
\label{appendix_D_interval_width_plots_DGM_5}
\begin{figure}[H]
      \centering
    \includegraphics[width=16cm]{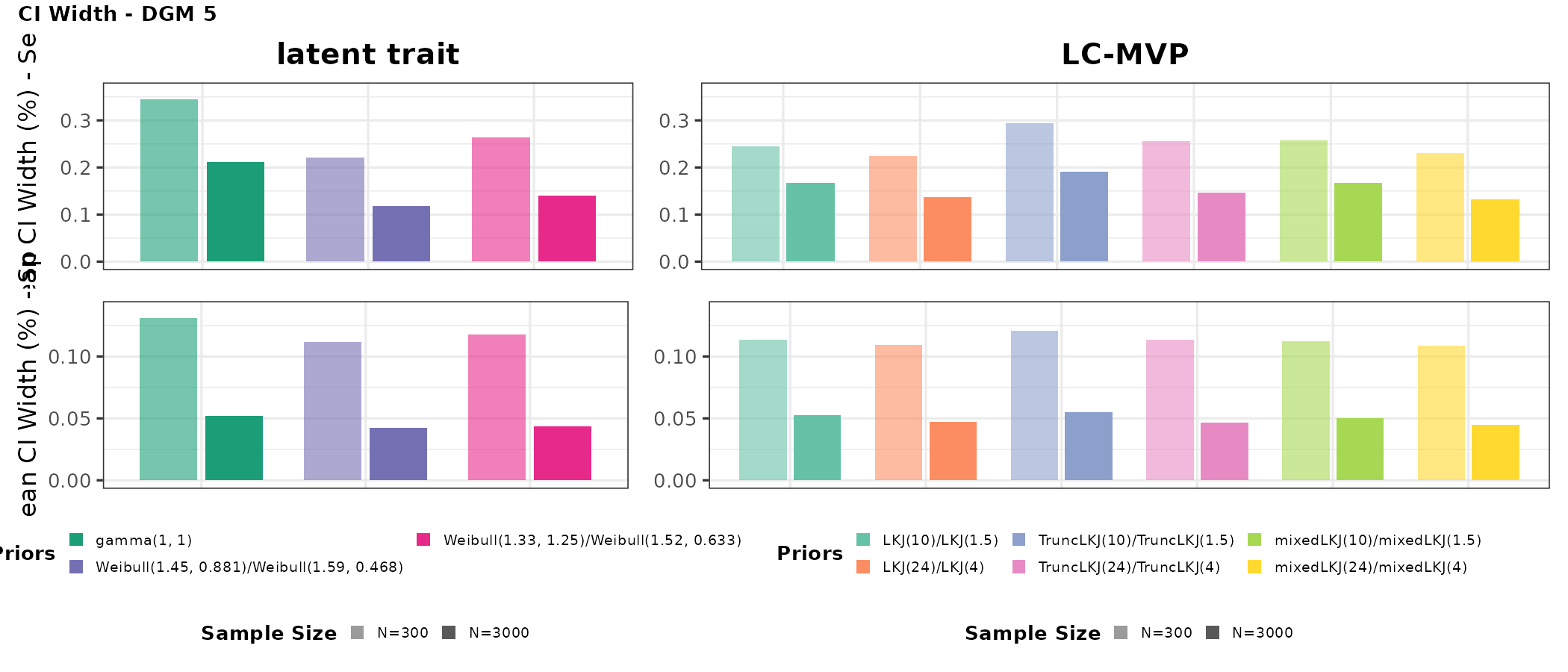}
    \caption{ Simulation study results (Se and Sp) for DGM \#5 - Interval width }
    \label{Appendix_figure:Sim_study_Width_DGM_5_}
\end{figure}

\newpage
\setcounter{figure}{0}
\setcounter{table}{0}
\renewcommand{\thefigure}{E.\arabic{figure}}
\renewcommand{\thetable}{E.\arabic{table}}
\subsection{Appendix E - Efficiency analysis  }
\label{appendix_E_efficiency_analysis}
\begin{table}[H]
\centering
\caption{Computational efficiency for N=300: Median time to 1000 ESS (seconds)}
\footnotesize
\begin{tabular}{lcccc}
\toprule
\textbf{DGM} & \textbf{Model} & \textbf{Prior} & \textbf{Time (SD)} & \textbf{n} \\
\midrule
1 & LC-MVP & LKJ(24,4) & 6.1 (0.7) & 430 \\
4 & LC-MVP & LKJ(24,4) & 6.2 (0.4) & 214 \\
5 & LC-MVP & LKJ(24,4) & 6.4 (0.6) & 246 \\
2 & LC-MVP & LKJ(24,4) & 6.4 (0.7) & 599 \\
3 & LC-MVP & LKJ(24,4) & 6.6 (0.8) & 559 \\
1 & LC-MVP & LKJ(10,1.5) & 8.2 (1.0) & 426 \\
2 & LC-MVP & LKJ(10,1.5) & 8.7 (2.5) & 565 \\
4 & LC-MVP & LKJ(10,1.5) & 8.8 (1.6) & 213 \\
3 & LC-MVP & LKJ(10,1.5) & 9.1 (1.5) & 526 \\
3 & LC-MVP & mixedLKJ(24,4) & 9.2 (2.0) & 437 \\
5 & LC-MVP & LKJ(10,1.5) & 9.2 (1.6) & 247 \\
5 & LC-MVP & mixedLKJ(24,4) & 9.3 (1.4) & 251 \\
2 & LC-MVP & mixedLKJ(24,4) & 9.5 (2.1) & 536 \\
4 & LC-MVP & mixedLKJ(24,4) & 9.5 (1.5) & 231 \\
1 & LC-MVP & mixedLKJ(24,4) & 9.5 (2.7) & 485 \\
4 & Latent trait & Weibull(1.45,0.881) & 9.9 (3.9) & 320 \\
3 & LC-MVP & mixedLKJ(10,1.5) & 11.4 (14.5) & 406 \\
2 & Latent trait & Weibull(1.45,0.881) & 11.4 (4.0) & 509 \\
3 & LC-MVP & TruncLKJ(24,4) & 11.4 (7.6) & 547 \\
2 & LC-MVP & mixedLKJ(10,1.5) & 11.5 (13.1) & 479 \\
1 & Latent trait & Weibull(1.45,0.881) & 11.7 (6.2) & 583 \\
2 & LC-MVP & TruncLKJ(24,4) & 11.7 (5.9) & 575 \\
5 & Latent trait & Weibull(1.45,0.881) & 12.3 (68.0) & 189 \\
5 & LC-MVP & TruncLKJ(24,4) & 12.8 (18.0) & 354 \\
1 & LC-MVP & mixedLKJ(10,1.5) & 12.9 (13.3) & 518 \\
5 & LC-MVP & mixedLKJ(10,1.5) & 13.2 (5.7) & 276 \\
4 & LC-MVP & TruncLKJ(24,4) & 13.7 (29.0) & 298 \\
1 & LC-MVP & TruncLKJ(24,4) & 14.4 (17.0) & 588 \\
4 & LC-MVP & mixedLKJ(10,1.5) & 14.8 (8.0) & 264 \\
4 & Latent trait & Weibull(1.33,1.25) & 15.2 (30.3) & 444 \\
3 & Latent trait & Weibull(1.45,0.881) & 21.3 (189.8) & 756 \\
2 & Latent trait & Weibull(1.33,1.25) & 22.3 (115.1) & 506 \\
5 & Latent trait & Weibull(1.33,1.25) & 25.8 (236.5) & 276 \\
5 & LC-MVP & TruncLKJ(10,1.5) & 34.2 (65.8) & 388 \\
2 & LC-MVP & TruncLKJ(10,1.5) & 40.1 (64.4) & 522 \\
3 & LC-MVP & TruncLKJ(10,1.5) & 41.2 (99.5) & 532 \\
1 & Latent trait & Weibull(1.33,1.25) & 41.5 (426.4) & 627 \\
3 & Latent trait & Weibull(1.33,1.25) & 50.9 (408.7) & 686 \\
4 & LC-MVP & TruncLKJ(10,1.5) & 73.9 (73.8) & 352 \\
5 & Latent trait & Gamma(1,1) & 80.2 (214.5) & 512 \\
1 & LC-MVP & TruncLKJ(10,1.5) & 87.6 (185.8) & 600 \\
3 & Latent trait & Gamma(1,1) & 93.2 (176.0) & 616 \\
4 & Latent trait & Gamma(1,1) & 112.9 (557.3) & 407 \\
2 & Latent trait & Gamma(1,1) & 119.8 (267.6) & 476 \\
1 & Latent trait & Gamma(1,1) & 145.1 (606.4) & 601 \\
\bottomrule
\end{tabular}
\begin{tablenotes}
\footnotesize
\item Note: 
"Time" here means the estimated time to which the minimum number of effective samples (ESS) - min(ESS) - is equal to 1000.
\end{tablenotes}
\label{Appendix_table:efficiency_N300_full}
\end{table}

\begin{table}[H]
\centering
\caption{Computational efficiency for N=3000: Median time to 1000 ESS (seconds)}
\footnotesize
\begin{tabular}{lcccc}
\toprule
\textbf{DGM} & \textbf{Model} & \textbf{Prior} & \textbf{Time (SD)} & \textbf{n} \\
\midrule
5 & Latent trait & Weibull(1.45,0.881) & 78.3 (101.5) & 165 \\
4 & LC-MVP & LKJ(24,4) & 86.5 (28.7) & 70 \\
2 & LC-MVP & LKJ(24,4) & 91.1 (19.7) & 193 \\
3 & LC-MVP & LKJ(24,4) & 93.7 (24.6) & 159 \\
5 & LC-MVP & LKJ(24,4) & 102.6 (22.1) & 74 \\
3 & Latent trait & Weibull(1.45,0.881) & 107.1 (145.9) & 556 \\
4 & LC-MVP & mixedLKJ(24,4) & 107.4 (303.0) & 64 \\
4 & Latent trait & Weibull(1.45,0.881) & 109.1 (169.9) & 68 \\
5 & LC-MVP & mixedLKJ(24,4) & 115.1 (133.1) & 65 \\
5 & Latent trait & Weibull(1.33,1.25) & 119.6 (70.4) & 92 \\
1 & Latent trait & Weibull(1.45,0.881) & 122.7 (23.9) & 64 \\
1 & LC-MVP & mixedLKJ(24,4) & 137.5 (48.1) & 95 \\
1 & LC-MVP & LKJ(24,4) & 155.2 (22.8) & 87 \\
3 & LC-MVP & mixedLKJ(24,4) & 162.5 (128.6) & 147 \\
3 & Latent trait & Weibull(1.33,1.25) & 176.7 (300.1) & 446 \\
2 & Latent trait & Weibull(1.45,0.881) & 182.1 (81.0) & 157 \\
2 & LC-MVP & mixedLKJ(24,4) & 187.0 (82.2) & 165 \\
3 & LC-MVP & LKJ(10,1.5) & 205.5 (64.1) & 162 \\
1 & Latent trait & Weibull(1.33,1.25) & 209.0 (44.6) & 64 \\
5 & LC-MVP & LKJ(10,1.5) & 210.7 (82.1) & 77 \\
2 & LC-MVP & LKJ(10,1.5) & 224.3 (65.7) & 200 \\
1 & LC-MVP & LKJ(10,1.5) & 231.2 (53.0) & 124 \\
4 & Latent trait & Weibull(1.33,1.25) & 244.1 (198.5) & 152 \\
4 & LC-MVP & mixedLKJ(10,1.5) & 248.7 (417.8) & 64 \\
4 & LC-MVP & LKJ(10,1.5) & 268.8 (85.6) & 42 \\
5 & LC-MVP & mixedLKJ(10,1.5) & 278.1 (397.6) & 70 \\
2 & Latent trait & Weibull(1.33,1.25) & 281.9 (94.9) & 200 \\
1 & LC-MVP & mixedLKJ(10,1.5) & 310.3 (186.5) & 97 \\
3 & LC-MVP & mixedLKJ(10,1.5) & 328.7 (214.3) & 97 \\
5 & Latent trait & Gamma(1,1) & 350.1 (695.6) & 226 \\
2 & LC-MVP & mixedLKJ(10,1.5) & 356.6 (236.6) & 172 \\
1 & LC-MVP & TruncLKJ(10,1.5) & 459.5 (1676.8) & 121 \\
4 & Latent trait & Gamma(1,1) & 477.1 (1150.5) & 125 \\
3 & Latent trait & Gamma(1,1) & 554.5 (653.7) & 520 \\
1 & Latent trait & Gamma(1,1) & 575.6 (1039.9) & 53 \\
3 & LC-MVP & TruncLKJ(24,4) & 702.9 (533.4) & 157 \\
2 & LC-MVP & TruncLKJ(24,4) & 736.7 (626.7) & 165 \\
1 & LC-MVP & TruncLKJ(24,4) & 753.3 (943.9) & 116 \\
5 & LC-MVP & TruncLKJ(24,4) & 762.8 (648.6) & 75 \\
4 & LC-MVP & TruncLKJ(24,4) & 844.1 (711.4) & 68 \\
2 & LC-MVP & TruncLKJ(10,1.5) & 858.7 (958.9) & 140 \\
4 & LC-MVP & TruncLKJ(10,1.5) & 956.6 (731.8) & 68 \\
2 & Latent trait & Gamma(1,1) & 955.8 (643.9) & 262 \\
3 & LC-MVP & TruncLKJ(10,1.5) & 972.9 (881.7) & 142 \\
5 & LC-MVP & TruncLKJ(10,1.5) & 1229.8 (1707.0) & 81 \\
\end{tabular}
\begin{tablenotes}
\footnotesize
\item Note: 
"Time" here means the estimated time to which the minimum number of effective samples (ESS) - min(ESS) - is equal to 1000.
\end{tablenotes}
\label{Appendix_table:efficiency_N3000_full}
\end{table}


\printbibliography

\end{document}